\documentclass[prd,twocolumn,nofootinbib,showpacs]{revtex4-1}

\usepackage{amsfonts}
\usepackage{amsmath}
\usepackage{amssymb}
\usepackage{bm}
\usepackage{dcolumn}
\usepackage{graphicx}
\usepackage{graphics}
\usepackage[latin1]{inputenc}
\usepackage{latexsym}
\usepackage{rotating}
\usepackage{hyperref}
\usepackage[all]{hypcap} 
\usepackage{xspace} 
\usepackage[usenames,dvipsnames]{color}
\usepackage{mathrsfs}
\usepackage{subfigure}
\usepackage{aas_macros}
\usepackage{amsfonts}
\usepackage{booktabs}
\usepackage{siunitx}

\usepackage{ulem}
\usepackage{outlines}
\usepackage{enumitem}
\setenumerate[1]{label=\Roman*.}
\setenumerate[2]{label=\Alph*.}
\setenumerate[3]{label=\roman*.}
\setenumerate[4]{label=\alph*.}

\definecolor {darkgreen}{rgb}{0.2,0.7,0.2}

\newcommand\be{\begin{equation}}
\newcommand\ba{\begin{eqnarray}}
\newcommand\ee{\end{equation}}
\newcommand\ea{\end{eqnarray}}

\newcommand\bw{\begin{widetext}}
\newcommand\ew{\end{widetext}}

\newcommand{\nn}{\nonumber}

\newcommand{\maxtxt}{\text{\rm \tiny max}}

\newcommand{\implicit}{\text{\rm \tiny implicit}}
\newcommand{\explicit}{\text{\rm \tiny explicit}}
\newcommand{\SPA}{\text{\rm \tiny SPA}}
\newcommand{\exact}{\text{\rm \tiny exact}}

\begin{document}
\title{A Fourier Domain Waveform for Non-Spinning Binaries with Arbitrary Eccentricity}

\author{Blake Moore}
\affiliation{eXtreme Gravity Institute, Department of Physics, Montana State University, Bozeman, MT 59717, USA.}

\author{Travis Robson}
\affiliation{eXtreme Gravity Institute, Department of Physics, Montana State University, Bozeman, MT 59717, USA.}

\author{Nicholas Loutrel}
\affiliation{eXtreme Gravity Institute, Department of Physics, Montana State University, Bozeman, MT 59717, USA.}

\author{Nicol\'as Yunes}
\affiliation{eXtreme Gravity Institute, Department of Physics, Montana State University, Bozeman, MT 59717, USA.}

\date{\today}

\begin{abstract} 

Although the gravitational waves observed by advanced LIGO and Virgo are consistent with compact binaries in a quasi-circular inspiral prior to coalescence, eccentric inspirals are also expected to occur in Nature.  
Due to their complexity, we currently lack ready-to-use, analytic waveforms in the Fourier domain valid for sufficiently high eccentricity, and such models are crucial to coherently extract weak signals from the noise.    
We here take the first steps to derive and properly validate an analytic waveform model in the Fourier domain that is valid for inspirals of arbitrary orbital eccentricity. 
As a proof-of-concept, we build this model to leading post-Newtonian order by combining the stationary phase approximation, a truncated sum of harmonics, and an analytic representation of hypergeometric functions. 
Through comparisons with numerical post-Newtonian waveforms, we determine how many harmonics are required for a faithful (matches above 99\%) representation of the signal up to orbital eccentricities as large as 0.9.
As a first byproduct of this analysis, we present a novel technique to maximize the match of eccentric signals over time of coalescence and phase at coalescence.
As a second byproduct, we determine which of the different approximations we employ leads to the largest loss in match, which could be used to systematically improve the model because of our analytic control.  
The future extension of this model to higher post-Newtonian order will allow for an accurate and fast phenomenological hybrid that can account for arbitrary eccentricity inspirals and mergers.

\end{abstract}

\pacs{04.30.-w,04.25.-g,04.25.Nx}


\maketitle

\section{Introduction}
\label{intro}

Eccentric binaries circularize rapidly as their orbital separation shrinks due to the emission of gravitational waves (GWs). Since the target sources of ground-based detectors, such as the Advanced Laser Interferometer Gravitational-Wave Observatory (aLIGO) \cite{2015CQGra..32g4001L}, advanced Virgo (aVirgo)~\cite{TheVirgo:2014hva}, LIGO-India~\cite{indigo}, and KAGRA~\cite{Aso:2013eba}, are thought to form at large initial separations, one expects their orbital separation and eccentricity will have decreased considerably by the time they are detectable. As such, the modeling of GWs has focused on quasi-circular binaries, and indeed all current aLIGO/aVirgo detections can be captured well with quasi-circular GW models \cite{GW150914, GW151226, GW170104, GW170814, GW170817, GW170608}.

Several astrophysical scenarios, however, suggest that some small number of binaries could have moderate eccentricities while emitting GWs at frequencies in the sensitivity band of ground-based detectors, and these different formation scenarios can be constrained through detection of eccentric signals~\cite{Nishizawa:2016}. A very small number of weakly eccentric sources emitting detectable GWs are expected to be formed through isolated stellar evolution (field binaries). Kowalska et~al.~\cite{2011A&A...527A..70K} simulated field binary evolution and found typical eccentricities of $\sim 10^{-4}$, with roughly $1\%$ of binaries having eccentricities greater than 0.01 when emitting GWs detectable by ground based networks.

In contrast to binaries formed through isolated stellar evolution, binaries formed in dense stellar regions, such as globular clusters, are significantly more likely to be eccentric due to many-body interactions, such as the Kozai-Lidov mechanism~\cite{2013PhRvL.111f1106S, 2013arXiv1308.5682A, 2013ApJ...773..187N, 2014ApJ...781...45A}. The latter is a form of orbital resonance where oscillations in inclination and eccentricity are induced in a hierarchical triple~\cite{1962AJ.....67..591K}. Recently, Rodriguez et~al.~\cite{2018PhRvL.120o1101R} (see also Samsing~\cite{2018PhRvD..97j3014S}) incorporated post-Newtonian (PN) effects\footnote{The post-Newtonian approximation is one in which the field equations are solved assuming small velocities and weak gravitational fields in an expansion in powers of $(\frac{v}{c})$, where $v$ is the orbital velocity and $c$ is the speed of light \cite{blanchet-review}. By $n$PN order we mean an expansion to order $({v}/{c})^{2 n}$.} in orbital dynamics and found that $10\%$ of binaries in globular clusters emitting GWs in the sensitivity band of ground based detectors will have eccentricities greater than $0.1$. For more studies which focus on the eccentricity distribution of sources for ground-based detectors see~\cite{2009MNRAS.395.2127O, Samsing:2013kua, 2015arXiv150905080A}.

Since there may be some small number of detectable binaries with non-negligible eccentricity, it is natural to consider what error is incurred by neglecting eccentricity in the modeling.  Martel and Poisson \cite{1999PhRvD..60l4008M} computed the fitting factor (FF), i.e.~the overlap maximized over all parameters of the model, between quasi-circular templates and eccentric signals for a variety of sources at leading PN order. This study showed that as the eccentricity of the signal increases and the total mass decreases, the FF decreases. Since the percent loss in detection rate scales like $1-FF^{3}$ \cite{2008CQGra..25k4033A}, neglecting a moderate eccentricity in source modeling can lead to a significant loss in detection rate.  Loss in detection rate has also been the focus of several other studies \cite{Huerta:2013qb, 2010PhRvD..81b4007B, 2009CQGra..26d5013C}, which varied the range of masses considered and the PN order, all leading to similar conclusions: for low mass systems, the loss in match due to sub-optimal templates is significant when $e\geq 0.05$, and for higher mass systems when  $e\geq 0.1$. 

But even if an eccentric signal is detected with quasi-circular templates, the lack of eccentricity modeling will nevertheless lead to an associated parameter bias. Favata \cite{2013arXiv1310.8288F} showed that the systematic error in the symmetric mass ratio incurred by neglecting eccentricity in the model exceeds the statistical error of aLIGO for initial eccentricities as small as $e \sim 2 \times 10^{-3}$ for a binary neutron star system. For the third-generation Einstein Telescope \cite{et}, there is significant parameter error for even smaller orbital eccentricities. Schematically, this is because parameter biases become important when the match (M), i.e. the overlap without maximizing over intrinsic parameters, between quasi-circular templates and eccentric signals drops below $1-D/(2 \rho^{2})$~\cite{2017PhRvD..95j4004C}, where $D$ is the effective dimensionality of the model and $\rho$ is the signal-to-noise ratio of the detection. Thus, even though the FF may be high, the M may still not be high enough for high signal-to-noise ratio events.   

The need for eccentric waveforms has therefore encouraged some research in eccentric modeling. The Post-Circular formalism (PC), introduced at Newtonian order in \cite{PhysRevD.80.084001}, was one of the first attempts to provide an analytic frequency domain waveform that incorporated eccentricity. The philosophy of this formalism is to expand all relevant quantities in the small eccentricity limit. Tanay, et~al.~\cite{Tanay:2016zog} extended this work to 2PN order, and Ref.~\cite{Moore:2016qxz} extended it to 3PN order, but keeping only leading-order in eccentricity corrections. While the PC models are computationally fast, they are not able to handle moderate eccentricities, and so other modeling efforts have combined analytic and numerical methods to arrive at a more accurate model. Pierro et~al.~\cite{1996NCimB.111..631P} solved the equations of motion necessary to build the Fourier domain model of~\cite{2012PhRvD..86j4027M} without making a small eccentricity approximation by combining special functions (hypergeometric functions and Bessel functions) with certain numerical inversions. While exact in some regards within the PN approximation, this model is computationally expensive and it has only been extended to 1PN order so far \cite{2015PhRvD..92d4038M}. 

We here take the first steps toward the construction and validation of a ready-to-use and computationally-efficient waveform model in the Fourier domain that is valid to arbitrary eccentricity. The new model combines the accuracy of~\cite{1996NCimB.111..631P} with the efficiency of the PC models, without requiring the evaluation of costly special functions in the Fourier phase of the frequency response. Instead, the model is constructed by combining elements of the stationary-phase approximation, a truncated sum over harmonics and an analytic representation of hypergeometric functions. Schematically, the Fourier transform of the plus- and cross-polarizations in the new model is
\begin{align}
\label{eq:SPA-schematic}
\tilde{h}_{+,\times}(f) &= \frac{m \eta}{R} \sum_{j=1}^{N} \tilde{{\cal{A}}}^{(j)}_{+,\times}(f) \;
 e^{i\psi_j(f)} \,\,.
\end{align}
where $\tilde{{\cal{A}}}^{(j)}_{+,\times}$ is an analytic, slowly-varying, complex Fourier amplitude and $\psi_{j}$ is an analytic, rapidly-varying, real Fourier phase, where $j$ is the harmonic index and $N$ is the truncation index, with $m$, $\eta$ and $R$ the total mass, symmetric mass ratio and luminosity distance to the source respectively. 

\begin{figure}[htp]
\includegraphics[clip=true,angle=0,width=0.475\textwidth]{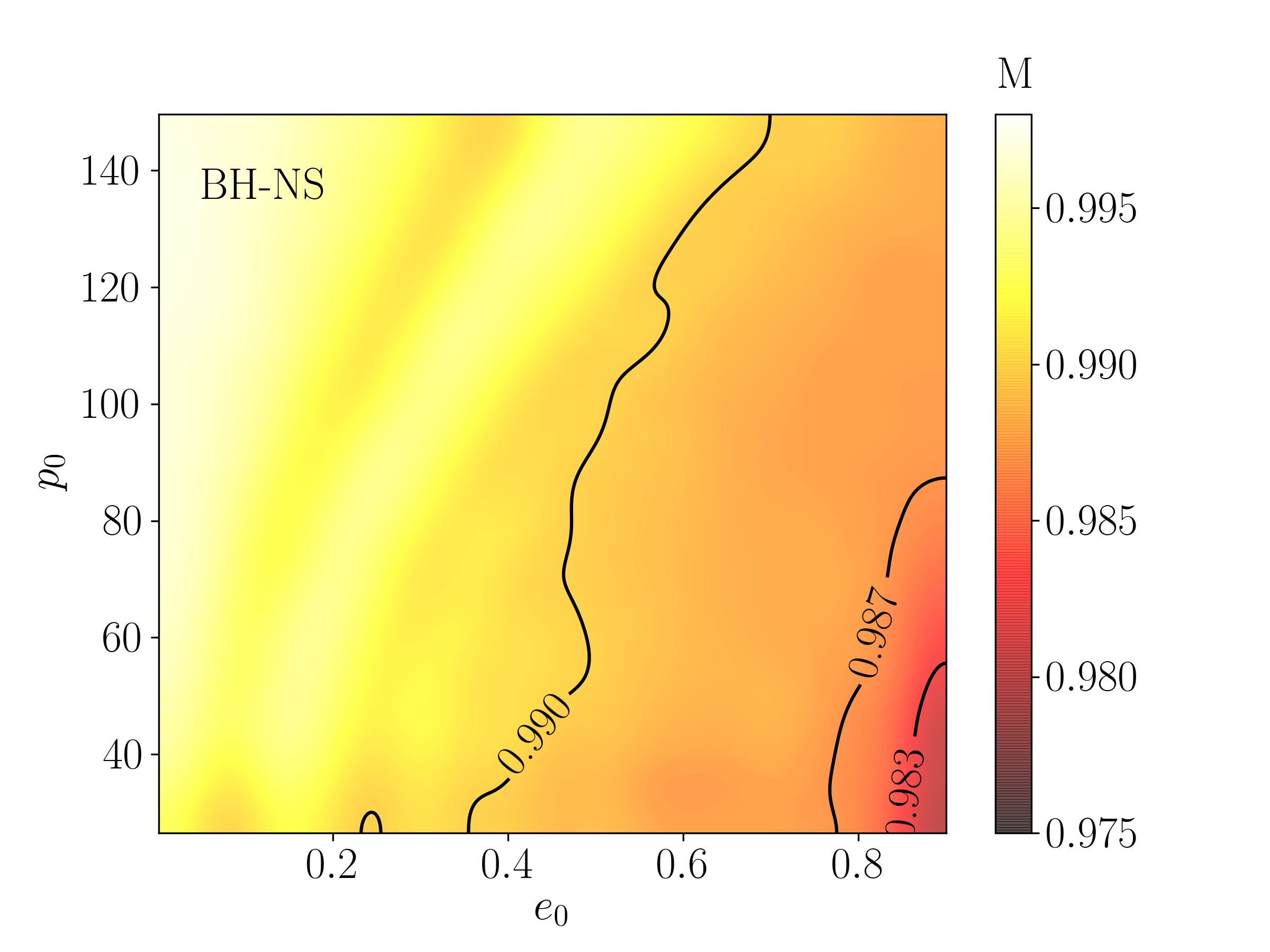}
\caption{\label{fig:nsbhmatch} Match between our Fourier domain model and a fully numerical PN waveform as a function of initial orbital eccentricity, $e_0$, and initial dimensionless semilatus rectum, $p_0$, for a $(10,1.4)M_{\odot}$ black hole - neutron star binary (BH-NS). The match is greater than 99\% for more than half of the explored parameter space. The decay in match at high initial eccentricity and small initial semilatus rectum is due to finite time effects.}
\end{figure}

The key of the new model is an analytic prescription for the Fourier amplitude and for the Fourier phase, with the truncation index determined from a match analysis. Although the Fourier phase can be solved for exactly in terms of hypergeometric functions, this representation is not computationally efficient. Instead, we explored different analytic representations of hypergeometric functions, and found that Taylor expansions about small eccentricity do an exceptional job at capturing the exact result. The Fourier amplitude, on the other hand, is not expanded in small eccentricity, and it is instead kept in its exact PN form. The truncation index is determined by requiring that the match between the truncated series and an infinite series be at least 99\%. We find in practice that for most initial eccentricity cases the sum need only be taken to the tenth term or less. We then verify that the resulting waveforms are faithful (with matches $\sim 99$\% to numerical PN waveforms) for aLIGO sources with initial orbital eccentricities as high as 0.9, as shown in Fig.~\ref{fig:nsbhmatch} for a black hole--neutron star binary (BH-NS). All throughout, we work to leading PN order, focusing mostly on faithfulness measures for the aLIGO detector, but the approach can easily be extended to higher PN order and to other detectors.  

Two main byproducts are also generated from this analysis. First, we develop a new method to efficiently maximize the overlap over the time and phase of coalescence of the new eccentric model. Maximization over these extrinsic parameters is a solved problem for quasi-circular binaries, but the later must be generalized non-trivially when including several harmonics with comparable power. Second, we investigate which elements of the approximations that make up our new model leads to the largest loss in accuracy. This error analysis therefore allows us to identify which elements should be taken to higher order if a higher match is desired. Due to the analytic control of the waveform model, such extensions to higher order are straightforward. 

The remainder of this paper presents the details of the results described above. 
Section~\ref{sec:funds} reviews the fundamentals of eccentric GW emission, including the parameterization of the orbit and its time evolution, as well as the decomposition of the signal into a sum of harmonics of the mean orbital frequency. 
Section~\ref{sec:maximization} reviews some basic data-analysis measures to compare waveform models, while presenting the new method to efficiently maximize over the time and phase of coalescence of eccentric templates.  
Section~\ref{sec:models} discusses the details of previous models and it introduces the new models we develop in this paper.
Section~\ref{sec:Err} studies which of the different approximations used in the new model leads to the largest loss in match. 
Section~\ref{sec:faithharms} carries out a faithfulness study of the new analytic model. 
Section~\ref{sec:futwork} concludes and points to future research. 
Throughout this work we use geometric units ($c=1=G$).

\section{Fundamentals of Eccentric Binary GW Emission}
\label{sec:funds}
In this section we begin by reviewing the Newtonian parameterization of the Kepler problem in the absence of radiation reaction. We show that the associated time domain GW waveform can be decomposed into a sum of harmonics of the mean orbital frequency. We then review how radiation reaction affects the orbital dynamics and how the application of the SPA leads to a Fourier response with similar structure to the time domain harmonic decomposition. 
\subsection{Newtonian Emission in the Absence of Radiation Reaction}
\label{sub:funds}
 In the Newtonian treatment of the two-body problem, the dynamics of an elliptical orbit restricted to a plane are described by: 
 \allowdisplaybreaks[4] 
 \begin{subequations}
  \label{eq:newtellipse}
 \begin{align}
 r &= a(1-e\cos u), \label{eq:orbsep}\\
 \phi-\phi_{0} &= 2\arctan\left[\left(\frac{1+e}{1-e}\right)^{1/2}\tan \frac{u}{2}\right], \\
  l &= 2\pi F(t-t_{0})=u-e\sin u. \label{eq:keps}
 \end{align}
 \end{subequations} 
Here $r$ is the magnitude of the relative separation vector, which we choose to be on the $x$--$y$ plane $\vec{r}=(r\cos \phi,r\sin \phi,0)$, $\phi$ is the orbital phase, $e$ is the orbital eccentricity, $m$ is the total mass, and $F$ is the mean orbital frequency defined by $F=1/P$, where $P$ is the orbital period. The semi-major axis, $a$, is related to the mean orbital frequency via Kepler's third law: $(2\pi F)^2a^3=m$. The angles $l$ and $u$ are the mean anomaly and eccentric anomaly, respectively, while the constants $t_{0}$ and $\phi_{0}$ arise from integration and specify the initial orientation at some time $t_{0}$. From the above equations, one can also easily derive the following differential equations 
 \begin{subequations}
  \label{eq:newtellipse-2}
 \begin{align}
 \dot{r} &= (2\pi mF)^{1/3} \frac{e \sin u}{(1-e \cos u)}, \\
 \dot{\phi} &= \frac{2\pi F (1+e \cos \phi)^{2}}{(1-e^2)^{3/2}}. \label{eq:phidot}
 \end{align}
 \end{subequations}
Even in the Newtonian treatment, one is unable to analytically solve for the orbital separation and phase as explicit functions of time. Instead, one is forced to numerically invert Kepler's equation, Eq.~\eqref{eq:keps}, in order to obtain the eccentric anomaly as a function of time, and thus the orbital separation and phase as functions of time. Alternatively, one can solve the differential equations presented above to obtain the orbital phase and the separation distance as a function of time. 

In General Relativity, the accelerated motion of massive bodies leads to the emission of GWs. Following Martel and Poisson~\cite{1999PhRvD..60l4008M}, the GW polarizations are given by
\allowdisplaybreaks
\begin{align}
h_{+} &= -\frac{m \eta}{p R}\bigg\lbrace\left[2\cos(2\phi - 2\beta) + \frac{5}{2}e\cos(\phi-2\beta) \right.\nonumber\\
&\left.+ \frac{1}{2}e\cos(3\phi-2\beta)+e^{2}\cos 2\beta\right](1+\cos^{2}\iota) \nonumber\\
&+\left[e\cos\phi + e^{2}\right]\sin^{2}\iota \bigg\rbrace\,\,,
\label{eq:plus_strain}
\end{align}
and
\begin{align}
h_{\times} &= -\frac{m \eta}{p R} \bigl[4 \sin(2\phi-2\beta)+5e\sin(\phi-2\beta) \bigr. \nonumber\\
&\bigl.+e\sin(3\phi-2\beta)-2e^{2}\sin 2\beta \bigr]\cos\iota \,\,,
\label{eq:cross_strain}
\end{align}
where $R$ is the luminosity distance and $\eta =  m_{1} m_{2}/m^{2}$ is the symmetric mass ratio, with $m_{1,2}$ the component masses. The angles $\beta$ and $\iota$ are the polar angles describing the polarization axes. The \textit{dimensionless} semilatus rectum, $p$, is related to the eccentricity and semi-major axis by $a=pm/(1-e^2)$. Following Moreno-Garrido, et~al.~\cite{1995MNRAS.274..115M} we decompose the time domain signal into harmonics of the mean orbital frequency such that the signal takes the form
\begin{equation}
\label{eq:timedomdecomp}
h_{+,\times}(t) = -\frac{m\eta}{R}\left(2\pi m F\right)^{2/3}\sum_{j=1}^{\infty}\left[C^{(j)}_{+,\times}\cos jl + S^{(j)}_{+,\times}\sin jl \right]  \, .
\end{equation}
The harmonic coefficients $C^{(j)}_{+,\times}$ and $S^{(j)}_{+,\times}$ are functions of the orbital eccentricity $e$ and the polarization angles $(i, \beta)$.

\begin{figure*}[htp]
\includegraphics[clip=true,angle=0,width=0.475\textwidth]{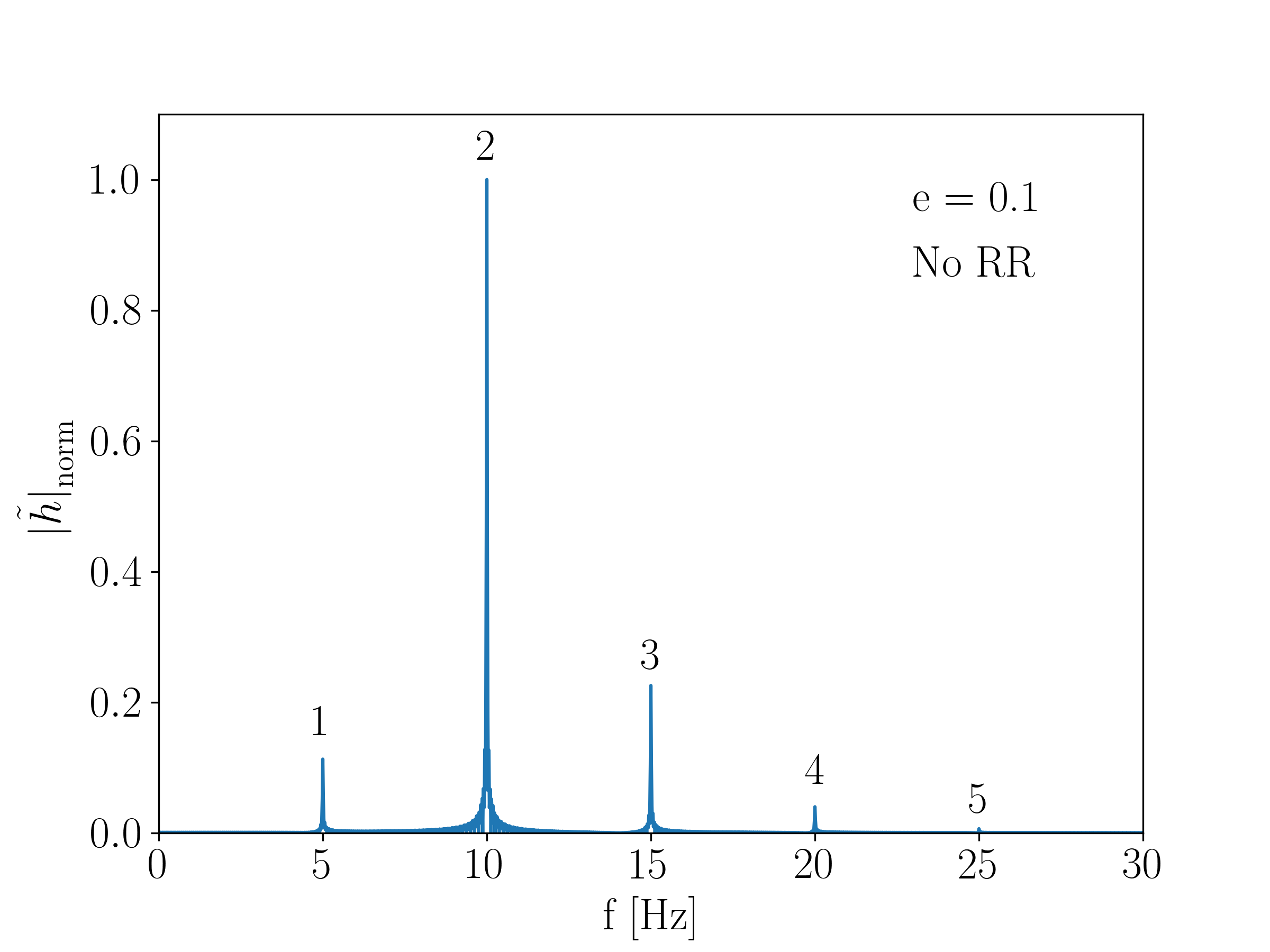} 
\includegraphics[clip=true,angle=0,width=0.475\textwidth]{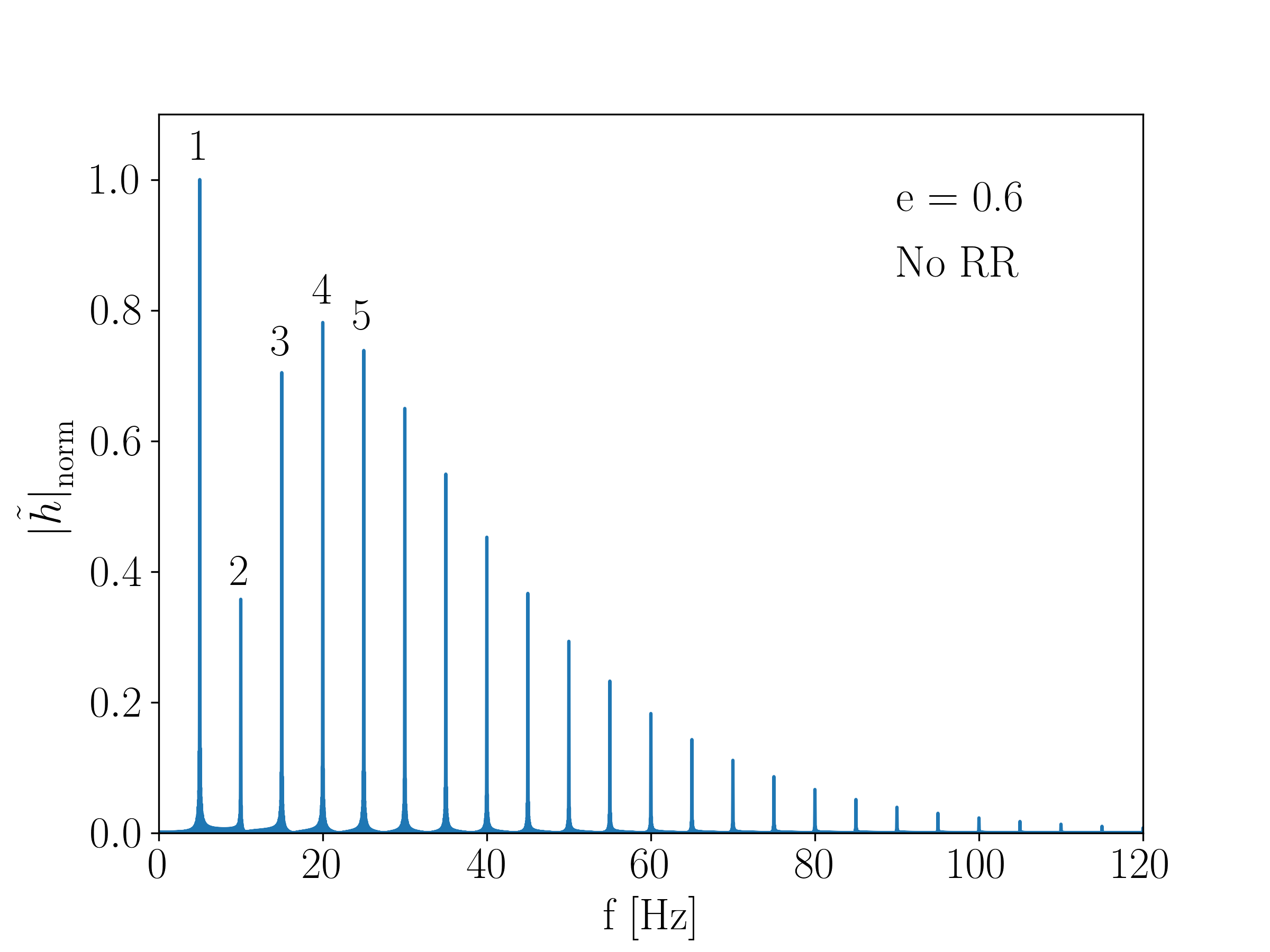} 
\caption{\label{fig:no_RR}(Color Online) The normalized Fourier amplitude of the numerically evolved GW signal in the absence of radiation reaction, obtained by numerically solving Eq.~\eqref{eq:phidot} and discretely Fourier transforming Eq.~\eqref{eq:cross_strain} for a $(10,10)M_{\odot}$ binary black hole (BBH) system with a mean orbital frequency of $5$~Hz and orbital eccentricity of $0.1$ (left) and $0.6$ (right). Observe that the Fourier amplitude naturally splits into harmonics of the mean orbital frequency, where we have labeled the first five. For systems with larger orbital eccentricities, there are many harmonics of comparable strength present.}
\end{figure*}

We now briefly review how these coefficients are obtained. Making use of the relation $\cos\phi = e^{-1}\left[a(1-e^{2})/r-1\right]$, we express the strain polarizations in Eqs.~\eqref{eq:plus_strain} and~\eqref{eq:cross_strain} in the form 
\begin{align}
\label{eq:preexp}
h_{+,\times} &= B_1\cos\phi + B_2\left(\frac{a}{r} \right)^{2}\cos\phi   +  B_3\sin\phi 
\nonumber \\ &
+ B_4\left(\frac{a}{r} \right)^{2}\sin\phi \,\,.
\end{align}
Here the $B_i$ are functions of the orbital eccentricity, mean orbital frequency, and the angles $\iota$ and $\beta$, which can be found in Appendix A of \cite{1995MNRAS.274..115M}. Neglecting radiation-reaction, we have access to the following Fourier series~\cite{Kovalevsky}\,,
\begin{subequations}
\label{eq:bessexp}
	\begin{align}
	\cos\phi &= -e + \frac{2}{e}(1-e^{2})\sum_{j=1}^{\infty}J_{j}(je)\cos jl \,\,, \\
	\sin\phi &= \sqrt{1-e^{2}} \sum_{j=1}^{\infty}\left[J_{j-1}(j e) - J_{j+1}(j e)\right] \sin j l \,\,,\\
	\left(\frac{a}{r}\right)^{2}\cos\phi&=\sum_{j=1}^{\infty}j \left[J_{j-1}(j e) - J_{j+1}(j e)\right] \cos j l \,\,,\\
	\left(\frac{a}{r}\right)^{2}\sin\phi &=\sum_{j=1}^{\infty}j \left[J_{j-1}(j e) + J_{j+1}(j e)\right] \sin j l  \,\,,
	\end{align}
\end{subequations}
where $J_{j}(x)$ are Bessel functions of the first kind. Combining Eqs.~\eqref{eq:preexp}
and~\eqref{eq:bessexp} and rearranging to match the form given in Eq.~\eqref{eq:timedomdecomp} yields the harmonic amplitudes:
\begin{subequations}
\label{eq:amps}
	\begin{align}
	C^{(j)}_{+} &= \frac{2}{e^{2}}\bigg\lbrace c_{2\beta}(1+c_{\iota}^{2})e(1-e^{2})j(J_{j+1}(je)
\nonumber \\	
	& - \left[c_{2\beta}(1+c_{\iota}^{2})(e^{2}-2)+e^{2}s_{\iota}^{2}\right]J_{j}(je) \bigg\rbrace \,\,, \\
	S^{(j)}_{+} &= \frac{4}{e^{2}}(1+c_{\iota}^{2})s_{2\beta}\sqrt{1-e^{2}}\bigg\lbrace e J_{j-1}(je) 
	\nn \\
	&- \left[1+(1-e^{2})j\right]J_{j}(je)\bigg\rbrace \,\, , \\
	C^{(j)}_{\times} &= \frac{4}{e^{2}}s_{2\beta}c_{\iota}\bigg\lbrace2e(1-e^{2})j J_{j-1}(je) 
	\nn \\
	&- 2\left[1 + (1-e^{2})j - \frac{e^{2}}{2} \right]J_{j}(je)\bigg\rbrace \,\,, \\
	S^{(j)}_{\times} &= \frac{8}{e^{2}}c_{2\beta}c_{\iota}\sqrt{1-e^{2}}\bigg\lbrace e J_{j-1} - \left[1 + (1-e^{2})j\right]J_{j}(je)\bigg\rbrace \,\,,
	\end{align}
\end{subequations}
with the notation $c_{\theta} \equiv \cos\theta$ and $s_{\theta} \equiv \sin\theta$. These expressions are exact, and thus, the waveform in Eq.~\eqref{eq:timedomdecomp} is valid to all eccentricities.

Figure \ref{fig:no_RR} shows the normalized amplitude of the Fourier transform of the GW signal in the absence of radiation reaction (i.e.~for a system whose mean orbital frequency and eccentricity remain constant). As the figure shows, the Fourier amplitude is composed of harmonics of the mean orbital frequency $F$. For the small eccentricity case shown on the left panel, the second harmonic is clearly dominant. However, as the eccentricity is increased, as shown on the right panel, the first harmonic of the mean motion dominates and many harmonics are of comparable strength.

Figure \ref{fig:no_RR} demonstrates that one \it cannot \rm specify a time domain quantity, such as the orbital eccentricity or the mean orbital frequency at a unique GW frequency. The presence of multiple harmonics demands that at any given time an eccentric binary emits GWs at several different GW frequencies. For example, although the eccentricity of the emitting binary is 0.6 at all times on the right panel of Fig.~\ref{fig:no_RR}, this system emits GWs with significant power at 5, 10, 15, 20~Hz, etc. As such, there is no one-to-one mapping between eccentricity and GW frequency, and one cannot unambiguously define an eccentricity as a signal ``enters band." A much more sensible statement is to refer to the orbital eccentricity at a given value of the mean orbital frequency, which is uniquely defined.

\subsection{Radiation Reaction and GW Fourier Response}
\label{sub:RR&struct}
\begin{figure*}[htp]
\includegraphics[clip=true,angle=0,width=0.475\textwidth]{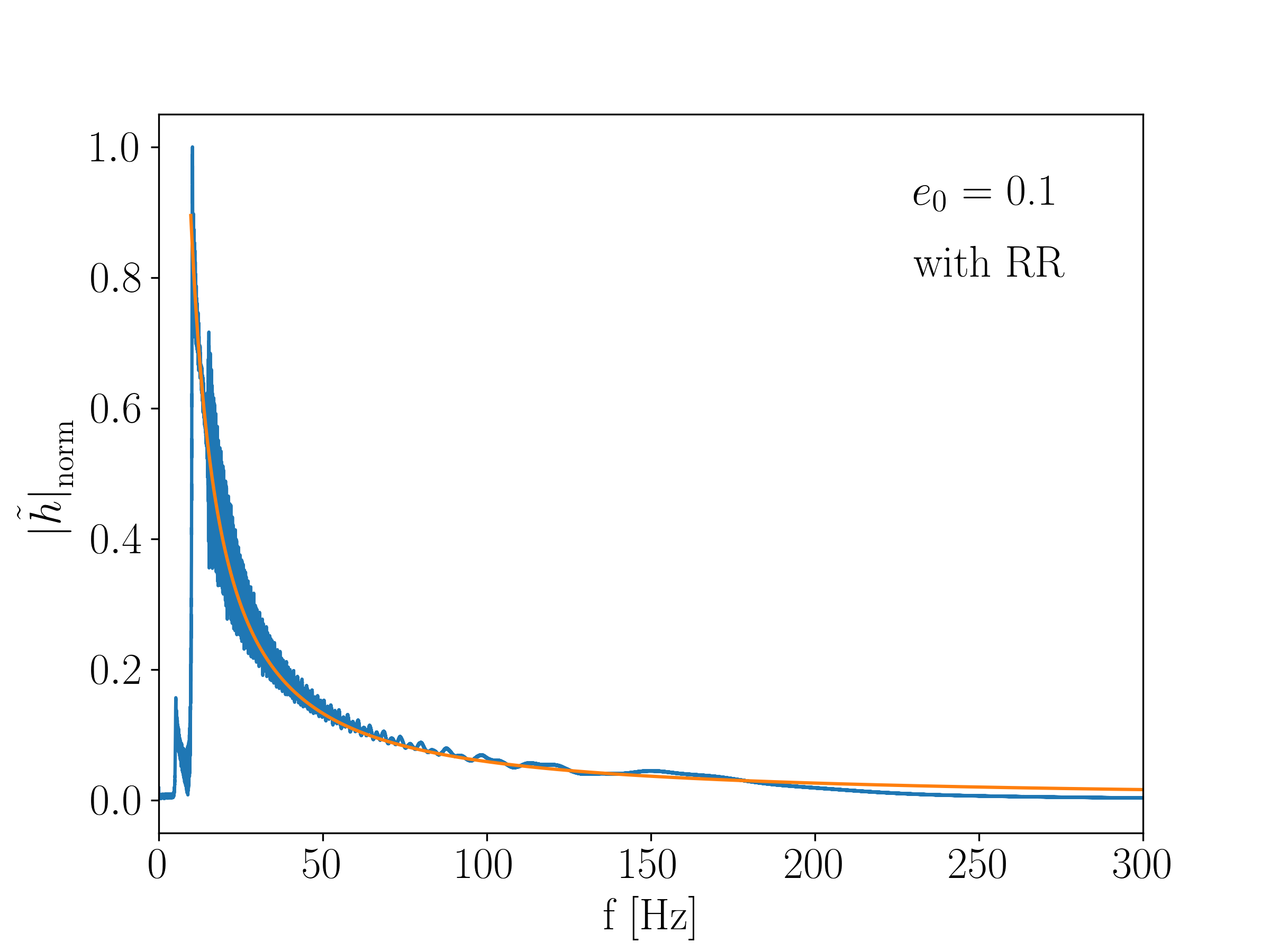} 
\includegraphics[clip=true,angle=0,width=0.475\textwidth]{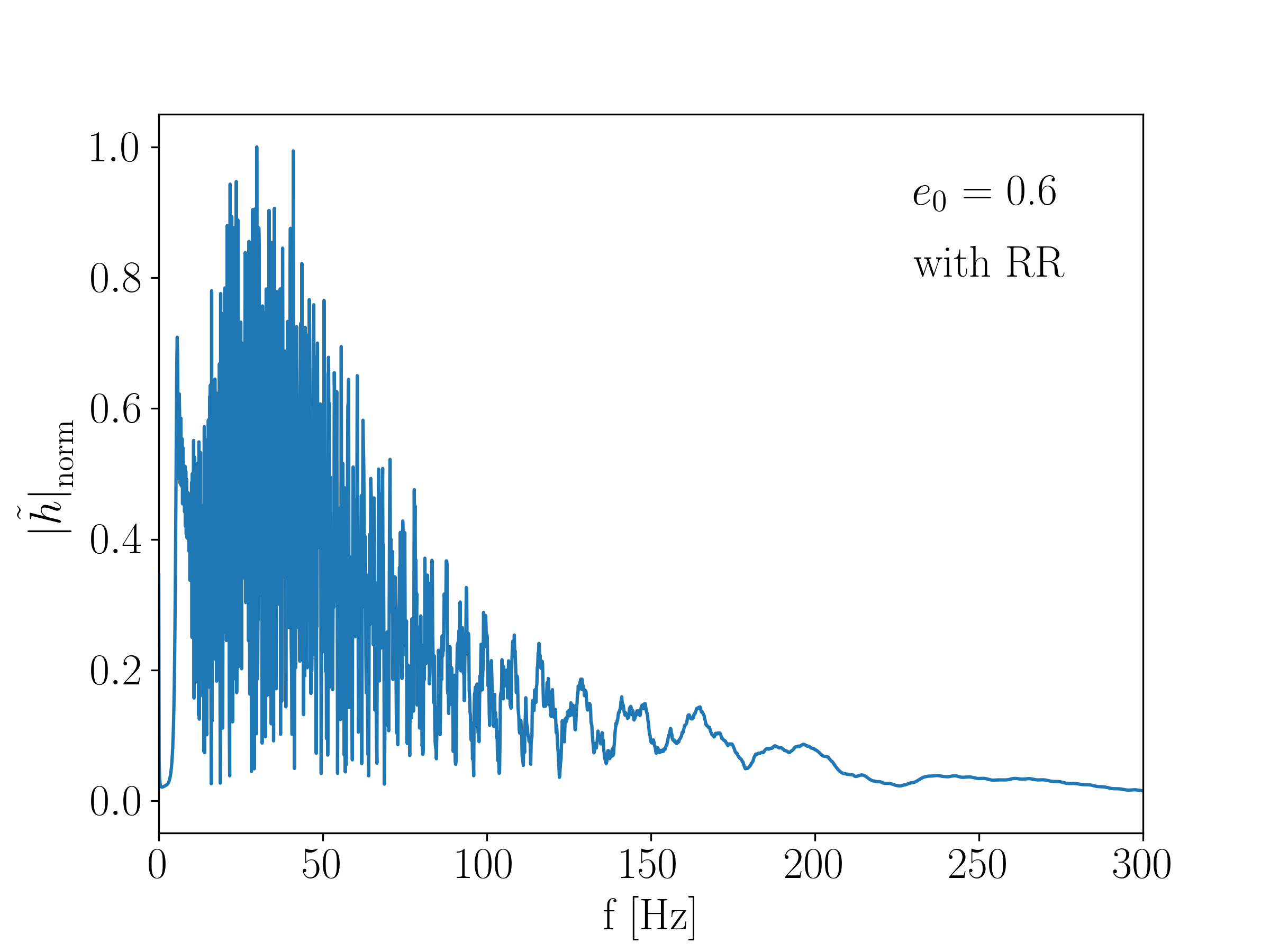} 
\caption{\label{fig:RR_on} (Color Online) The normalized Fourier amplitude of the numerically evolved GW signal in the presence of radiation reaction obtained by numerically solving Eqs.~\eqref{eq:phidot}, \eqref{eq:dFdt}, and \eqref{eq:dedt} for a $(10,10)M_{\odot}$ binary black hole (BBH) system with an initial mean orbital frequency of 5~Hz at an initial orbital eccentricity of $0.1$ (left) and $0.6$ (right). In the low eccentricity case on the left panel, the amplitude is similar to that of a quasi-circular GW (shown in orange), but there is some interference between harmonics above $15$~Hz that leads to small oscillations about the quasi-circular spectrum. In the moderate eccentricity case shown on the right panel, the amplitude displays considerable oscillations from the interference of many harmonics of comparable strength.}
\end{figure*}
Let us now consider the effect of radiation reaction on the emitted GWs in the frequency domain. In General Relativity, GWs carry away energy and momentum from the binary, and in response, $e$ and $F$ vary with time. At leading-order in the PN expansion, the equations for $r$, $\dot{r}$, $\phi$, and $\dot{\phi}$, Eqs.~\eqref{eq:newtellipse}-\eqref{eq:newtellipse-2}, remain the same, as does Kepler's equation, but the mean anomaly $l$ now obeys
 \begin{equation}
 \label{eq:meananom}
 l=\int^t 2\pi F(t')dt'.
 \end{equation} 
The time evolution of the mean orbital frequency and eccentricity were first derived in \cite{PetersMathews}, and are given by
\begin{equation}
\label{eq:dFdt}
\frac{dF}{dt}=\frac{\eta}{2\pi m^2}(2\pi mF)^{11/3}\left(\frac{96+292e^2+37e^4}{5(1-e^2)^{7/2}}\right)
\end{equation}
and
\begin{equation}
\label{eq:dedt}
\frac{de}{dt}=-\frac{\eta}{m}(2\pi mF)^{8/3}e\left(\frac{304+121e^2}{15(1-e^2)^{5/2}}\right).
\end{equation}
These equations can be combined to form
\begin{equation}
\frac{dF}{de}=-3\frac{F}{e}\left[\frac{96+292e^2+37e^4}{(1-e^2)(304+121e^2)}\right],
\end{equation}
which is separable and easily solved
\begin{equation}
F \sigma(e)^{3/2} = C_0 \,\,
\label{eq:e_F}
\end{equation}
with the definition
 \begin{equation}
 \label{eq:sigma}
\sigma(e) = \frac{e^{12/19}}{1-e^{2}}\left(1+\frac{121}{304}e^{2}\right)^{870/2299}\,.
\end{equation}
The constant $C_0$ is set by the initial conditions $F_0 \sigma(e_0)^{3/2}=C_0$, where $e_0$ is the orbital eccentricity when the mean orbital frequency is $F_{0}$.

Applying the stationary phase approximation (SPA), reviewed in Appendix \ref{app:SPA}, to the time domain harmonic decomposition of the GW signal, Eq.~\eqref{eq:timedomdecomp}, yields
\begin{align}
\label{eq:SPA}
\tilde{h}_{+,\times}^{\SPA} \! &= \! -\frac{m \eta}{2 R} \sum_{j=1}^{\infty} \frac{(2\pi m F(t^{\ast}_j))^{2/3}}{\sqrt{j \dot{F}(t_{j}^{\ast})}} \! \left[C^{(j)}_{+,\times}(t_{j}^{\ast}) + i S^{(j)}_{+,\times}(t_{j}^{\ast})\right] \nonumber \\
& \times e^{i\psi_j}\Theta\left(f-jF_0\right)\Theta\left(jF_{\text{\rm \tiny LSO}}-f\right),
\end{align}
where the Fourier phase for each harmonic, $\psi_j$, is given by 
\begin{equation}
\label{eq:psijj}
\psi_j=2\pi ft_{j}^{\ast}-jl(t_{j}^{\ast})-\pi/4 \, \, .
\end{equation}
Here $t^{\ast}_j$ is the ``stationary'' time for the $j^{\rm th}$ harmonic which relates the mean orbital frequency ($F$) to the Fourier frequency ($f$) through the stationary phase condition
\begin{equation}
j F(t_{j}^{\ast}) = f \,\, .
\label{eq:stationary_time}
\end{equation}
Equation \ref{eq:stationary_time} supports the last conclusion of Sec.~\ref{sub:funds}: an eccentric binary emits GWs at all integer multiples of its mean orbital frequency, and the mapping between time and GW frequency is harmonic dependent, and thus, not one-to-one.

The SPA waveform model of Eq.~\eqref{eq:SPA} contains a Heaviside function, $\Theta(x)$, because of the finiteness of GW emission in the time domain. A binary that is formed at $t_0$ will emit GWs until it merges, but the PN model is not valid once the orbital velocities become a non-negligible fraction of the speed of light. As is customary in the GW literature, we thus terminate the time-domain PN waveforms at the eccentric analogue of the innermost stable circular orbit of a point-particle in a Schwarzschild spacetime: the Last Stable Orbit (LSO). The mean orbital frequency at the LSO, $F_{\text{\rm \tiny LSO}}$, is defined by \cite{Cutler:1994pb, 2008PhRvD..77j3005L, 2009CQGra..26h5001K}
\begin{equation}
\label{eq:FLSO}
F_{\text{\rm \tiny LSO}}=\frac{1}{2\pi m}\left(\frac{1+e_{\text{\rm \tiny LSO}}}{6+2e_{\text{\rm \tiny LSO}}}\right)^{3/2}.
\end{equation}
When one computes the Fourier transform of this time-domain PN model in the SPA, the Heaviside function persists, as we review in Appendix \ref{app:SPA}. 

Figure~\ref{fig:RR_on} shows the normalized Fourier amplitude of the GW signal when radiation reaction is included for systems with the same initial condition as shown in Fig.~\ref{fig:no_RR}. In the low eccentricity case ($e_0=0.1$) shown on the left panel, the amplitude is very close to the well known $f^{-7/6}$ trend of quasi-circular GW models. The rapid oscillations appearing past 15~Hz are due to interference between different harmonics. In the high eccentricity case ($e_{0}=0.6$) shown on the right panel, this trend is lost, and instead one finds rapid and large oscillations due to many harmonics of comparable strength interfering with one another. This result makes it clear that a faithful representation of the Fourier transform of eccentric signals must necessarily include several harmonic terms oscillating at different Fourier frequencies.

\begin{figure}
\includegraphics[clip=true,angle=0,width=0.475\textwidth]{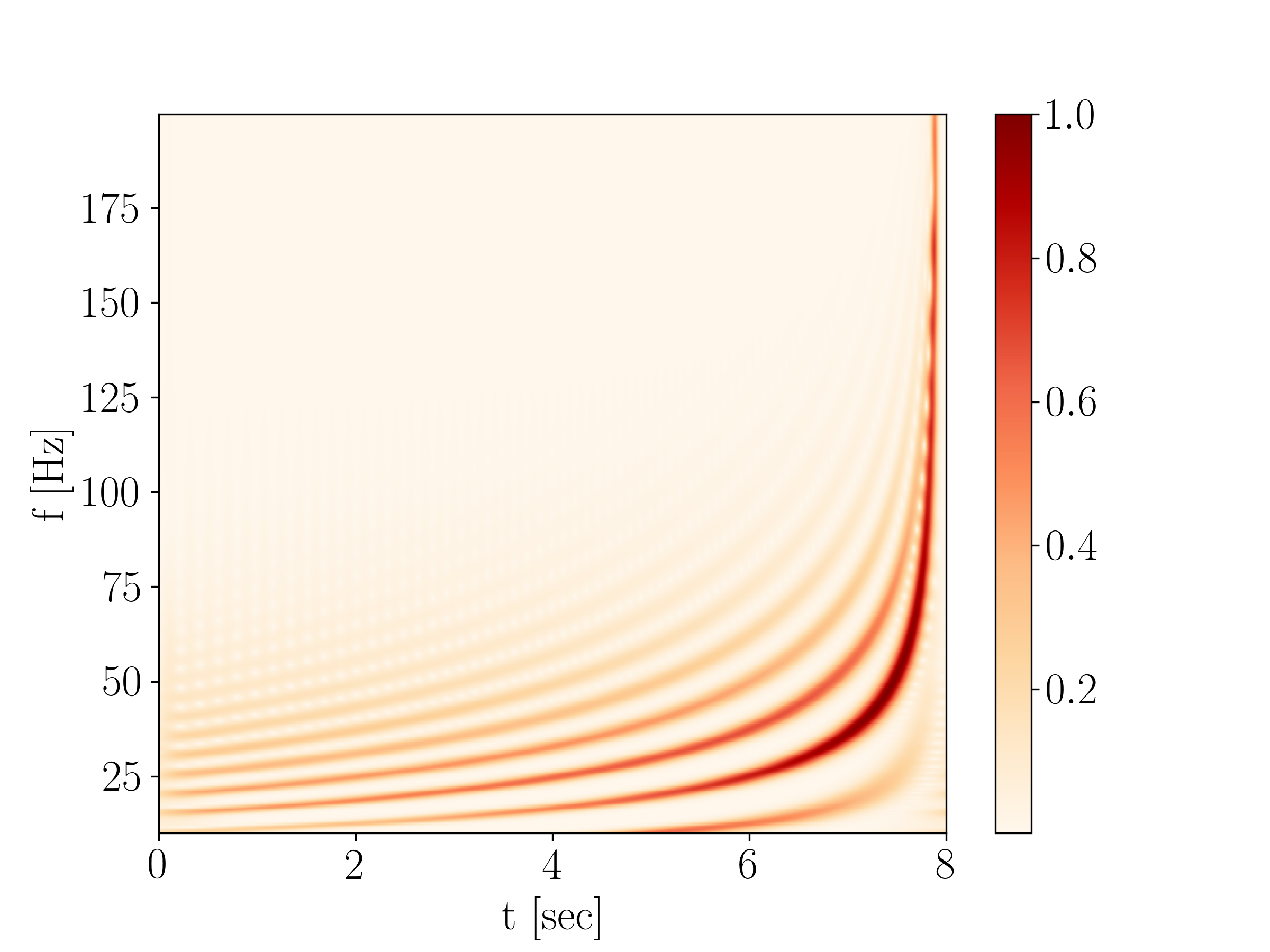} 
\caption{\label{fig:time-freq}(Color Online) The normalized amplitude in a time-frequency representation (Q-transform) of the same system as that used for the right panel of Fig.~\ref{fig:RR_on}. The presence of many harmonics in the signal leads to several tracks in time-frequency space. At later times the higher harmonics have considerably smaller amplitude, as a consequence of eccentricity decay.}
\end{figure}

The harmonic structure of the signal can be more easily appreciated through a Q-transform, as shown for example in Fig.~\ref{fig:time-freq} for the same BBH system as that used in the right panel of Fig.~\ref{fig:RR_on}. The Q-transform is a wavelet transform where the basis wavelets are Gaussian-windowed complex exponentials. Since these wavelets are localized both in time and frequency, the Q-transform produces a time-frequency representation of the GW signal. A large value of $Q$ localizes the wavelets more in frequency, while a low $Q$ localizes the wavelets more in time, and so in Fig.~\ref{fig:time-freq} we use a $Q$ of $40$. The harmonic structure shown in Eq.~\eqref{eq:SPA} manifests itself in Fig.~\ref{fig:time-freq} as several different tracks in time-frequency. At later times, higher harmonics become subdominant as their amplitude is proportional to the orbital eccentricity, which has significantly decayed.

The general structure of the Fourier-domain waveform in Eq.~\eqref{eq:SPA} is common across all current eccentric models. The main differences arise in how one treats (i) the mean anomaly $l$ and the stationary time $t^{\ast}_j$ as functions of frequency, which appear in the Fourier phase, as well as (ii) the harmonic amplitudes and the choice of truncation of the sum. At higher PN order, other differences arise, such as the precise way in which periastron precession is modeled and the inclusion of modifications to Kepler's equations at 2PN order. As a first step toward the construction of a new, analytic Fourier-domain waveforms for arbitrarily eccentric binaries, we will refrain from going to higher PN order here, but we will review current analytic models in Sec.~\ref{sub:prevwork}.

\section{Measures to Compare Eccentric Waveforms}
\label{sec:maximization}

Before proceeding with a description of current analytic models, and the development of a new one, it will be useful to first describe how to validate and compare different models. A useful data analysis measure to quantify the agreement between two waveforms $h_1$ and $h_2$ is through the match
\begin{equation}
\label{eq:match}
M=\max \limits_{t_c,l_c}\frac{(h_1|h_2)}{\sqrt{(h_1|h_1)(h_2|h_2)}}\,.
\end{equation}
This statistic is the normalized inner product between two waveforms, a ``signal'' $h_{1}$ and a ``model'' $h_{2}$, maximized over a time shift $t_{c}$ and phase shift $l_{c}$, which in the models we consider here arise in the phase functions $t$ and $l$ of Eq.~\eqref{eq:psijj} as constants of integration. The inner products are defined by
\begin{equation}
(h_1|h_2)=4 \text{Re}\int_{0}^{\infty}\frac{\tilde{h}_1^{\ast}\tilde{h}_2}{S_n(f)}df\,,
\end{equation}
where $S_n(f)$ is the noise power spectral density of the detector and Re is short-hand for the real part. In this paper, we use the design-aLIGO spectral noise density (zero-detuned, high-power) noise curve, which assumes stationary Gaussian noise \cite{2017CQGra..34d4001A}.

When the value of the match is unity, the model perfectly represents the signal to within a time and phase offset, while the more different they are, the lower the match becomes. What value of the match is then high enough for the model to be a ``faithful'' representation of the signal? To set this threshold, we demand that the systematic error from mismodeling is smaller than the statistical error. Following the detailed discussion in Appendix G of \cite{2017PhRvD..95j4004C}, this requirement translates to 
\begin{equation}
1-M<\frac{D}{2\rho^2}\,,
\end{equation}
where $D$ is the (effective) dimension of the model, roughly $10$ in our case, and $\rho$ is the signal to noise ratio (SNR) defined by $\rho^2=(h|h)$.

For quasi-circular GW templates, techniques have been developed to rapidly maximize Eq.~\eqref{eq:match} over a time and phase offset. For eccentric templates, however, this maximization is complicated by the harmonic dependence of the phase offsets arising from $l_c$. For the remainder of this section, we will thus first review the maximization techniques valid in the quasi-circular limit, and then we will extend them to the case of eccentric templates, assuming the different harmonics are mutually orthogonal. Lastly, we investigate the error incurred using this maximization scheme as we relax our assumption of orthogonality.

\subsection{Review of Quasi-Circular Match Maximization}

For quasi-circular GW templates, only the $j=2$ harmonic is non-vanishing at leading PN order. Let us then write the template model as $h=\hat{h}e^{i\phi_c-2\pi i ft_c}$, where $\hat{h}$ is a complex function of frequency and $\phi_c=2l_c$. Suppose now that we have some data $d$ that is perfectly represented by our model aside from an orbital phase shift and a time shift: $d=\hat{h}e^{i\phi_0-2\pi i ft_0}$, where $\phi_0$ and $t_0$ are inherent to the data and we do not have access to their values. Our task is then to develop an algorithm to find the values of $(\phi_{c},t_{c})$ that will maximize the overlap between $d$ and $h$, where in this case we know the solution is simply $(\phi_{c},t_{c}) = (\phi_{0},t_{0})$.

The inner product we wish to maximize is 
 \begin{equation}
 \label{eq:prod}
 (h|d)=4 \text{Re}\left[e^{-i\phi_c}\int_{0}^{\infty}\frac{|\hat{h}|^2}{S_n(f)}e^{-2\pi ift_0+i\phi_0}e^{2\pi ift_c}df\right].
 \end{equation}
Without specifying $t_c$ or $\phi_c$ we are able to construct the function
 \begin{equation}
 \tilde{G}(f)= \frac{|\hat{h}|^2}{S_n(f)}e^{-2\pi ift_0+i\phi_0}\,,
 \end{equation}
whose inverse Fourier transform ${\cal{F}}^{-1}[\cdot]$ with respect to $t_c$,
 \begin{align}
 \label{eq:circket}
 G(t_c)&=\mathcal{F}^{-1}[\tilde{G}(f)]=\int_0^{\infty}\frac{|\hat{h}|^2}{S_n(f)}e^{-2\pi ift_0+i\phi_0}e^{2\pi ift_c}df \,,
 \end{align}
 appears in the above inner product
\begin{equation}
\label{eq:prodket}
(h|d)=4\text{Re}\left[e^{-i\phi_c}G(t_c)\right].
\end{equation}
The quantity $G(t_c)$ is a complex number for any value of $t_c$ and the factor of $e^{-i\phi_c}$ rotates the argument of the real operator in Eq.~\eqref{eq:prodket} in the complex plane, but does not change the \it magnitude \rm of $G(t_c)$. Thus, we can maximize over $\phi_c$ by taking the magnitude of $G(t_c)$:
\begin{equation}
\max\limits_{\phi_c}(h|d)=4|G(t_c)|.
\end{equation}
The overlap maximized over both $\phi_{c}$ and $t_{c}$, the match $M$, is found by searching for the maximum value of the array returned by $G(t_{c})$ in the inverse Fourier transform over $t_{c}$, or simply
\be
M = 4 \max \limits_{t_{c}} |G(t_{c})|\,.
\ee
In this method the values of $t_c$ and $\phi_c$ that maximize the match need not be computed explicitly to evaluate the match, and thus, we refer to this method as \textit{implicit maximization}. 

In order to explicitly find the $(t_c,\phi_c)$ pair that maximizes the match, one can begin by identifying the time corresponding to the value of $t_{c} = t_{\maxtxt}$ that maximizes $|G(t_{c})|$. The quantity  $G(t_{\maxtxt})$ is a complex number that is rotated off the real axis by $\phi_0$ (easily verified upon inspection of Eq.~\eqref{eq:circket} with $t_c=t_0=t_{\maxtxt}$). Thus we find the value of the $\phi_c$ that maximizes the inner product ($\phi_{\maxtxt}$) via:
\begin{equation}
\phi_0=\phi_{\maxtxt}=\arctan \left[\frac{\text{Im}(G(t_{\maxtxt}))}{\text{Re}(G(t_{\maxtxt}))}\right].
\end{equation}
The pair that maximize the match is then $(t_\maxtxt,\phi_\maxtxt)$ and one calculates the match explicitly a posteriori.  Since in this case $t_{\maxtxt}$ and $\phi_{\maxtxt}$ need to be found explicitly to evaluate the match, we refer to this method as \textit{explicit maximization}. 

These two analytic methods to maximize the match are very similar in the quasi-circular case, but as we shall see, this is not the case for eccentric templates. Moreover, the implicit maximization method is computationally faster, as it does not require the evaluation of $t_{\maxtxt}$ or $\phi_{\maxtxt}$. We shall see next how all of this comes to play for eccentric templates.

\subsection{Eccentric Match Maximization}
The maximization of the match between an eccentric template and an eccentric signal is complicated by the harmonic dependence of the phase offsets arising from the mean anomaly at coalescence. Let us then consider the template $h$ and the data $d$ to be of the form
\begin{equation}
d=\sum_{k=1}^{\infty}\hat{h}_ke^{ikl_0-2\pi ift_0},
\end{equation}
and
\begin{equation}
h=\sum_{j=1}^{\infty}\hat{h}_je^{ijl_c-2\pi ift_c}.
\end{equation}
Where $\hat{h}_j$ and $\hat{h}_k$ are complex functions of frequency associated with the $j^{th}$ and $k^{th}$ harmonics of mean orbital frequency. We are then tasked with maximizing 
\begin{align}
\label{eq:eccprod}
 (h|d) &=4 \text{Re}\left[\int_{0}^{\infty}\sum_{j,k=1}^{\infty}\frac{\hat{h}_j^{\ast}\hat{h}_k}{S_n(f)}e^{(ikl_0-ijl_c)}e^{-2\pi ift_0}e^{2\pi ift_c}df\right] \nonumber \\
 &= \sum_{j,k=1}^{\infty} (h_j|d_k) \, .
 \end{align}
Without making any assumptions, the only way to maximize the above exactly is either through a grid search on $t_c$ and $l_c$ or through another numerical maximization scheme, such as a hill-climber algorithm. These methods are computationally expensive and slower than analytic techniques when the latter exist. 

The analytic maximization techniques used in the previous subsection do not immediately extend to maximization of the match for eccentric waveforms, but we generalize them under the assumption that the harmonics are mutually orthogonal. In the absence of radiation reaction, the different harmonics are exactly mutually orthogonal, $(h_j|d_k)=0$ for $j \neq k$, because they are delta functions centered at integer multiples of mean orbital frequency. In the presence of radiation-reaction, this is not exactly the case any longer, but let us continue to assume it is so, and then check at the end the amount of error introduced by this approximation.

Working in the mutual orthogonal approximation, Eq.~\eqref{eq:eccprod} becomes
\begin{align}
\label{eq:thogprod}
(h|d) &=\sum_{j,k=1}^{\infty} (h_j|d_k) \approx \sum_{j=1}^{\infty} (h_j|d_j)\,,
\nonumber \\
&\approx 4\text{Re}\left[\sum_{j=1}^{\infty}e^{ijl_c}\int_{0}^{\infty}\frac{|\hat{h_j}|^2}{S_n(f)}e^{-2\pi f t_0 + ijl_0}e^{2\pi f t_c}df\right]\,.
\end{align}
Without specifying $t_c$ or $l_c$, we are able to construct
\begin{equation}
\tilde{G}_j(f)=\frac{|\hat{h_j}|^2}{S_n(f)}e^{-2\pi i f t_0+ijl_0} \,,
\end{equation}
whose inverse Fourier transform ${\cal{F}}^{-1}[\cdot]$ with respect to $t_c$,
\begin{align}
\label{eq:eccket}
G_j(t_c)&=  \mathcal{F}^{-1}\!\left[\tilde{G}_j(f)\right] \! = \! \int_{0}^{\infty} \! \frac{|\hat{h_j}|^2}{S_n(f)}e^{-2\pi i f t_0+ijl_0}e^{2 \pi i ft_c} df  \, ,
\end{align}
appears in the inner product of Eq.~\eqref{eq:thogprod}
\begin{equation}
\label{eq:sum-on-j}
(h|d)=4\sum_{j=1}^{\infty}\text{Re}\left[e^{ijl_c}G_j(t_c)\right] \, .
\end{equation}

This expression can now be maximized over $l_{c}$ and $t_{c}$ in a way analogous to the quasi-circular case. Each individual term in the sum of Eq.~\eqref{eq:sum-on-j} can be individually maximized on $l_c$ by taking the absolute value of the argument of the real operator because the factor $e^{ijl_c}$ only rotates $G_j(t_c)$ in the complex plane, but leaves its \it magnitude \rm unchanged. Thus we can maximize over $l_c$ by computing
\begin{align}
\label{eq:impmax}
\max\limits_{l_c}(h|d) &= 4\sum_{j=1}^{\infty}\max\limits_{l_c}\left\lbrace\text{Re}\left[e^{ijl_c}G_j(t_c)\right]\right\rbrace\nonumber \\
&= 4\sum_{j=1}^{\infty}|G_j(t_c)|.
\end{align}
To maximize on $t_c$ we take the maximum value of the array that results from taking the sum of the absolute values of the inverse Fourier transforms appearing above. Again, in this method of maximization one never explicitly calculates $l_{\maxtxt}$ to compute the match, and thus, we refer to it as the \emph{eccentric implicit maximization} method. This method runs the risk of \it overestimating \rm the match, as we will show later in Sec.~\ref{sub:orthog} because of the mutual orthogonality assumption.  

Let us now consider the explicit maximization method. In order to explicitly find $l_{\maxtxt}$ and $t_{\maxtxt}$, we begin by associating $t_{\maxtxt}$ with the maximum value of the sum of the absolute values of the inverse Fourier transforms appearing in Eq.~\eqref{eq:impmax}. Inspection of Eq.~\eqref{eq:eccket} reveals that $G_j(t_{\maxtxt})$ is a complex number which is rotated off the real axis by $jl_0$ (again since $t_c=t_0=t_{\maxtxt}$, the integrand appearing in Eq.~\eqref{eq:eccket} is purely real aside from the factor of $e^{ijl_0}$, which rotates it off the real axis). The maximum $l_c$ value is then found using trigonometry:
\begin{equation}
\label{eq:lcmax}
jl_0=jl_{\maxtxt}=\arctan\left\lbrace\frac{\text{Im}[G_j(t_{\maxtxt})]}{\text{Re}[G_j(t_{\maxtxt})]}\right\rbrace.
\end{equation} 
Unlike in the quasi-circular case, however, the above procedure only produces $jl_{\maxtxt}$, which is degenerate with $l_{\maxtxt} \rightarrow l'_{\maxtxt}+2\pi/j$. One is thus forced to either break this degeneracy by computing Eq.~\eqref{eq:lcmax} explicitly with two different harmonics or try each of the $j$ degenerate guesses for $l_{\maxtxt}$ (Since the $j=1$ harmonic is not degenerate, one could just use this harmonic.) In practice, we have found that the most reliable method is to compute Eq.~\eqref{eq:lcmax} for 3 different harmonics ($j=2,3,4$), break the degeneracy, try each $l_{\maxtxt}$, and take the one that leads to the best match. As in the quasi-circular case, we here explicitly find the value of $(l_{\maxtxt},t_{\maxtxt})$ that maximize the match, and thus, we refer to it as the \emph{eccentric explicit maximization} method. As expected, this method is a bit slower than the implicit maximization, but is better behaved as we relax the assumption of orthogonality between harmonics.

\subsection{Orthogonality Investigation}
\label{sub:orthog}

Let us begin by examining how much orthogonality is broken when including radiation-reaction by computing the quantity
\begin{equation}
\label{eq:delij}
\Delta_{i,j}= \max \limits_{l_{c}} \frac{( h_i|h_j )}{\sqrt{(h_i|h_i)(h_j|h_j)}}.
\end{equation}
The above inner product between different harmonics $h_i$ and $h_j$ is maximized over any relative phase shift. If the value of $\Delta_{i,j}$ is 0 for all $i$ and $j$, then the different harmonics are orthogonal and the above approximate methods of maximization are both \it exact\rm. In the numerics that ensue, we use our new eccentric model to evaluate $\Delta_{ij}$, which we refer to as the \emph{Newtonian eccentric Fourier domain} model (NeF) and will be described in detail in Sec.~\ref{sec:models}. The results below, however, should be representative of the orthogonality between harmonics of other PN models as well, as long as they're being evaluated in their domain of validity in eccentricity.   

\begin{table*}
\begin{tabular}{||c|c|c|c|c||}
\hline
$\Delta_{i,j}$ & 2 & 3 & 4 & 5 \\
\hline
1 & 0.0035 & 0.00041 & 0.00023 & 0.0043 \\
\hline
2 & - & 0.0013 & 0.000046 & 0.00011 \\
\hline 
3 & - & - & 0.0010 & 0.00022 \\
\hline 
4 & - & -& - & 0.00039 \\
\hline 
\end{tabular}
\qquad
\begin{tabular}{||c|c|c|c|c||}
\hline
$\Delta_{i,j}$ & 2 & 3 & 4 & 5 \\
\hline
1 & 0.0043 & 0.00074 & 0.00016 & 0.0068 \\
\hline
2 & - & 0.0025 & 0.00024 & 0.000088 \\
\hline 
3 & - & - & 0.000046 & 0.000083 \\
\hline 
4 & - & -& - & 0.00015 \\
\hline 
\end{tabular}
\qquad
\begin{tabular}{||c|c|c|c|c||}
\hline
$\Delta_{i,j}$ & 2 & 3 & 4 & 5 \\
\hline
1 & 0.03 & 0.016 & 0.011 & 0.0080\\
\hline
2 & - & 0.090 & 0.037 & 0.022 \\
\hline 
3 & - & - & 0.033 & 0.013 \\
\hline 
4 & - & -& - & 0.0077 \\
\hline 
\end{tabular}
\caption{\label{tab:ept3del}The value of $\Delta_{i,j}$ for a $(10,10)M_{\odot}$ system with $e_0=0.3$ and $F_0=3$~Hz (left table), $e_0=0.6$ and $F_0=3$~Hz (center table) and $e_0=0.9$ and $F_0=3$~Hz (right table). The first row and column of each table give the values of $i$ and $j$, respectively. The diagonal terms are 1 and $\Delta_{i,j}$ is symmetric, thus redundant entries have been omitted. The frequency resolution of the inner product appearing in $\Delta_{i,j}$ is $\delta f=0.0078$~Hz (left), $\delta f=0.03125$~Hz (center) and $\delta f=2$~Hz (right). Observe that orthogonality is weakly violated in the low and moderate-eccentricity cases, and less weakly violated in the high-eccentricity case.}
\end{table*}

Table \ref{tab:ept3del} shows $\Delta_{i,j}$ for the first 5 harmonics of the GWs emitted by a $(10,10)M_{\odot}$ binary black hole (BBH) system with a few different initial conditions, where the harmonics are generated with the NeF model. We adopt slightly different frequency resolutions, with choices made to correspond to the resolution of a discrete Fourier transform of a waveform with a period determined by the initial conditions, sampled at 8192Hz in the time domain, and zero-padded such that its length is the nearest power of 2. For the systems with lower eccentricity $(e_0=0.3,0.6)$, the values of $\Delta_{i,j}$ are fairly small, of ${\cal{O}}(10^{-4})$. However, for the higher eccentricity system ($e_0=0.9$), $\Delta_{i,j}$ is larger with values of ${\cal{O}}(10^{-2})$. This high value is partly due to the low frequency resolution, but it is consistent with the resolution of some of the matches that one would realistically find using waveforms of short duration.

Let us now examine the error incurred by our analytic maximization over $t_c$ and $l_c$ due to the absence of orthogonality. Assuming the maximum inner product is obtained when $t_c=t_0$ and $l_c=l_0$, we find by inspection of Eq.~\eqref{eq:eccprod} that the maximized inner product takes the form
\begin{equation}
\label{eq:nothprod}
M_{\exact}=4\sum_{k,j=1}^{\infty}\text{Re}\left[\int_{0}^{\infty}\frac{\hat{h}_j^{\ast}\hat{h}_k}{S_n(f)}e^{il_0(k-j)}df\right].
\end{equation}
When we consider the inner product maximized via the eccentric implicit maximization, Eq.~\eqref{eq:impmax}, it is straightforward to show that we are calculating the following in the absence of orthogonality:
\begin{equation}
\label{eq:impprod}
M_{\implicit} =4\sum_{j=1}^{\infty}\text{Abs}\left[\sum_{k=1}^{\infty}e^{ikl_0}\int_{0}^{\infty}\frac{\hat{h}_j^{\ast}\hat{h}_k}{S_n(f)}df\right].
\end{equation}
Comparing Eqs.~\eqref{eq:impprod} and~\eqref{eq:nothprod} reveals that the implicit maximization can potentially  \it overestimate \rm the value of the match and that it can even become greater than 1. Therefore, in the high eccentricity limit this maximization technique becomes inaccurate, but is still a useful approximation. 

Let us now consider the error incurred by using the eccentric explicit maximization. In this case, there is no way to overestimate the match and the overlap is still properly normalized. However, we can still incur an error because the $l_{\maxtxt}$ derived from the explicit maximization method in Eq.~\eqref{eq:lcmax} assumes mutual orthogonality, and thus, it may not be an accurate estimate of the true $l_{\maxtxt}$. Since $G_j(t_{\maxtxt})$ appears in Eq.~\eqref{eq:lcmax}, it is useful to consider this quantity in the absence of orthogonality:
\begin{equation}
\label{eq:larang}
G_j^{\exact}(t_{\maxtxt})=\int_0^{\infty} \frac{|\hat{h}_j|^2}{S_n(f)}e^{ijl_0}df+\sum_{k\neq j}^{\infty}\int_0^{\infty} \frac{\hat{h}_j^{\ast}\hat{h}_k}{S_n(f)}e^{ikl_0}df.
\end{equation}
Adopting the notation
\begin{equation}
\alpha_{j,k} = \int_{0}^{\infty}\frac{\hat{h}^{\ast}_j\hat{h}_k}{S_n(f)}e^{ikl_0}df,
\end{equation}
we rewrite Eq.~\eqref{eq:larang} as 
\begin{equation}
G_j^{\exact}(t_{\maxtxt})=\alpha_{j,j} \left(1+\sum_{k \neq j }^{\infty}\frac{\alpha_{j,k}}{\alpha_{j,j}}\right).
\end{equation}
We can now identify a small parameter 
\begin{equation}
\epsilon \equiv \sum_{k \neq j }^{\infty}\frac{\alpha_{j,k}}{\alpha_{j,j}}  \ll 1\,,
\end{equation}
and expand Eq.~\eqref{eq:lcmax} to first order in $\epsilon$ to obtain
\begin{align}
j l_{\maxtxt}^{\exact} &= \arctan\left[\frac{\text{Im}[G_j^{\exact}(t_{\maxtxt})]}{\text{Re}[G_j^{\exact}(t_{\maxtxt})}\right] 
\nonumber \\
&\approx jl_0 +\sin (jl_0)\cos (jl_0)[\text{Im}(\epsilon)-\text{Re}(\epsilon)].
\end{align}
We then see that the error in $l_{\maxtxt}$ using the explicit maximization method is proportional to the difference in the imaginary and real parts of $\epsilon$, which is always much less than unity. 
\begin{figure*}[htp]
	\begin{centering}
		\includegraphics[clip=true,angle=0,width=0.475\textwidth]{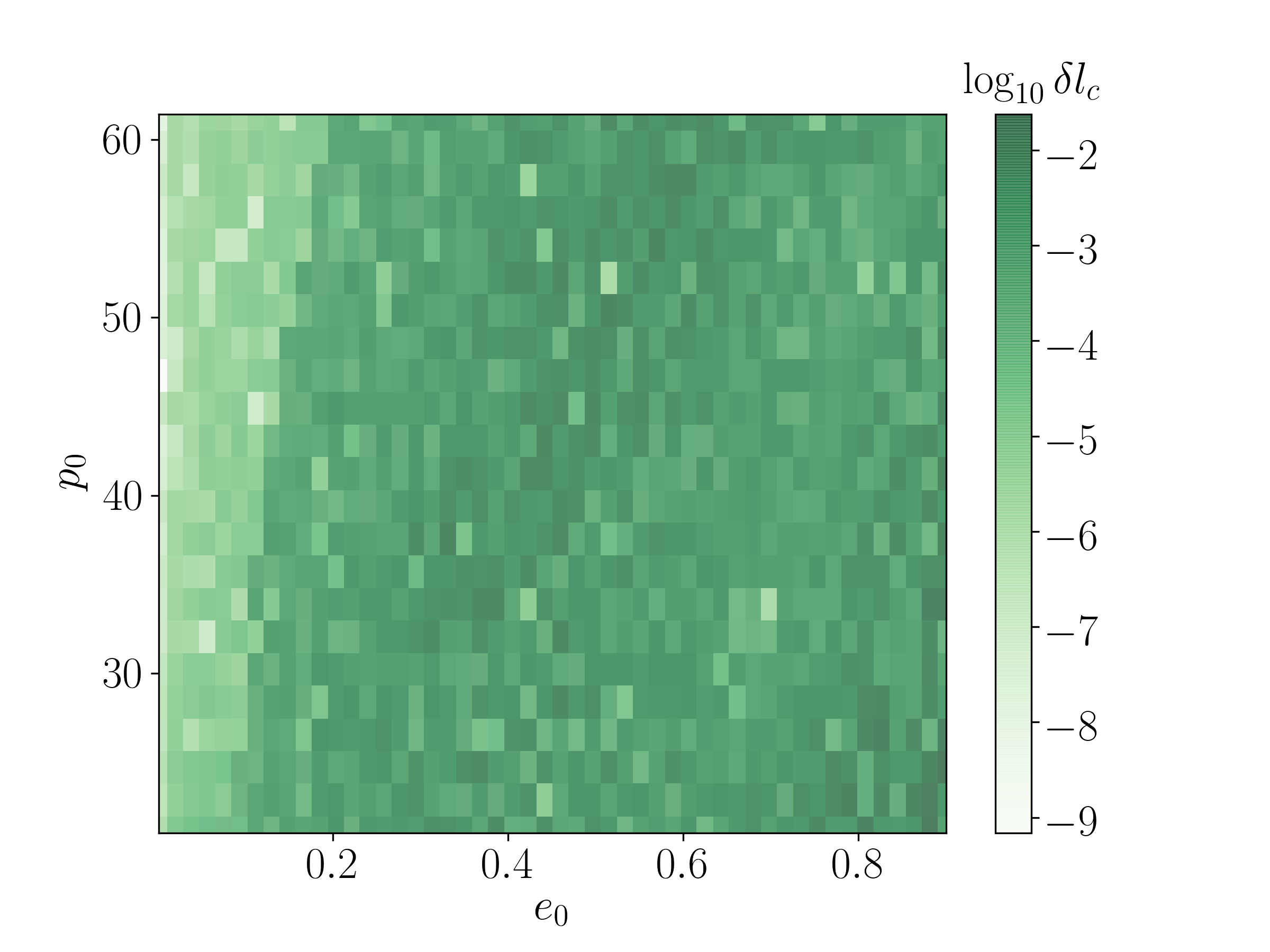} 
		\includegraphics[clip=true,angle=0,width=0.475\textwidth]{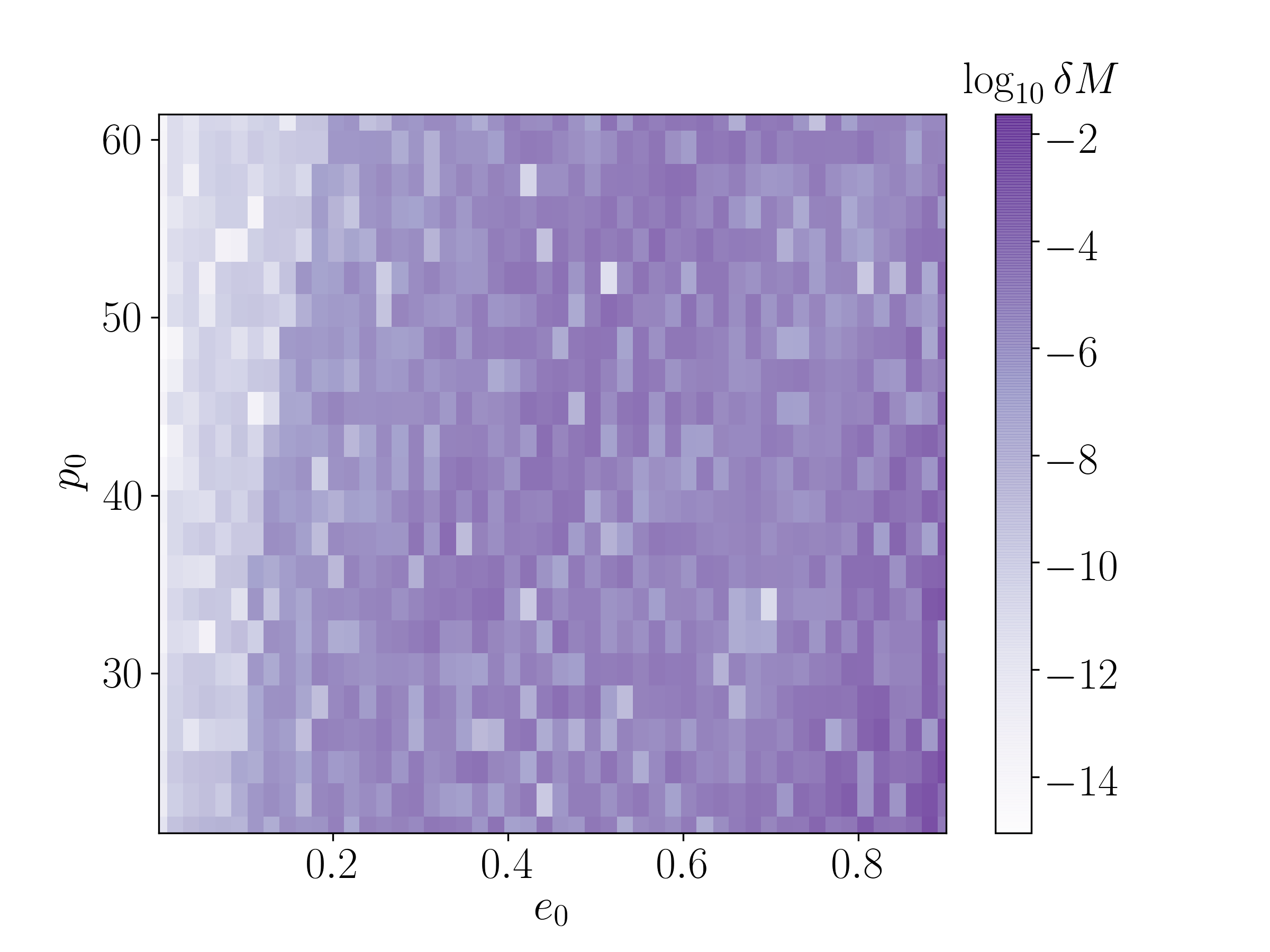} 
		\caption{\label{fig:lcerr}(Color Online) The log of the error between the $l_{\maxtxt}$ as predicted by Eq.~\eqref{eq:lcmax} and the maximum $l_c$ found by \texttt{Mathematica}'s built-in maximization on the left. On the right the error in the match as predicted by the explicit maximization technique and \texttt{Mathematica}'s built-in maximization is shown. The system is a BBH binary and the initial conditions are indicated on the x and y axis. The error in the match resulting from explicit maximization is greater as eccentricity increases, but the mean of the error in match is $\sim 10^{-6}$, which is much less than our estimated numerical accuracy. }
	\end{centering}
\end{figure*}

The left panel of Fig.~\ref{fig:lcerr} shows the error between $l_{\maxtxt}^{\explicit}$, as predicted by the explicit maximization in Eq.~\eqref{eq:lcmax}, and $l_{\maxtxt}^{\exact}$, as predicted by \texttt{Mathematica}'s built in maximization routine, while the right panel shows the resulting error in the match. The matches here are between NeF and an ``exact" time domain waveform (described in detail in Sec.~\ref{subsec:T4}) obtained by numerically solving the orbital dynamics, given in Eqs.~\eqref{eq:phidot}, \eqref{eq:dFdt}, and \eqref{eq:dedt}, and inserting these solutions into the plus and cross GW polarizations given in Eqs.~\eqref{eq:plus_strain} and~\eqref{eq:cross_strain}. For a majority of the initial conditions the explicit maximization is correct to a few ($3$-$5$) decimal places. The resulting error in match is also in about the $4^{\rm \tiny th}$-$8^{\rm \tiny th}$ digit, with a mean error of $10^{-6}$. We thus conclude that the explicit method of maximization is highly accurate.

\section{Eccentric Models}
\label{sec:models}

With all of this maximization discussion under control, let us now discuss eccentric waveform models. We begin by reviewing current models in the context of the general Fourier response presented in Eq.~\eqref{eq:SPA} of Sec.~\ref{sub:RR&struct}. The main result of this work, the derivation of our \emph{Newtonian eccentric Fourier domain model} (NeF), is presented in Sec.~\ref{sub:TayF2}. Lastly, we present the \emph{numerically evolved Newtonian time domain model} (NeT), which we will treat as ``exact" for the purpose of comparison with our frequency domain model.

\subsection{Previous Work}
\label{sub:prevwork}

In the Post-Circular formalism (PC), the waveform model is given by Eq.~\eqref{eq:SPA}, but expanding all quantities in a low eccentricity expansion. The functions appearing in the phase, $\psi_j$, are re-expressed via the chain rule as integrals over mean orbital frequency:
\begin{align}
\label{eq:pcphaseint}
t-t_{c} &=\int^{F}\frac{dF'}{\dot{F}(F',e(F'))} \,\,, \\
\label{eq:pcphaseint2}
l - l_{c} &= 2\pi \int^{F}\frac{F'}{\dot{F}(F',e(F'))}dF' \,\,.
\end{align}
The subscript $c$ denotes the respective quantity at the time of coalescence. Since the time derivative of the mean orbital frequency appearing in the denominator of the above integrands depends on the orbital eccentricity, one must obtain the orbital eccentricity as an explicit function of frequency, $e(F)$. Equation \eqref{eq:e_F} is transcendental for $e(F)$, but it can be inverted in the low eccentricity limit as a bi-variate expansion in $e_0$ and $\chi=F_0/F$. One then substitutes this inversion in the integrand appearing in Eqs.~\eqref{eq:pcphaseint} and~\eqref{eq:pcphaseint2} to analytically perform the integration.

In the PC formalism, the amplitudes of each harmonic appearing in Eq.~\eqref{eq:amps} are also expanded in the low eccentricity limit. Since the $j^{\rm th}$ harmonic amplitude scales, to leading order in a low eccentricity expansion, as $e^{j-2}$ except for  the $j=1$ harmonic, the number of harmonics kept in the sum appearing in Eq.~\eqref{eq:SPA} is controlled by the self consistency of the low eccentricity expansion. The end result is an analytic waveform of the form of Eq.~\eqref{eq:SPA} in which all pieces are series expansions in $e_0$ and $\chi$. The advantage of such models is their computational efficiency, but the main disadvantage is that they become inaccurate at moderate eccentricities.

The PC formalism has been implemented to several PN orders. The framework was introduced at Newtonian order in Ref. \cite{PhysRevD.80.084001}, keeping eccentric corrections up to $\mathcal{O}(e_0^8)$, and as a result, the sum in Eq.~\eqref{eq:SPA} was truncated at $j=10$. This work was extended to 2PN order in the phasing by \cite{Tanay:2016zog}, with the PN amplitude corrections kept at Newtonian order and the PN phase corrections truncated at $\mathcal{O}(e_0^6)$. A further extension to 3PN in the phasing was done in \cite{Moore:2016qxz}, keeping only the second ($j=2$) harmonic, the amplitude at Newtonian order in the quasi-circular limit and the PN phase corrections truncated at $\mathcal{O}(e_0^2)$. None of these models ought to be accurate for moderately eccentric systems, although a precise analysis in terms of the match and relative to exact, numerical PN waveforms have not yet been carried out. 

Several other extensions of the PC framework also exist. The enhanced PC (ePC) model introduced by \cite{Huerta:2014eca} leverages the results of \cite{PhysRevD.80.084001} and the quasi-circular part of the Fourier phase (known to 3.5PN order) in order to construct a 3.5PN eccentric model. However, this model is not constructed in a PN consistent manner.  Recently \cite{2018arXiv180108542K} incorporated spin into a PC-like model, using a semi-analytic approach by computing the phase functions numerically at 3PN order. As such, the Fourier phases are accurate for all eccentricities, but the amplitudes appearing in Eq.~\eqref{eq:SPA} are computed through a low-eccentricity expansion. As in the more vanilla PC models, these extensions are also valid only for small eccentricities, with e.g.~\cite{2018arXiv180108542K} claiming a matches greater than 0.99 provided $e_{0} \lesssim 0.3$.

One reason for the inaccuracy of all PC models is the low-eccentricity inversion of the orbital eccentricity as a function of mean orbital frequency, with comparison against numerical inversions sometimes used as a rough gauge of the regime of validity of the model. Section~\ref{sec:Err} studies this error quantitatively and in great detail, but let us here summarize the main results. Consider introducing inaccuracies in the exact solutions for $e(F)$ and $\psi_j$ parametrically, and then studying the loss in match as the magnitude of the inaccuracies and the orbital eccentricity is increased. Such a study would reveal that even for mildly eccentric binaries ($e_0 \sim 0.3$), the relative error in $e(F)$ must be below $\sim 10^{-4}$ and the relative error in $\psi_j$ must be below $\sim 10^{-3}$ for the match to remain above 99\%. These results strongly suggest that an accurate Fourier domain model must represent $e(F)$ and $\psi_j$ very accurately to avoid a large loss in match.

One potential source of confusion that occurs throughout the literature of the PC models is the interpretation of the parameter $\chi=F/F_{0}$. In order to evaluate the model at a given GW frequency, $\chi$ must be evaluated at the stationary point, $\chi(F(t^{\ast}_{j}))$, via the stationary phase condition $jF(t^{\ast}_j)=f$, yielding $\chi = f/(j F_{0})$.  What is often done is to choose $j=2$ in this relation so that $\chi = f/f_{0}$ with $f_{0} := 2 F_{0}$ identified as ``the frequency when the signal enters band." In reality, this is the GW frequency when the $j=2$ harmonic enters band, and although this harmonic dominates the entire signal when the eccentricity is truly small, it does not even when $e_{0} \sim 0.1$, as shown in Fig.~\ref{fig:no_RR}. This observation strongly suggests that an accurate Fourier domain model must represent the Fourier phase as a function of the harmonic index.  

As far as we know, there is only a single alternative to the small eccentricity approximation of the PC framework if one wishes to obtain analytic Fourier waveforms, but it comes at the cost of computational expense. In Ref.~\cite{1996NCimB.111..631P}, the stationary time and the orbital phase in Eq.~\eqref{eq:SPA} are solved at Newtonian order by changing the integration variable from time to eccentricity ($dt=de/\dot{e}$). The resulting integrals then yield hypergeometric functions which depend on the orbital eccentricity, and to invert these and express eccentricity as a function of orbital frequency, one resorts to numerical methods. Reference~\cite{1996NCimB.111..631P}, however, does not discuss how to truncate the sum appearing in Eq.~\eqref{eq:SPA} or how a formal model ought to be constructed. Reference~\cite{2015PhRvD..92d4038M} extends this method to 1PN order, but a 1PN term in the integral for the time to coalescence is there approximated as unity, when in reality it varies from 1 to 0.94. This leads to a $2\%$ error in $t(e)$, which Section~\ref{sub:eferr} shows is too large of an error for faithful modeling. Regardless of these issues, these alternative models are computationally expensive because the amplitudes are left exact as infinite sums, and costly hypergeometric functions appear in the Fourier phase.

\subsection{NeF Model}
\label{sub:TayF2}

Let us now introduce the main result of this work: an analytic Fourier-domain waveform model that is valid to eccentricities as high as 0.9 and that we will refer to as the \emph{Newtonian eccentric Fourier domain model} (NeF) throughout this work. This model is defined by the SPA of the harmonically-decomposed time-domain signal in Eq.~\eqref{eq:SPA}, where notice that the amplitude coefficients are \emph{not} expanded in small eccentricity, unlike what is done in the PC model. The model is then fully defined if we can 
\begin{itemize}
\item[(i)] Provide an accurate analytic representation for the harmonic-dependent Fourier phase $\psi_{j}(e)$,
\item[(ii)] Separately attempt to solve for the orbital eccentricity $e$ as a function of the mean orbital frequency $F$.
\end{itemize}
Once we have $\psi_{j}(e)$ and $e(F)$, we can then express the harmonic-dependent Fourier phase as a function of the GW frequency  through the stationary phase condition in Eq.~\eqref{eq:stationary_time}. 

\begin{figure*}[htp]
		\includegraphics[clip=true,angle=0,width=0.475\textwidth]{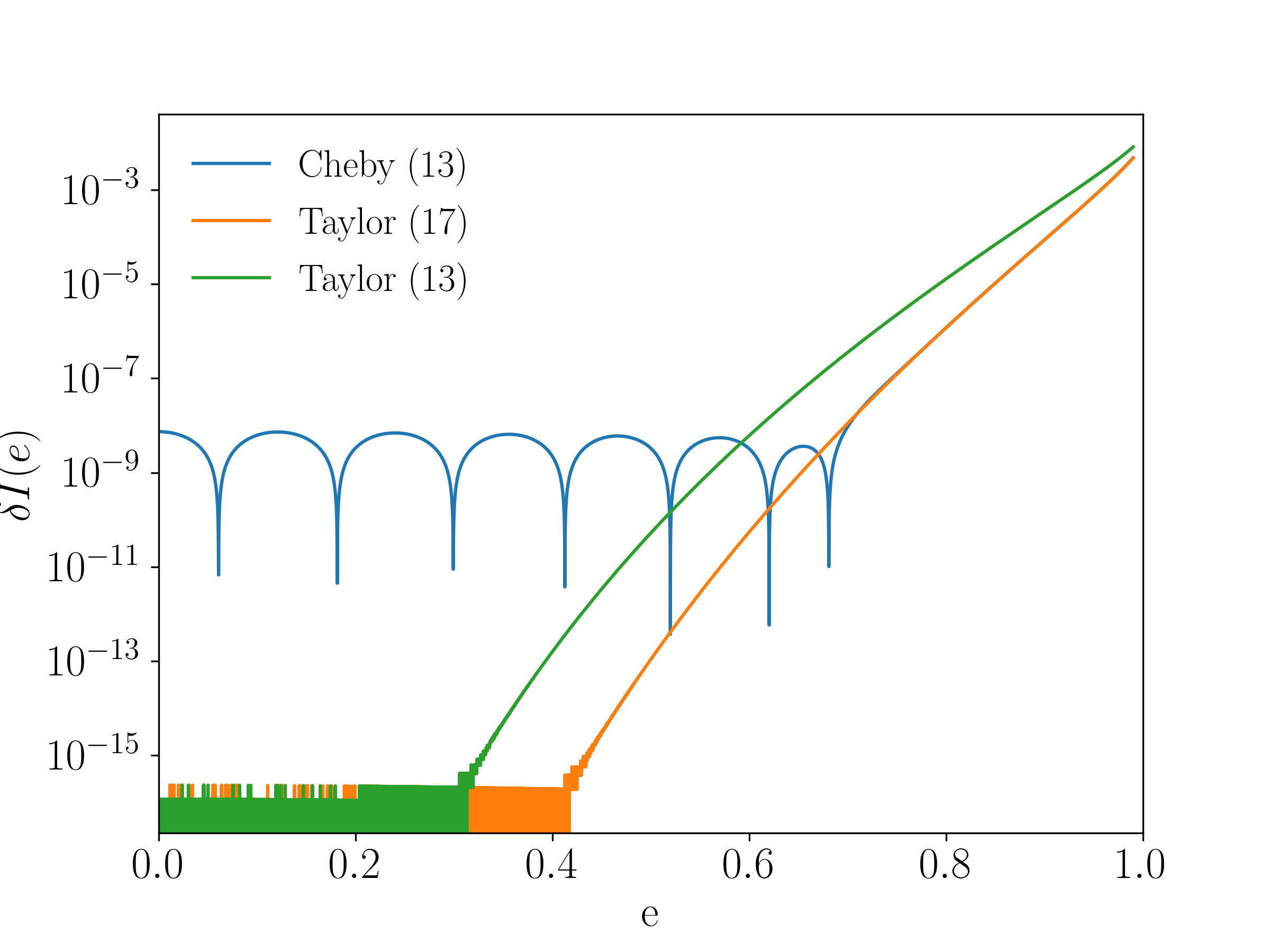} 
		\includegraphics[clip=true,angle=0,width=0.475\textwidth]{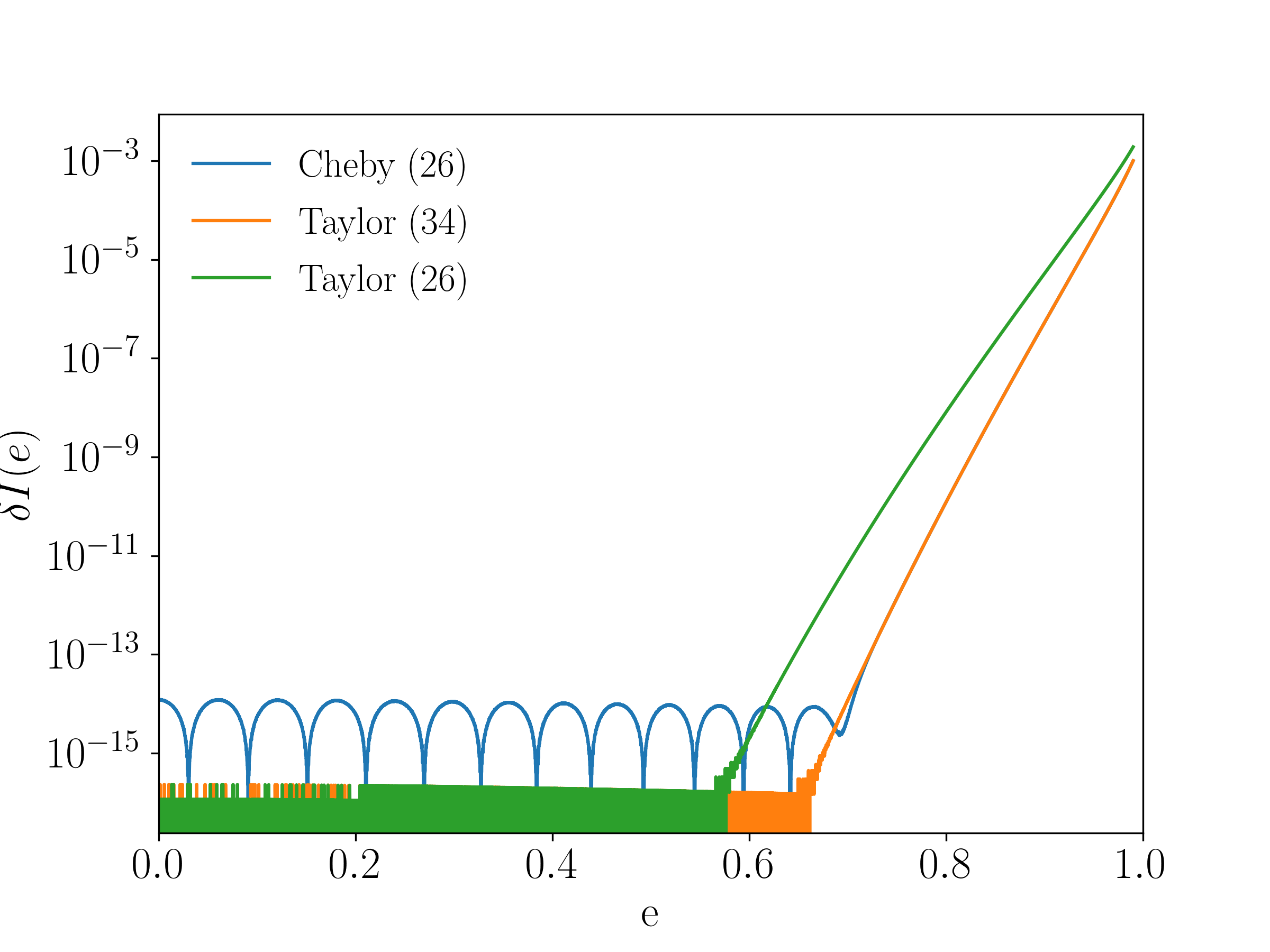} 
		\caption{\label{fig:phaserr}(Color Online) The error in $I(e)$ between the exact solution and the Chebyshev resummation, the Taylor expansion which is resummed, and a Taylor expansion which contains as many terms as the Chebyshev resummation. On the left, we keep 13 terms in the Chebyshev expansion, and 17 in the Taylor expansion which is resummed (as indicated in parenthesis in the legend). On the right we double the number of terms kept in both. We can achieve higher accuracy when keeping more terms, however, we find that a 13 term Chebyshev resummation is sufficient for our purposes.}
\end{figure*}

We begin by evaluating the different pieces that make up $\psi_{j}(e)$ in the SPA, as for example given in Eq.~\eqref{eq:psijj}. Following the work of~\cite{1996NCimB.111..631P}, we use the chain rule $dt=de/\dot{e}$ to write
\begin{align}
\label{eq:phaseints}
t-t_{c} &=\int_{0}^{e}\frac{de'}{\dot{e}(e')} \,\,, \\
l - l_{c} &= 2\pi \int_{0}^{e}\frac{F(e')}{\dot{e}(e')}de'  \,\,. 
\end{align}
Using the solution for $F(e)$ given in Eq.~\eqref{eq:e_F}, these integrals can be evaluated in closed form and they yield 
\begin{align}
t-t_{c} & = -\frac{m}{\eta}\frac{15}{(2\pi mF_0)^{8/3}\sigma(e_0)^4}\int_{0}^{e}\frac{\sigma(e')^{4}(1-e'^2)^{5/2}}{e'(304+121e'^2)}de'\nonumber \\
& = -\frac{m}{\eta}\frac{15}{(2\pi mF_0)^{8/3}\sigma(e_0)^4} I_t(e) \,\,, 
\end{align}
and 
\begin{align}
l-l_c  & = -\frac{30\pi}{(2\pi mF_0)^{5/3}\sigma(e_0)^{5/2}}\int_{0}^{e}\frac{[\sigma(e')(1-e'^2)]^{5/2}}{e'(304+121e'^2)}de'\nonumber \\
& =  -\frac{30\pi}{(2\pi mF_0)^{5/3}\sigma(e_0)^{5/2}}I_l(e) \,\,, 
\end{align}
where we have defined
\begin{align}
I_t(e) &=\frac{19}{48}e^{48/19}F_{1}\left(\frac{24}{19};-\frac{1181}{2299},\frac{3}{2};\frac{43}{19};-\frac{121}{304}e^2,e^2\right)\,,
\\
I_l(e) &=\frac{19}{30}e^{30/19} {}_{2}F_{1}\left(\frac{124}{2299},\frac{15}{19};\frac{34}{19};-\frac{121}{304}e^2\right).
\end{align}
Here $F_{1}$ in $I_{t}$ is the ApellF1 hypergeometric function and ${}_{2}F_{1}$ in $I_{l}$ is the generalized hypergeometric function. 

The Fourier phase of the $j^{\rm \tiny th}$ harmonic is then
\begin{align}
\psi_j(e) &= jl_c-2\pi ft_c-j\frac{15}{304\eta}\left(\frac{1}{2\pi m F_0}\right)^{5/3}\nonumber \\& \times\sigma(e_0)^{-5/2}\left\lbrace\frac{1}{\sigma[e(t^{\ast}_j)]^{3/2}}I_t(e)-I_l(e)\right\rbrace.
\end{align}
We can simplify the Fourier phase using the identity for ApellF1 hypergeometric functions 
\begin{align}
F_1(\alpha;\beta,\beta ';\gamma;x,y)=(1-x)^{-\beta}(1-y)^{-\beta '} \nonumber \\
\times F_1\left(\gamma -\alpha;\beta, \beta'; \gamma; \frac{x}{x-1},\frac{y}{y-1}\right).
\end{align}
which then yields
 \begin{align}
 \label{eq:final-answer}
   \psi_j &=jl_c-2\pi ft_c \nonumber \\
   &-j\frac{15}{304\eta}\left(\frac{1}{2\pi m F_0}\right)^{5/3}\sigma(e_0)^{-5/2} e^{30/19} I(e) \, \,  ,
\end{align}  
where
 \begin{align}
 \label{eq:psij}
 I(e) & =  \frac{19}{48(1+\frac{121e^2}{304})^{124/2299}}
 \nn \\ 
 &\times F_1\left(1;-\frac{1181}{2299},\frac{3}{2};\frac{43}{19};\frac{121e^2}{304+121e^2},\frac{e^2}{e^2-1}\right) 
 \nn \\ 
 &-\frac{19}{30} \; {}_{2}F_{1}\left(\frac{124}{2299},\frac{15}{19};\frac{34}{19};-\frac{121e^2}{304}\right) \, \, .
 \end{align}

The above equation for the phase is exact, and thus valid for all eccentricities, but it suffers from the fact that the hypergeometric functions are computationally costly to evaluate, especially $F_{1}$. One could of course create a look-up table for these functions to remove the computational cost, but as we shall see, there is a better, analytic approach. After exploring different representations of hypergeometric functions, we find that $I(e)$ is well-approximated by a Chebyshev resummation of a Taylor expansion about small eccentricity. After the collecting of like terms in eccentricity in the Chebyshev resummation, we are able to approximate the frequency dependence of the phase as
 \begin{align}
 \label{eq:chebapprox}
 & I(e) \approx \sum_{n=0}^{n=12}C_n e^{2n},
 \end{align}
where the $C_n$ coefficients and a brief review of Chebyshev resummations is provided in Appendix \ref{app:phasecoef}. The first few $C_n$, evaluated to double precision, are 
 \begin{subequations}
 \label{eq:C-coeffs}
\begin{align}
C_0 &=-0.2375000017589697 \, \,, \\
C_1 &=-0.3006152896315175 \, \,, \\
C_2 &=-0.009023468528641264\, \,, \\
C_3 &=-0.03715292002990571 \, \,, \\
C_4 &=-0.01543183701515415 \, \,.
\end{align}
\end{subequations} 
Using a simple timing study implemented in \texttt{Mathematica}, we find that the resulting approximation is $\sim 10^4$ times faster than evaluating the exact result in Eq.~\eqref{eq:psij}.

Figure~\ref{fig:phaserr} shows the relative error between the exact form of $I(e)$ and the Chebyshev resummation keeping 13 terms (left panel) and 26 terms (right panel), and between the exact form and the Taylor series which is Chebyshev resummed or the Taylor series with the same number of terms as what is kept in the Chebyshev resummation. By relative error, we explicitly mean $\delta I(e) = 1-I_{\text{\rm \tiny app}}(e)/I(e)$, where $I_{\text{\rm \tiny app}}(e)$ is the approximate solution for $I(e)$, which is the exact solution. Observe that the simple Taylor expansion does surprisingly well at representing the exact function. Observe also that the Chebyshev resummation is capable of representing the exact function in a wider range of eccentricity to a relative accuracy better than $10^{-8}$ and $10^{-13}$ in the $13$-term and $26$-term case respectively. We will see in Secs.~\ref{sec:Err} and~\ref{subsec:faith} that the level of error in the left panel is tolerable to obtain matches above 99\%. Nonetheless, the method can be easily extended to higher order if higher accuracy is desired for implementation in more sensitive, third-generation detectors.

The Fourier phase in Eq.~\eqref{eq:final-answer} is a function of the orbital eccentricity, but the latter must be mapped to  mean orbital frequency $F$, so that it can be further mapped to GW frequency $f$ via the stationary phase condition $jF[e(t_j^{\ast})]=f[e(t_j^{\ast})]$. Reasonably, one is tempted to invert the transcendental equation for $F(e)$ in Eq.~\eqref{eq:e_F}  in a low eccentricity approximation. However, Section~\ref{sub:lowefinv} will show that the low eccentricity inversion of $F(e)$ fails at $e_0 \gtrsim 0.3$, regardless of the number of terms kept. In fact, it is this inversion for $F(e)$ which is probably responsible for the failure of all PC models. 

Let us then discuss the analytic inversion of this function. The condition we must invert is 
\be
\label{eq:zetae}
\sigma(e) = \zeta\,,
\ee
where recall that $\sigma(e)$ is defined in Eq.~\eqref{eq:sigma}, we have defined $\zeta := (j C_{0}/f)^{2/3} = \sigma(e_{0}) (j F_{0} /f)^{2/3} = \sigma(e_0)(F_0/F)^{2/3}$, and we have used that the constant $C_0 =F_0 \sigma(e_0)^{3/2}$ ensures that $F(e_0)=F_0$. Defining the inverse function $\kappa$ such that $\kappa[\sigma(e)] = e$, the solution is then simply
\be
e(f) = \kappa(y) = \kappa\left[\left(\frac{f}{j C_{0}}\right)^{3/2}\right]\,.
\ee
Since the inverse function $\kappa$ is \emph{system-independent}, this function can be obtained once and only once by any means at our disposal.

Let us first discuss analytic inversions. We were able to obtain two analytic representations of $e(\zeta)$ that meet the error tolerance that will be laid out in Sec.~\ref{sec:Err} (relative error of $\mathcal{O}(10^{-6})$). The first is obtained by introducing a $\bar{\sigma}_{3}(e) \approx \sigma(e)$, whose inverse $\bar{\kappa}_{3}(\zeta)$ can be found algebraically (the subscript ``3" arises from the construction of other approximate $\bar{\sigma}_{n}$ which are reviewed in Appendix \ref{app:F(e)}). The approximate inverse function $\bar{\kappa}_{3}(\zeta)$ is given by
\begin{align}
\label{eq:kappa3}
\bar{\kappa}_{3}(\zeta) &= \left\lbrace -\frac{1}{43923+277248\zeta^{19/6}}\left[1824\zeta^{19/6} 
\nonumber \right. \right.\\ 
& \times \left(\frac{283475(2)^{2/3}}{(38)^{1/3}\alpha} - 152\right) \nonumber \\ 
& \left. \left. +4\left(18392-\left(\frac{2}{19}\right)^{1/3}\frac{2225432}{\alpha}-19^{2/3}\alpha \right) \right]  \right\rbrace^{1/2} \, ,
\end{align}
where we have defined
\begin{align}
\alpha &= \left\lbrace 2154218176 + 12750 \zeta^{19/12}\left(14641 + 92416 \zeta^{19/6}\right) 
\right. \nonumber \\
& \times \left[51\left(1216+11475 \zeta^{19/6}\right)\right]^{1/2} - 7650\zeta^{19/6} \nonumber \\
& \left. \times \left(-41031947+117830400\zeta^{19/6}\right)\right\rbrace^{1/3} \, .
\end{align}
Using this $\bar{\kappa}_{3}(\zeta)$, we can approximate $e(\zeta)$ to relative error of $\mathcal{O}(10^{-3})$ for sources with $e_0$ as high as 0.9. In order to further decrease the error we numerically fit the difference using a function of the form
\begin{equation}
\label{eq:legendrefit}
e(\zeta) \approx \bar{\kappa}_{3}(\zeta)\left(1+\frac{ae^{-b\zeta^{-c}}}{\zeta^d} + \sum_{n}^{30}H_{n}L_n(\zeta)\right) \,,
\end{equation}
where $L_n(\zeta)$ are Legendre polynomials and the constants $a$, $b$, $c$, $d$, $H_n$ are fitted for and presented in Appendix \ref{app:F(e)}. 
\begin{figure}
\includegraphics[clip=true,angle=0,width=0.475\textwidth]{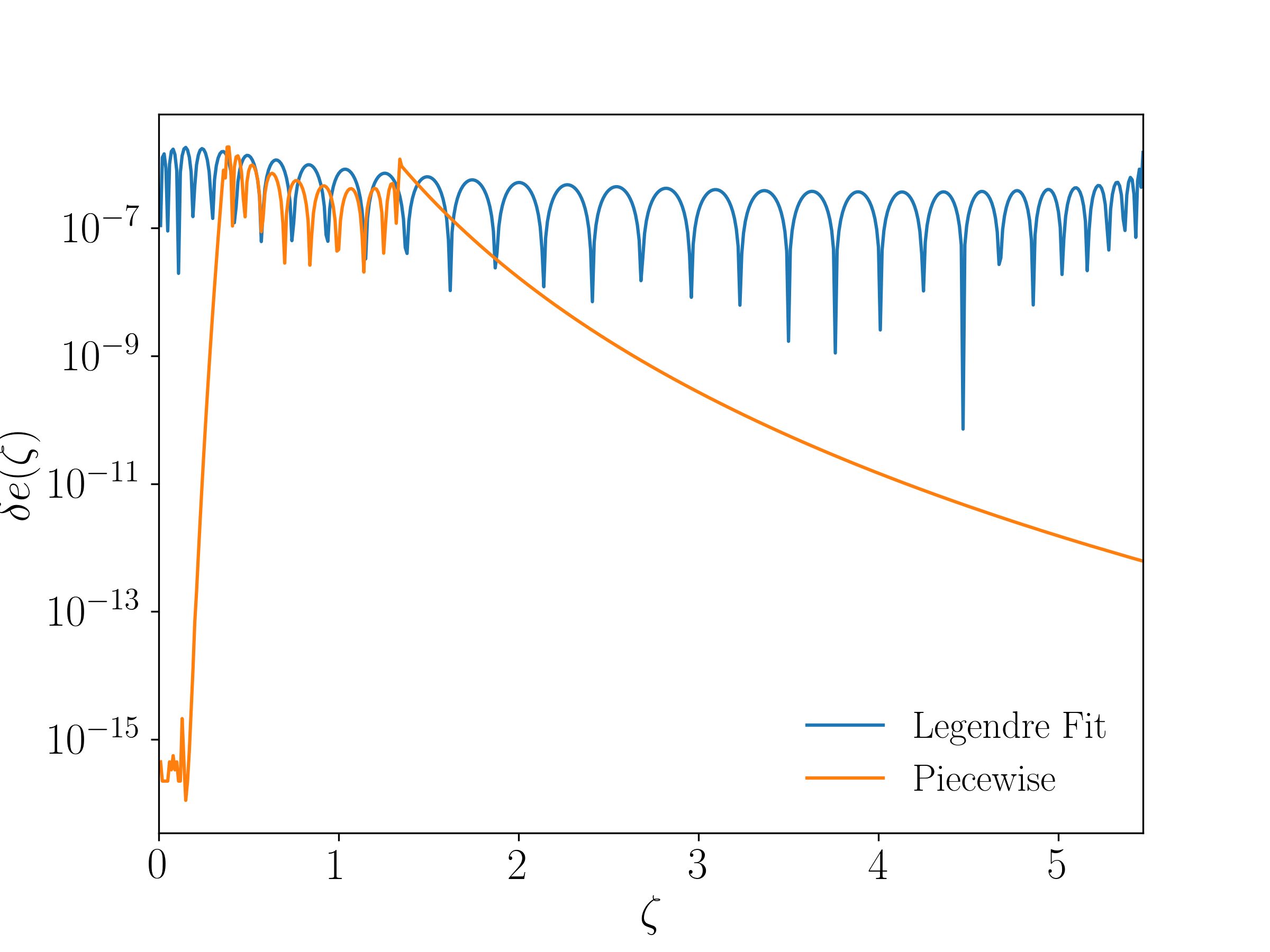}  
		\caption{\label{fig:fits}(Color Online) The relative error, $\delta e(\zeta)$, between a numerical solution and the Legendre fit presented in Eq.~\eqref{eq:legendrefit} (shown in blue) and the piecewise function in Eq.~\eqref{eq:piecewise1}. The piecewise representation is $\sim 30$ times faster than the Legendre fit and covers the entire domain of $\zeta \in [0, \infty]$, with less error.}
\end{figure}

A second analytic inversion that is faster than the first can be obtained through a piecewise ansatz
\begin{align}
  \label{eq:piecewise1}
e(\zeta) \approx \left\{ \begin{array}{cc} 
                e_{\text{\rm \tiny Low}}(\zeta) & \hspace{5mm} \zeta \leq \zeta_{\text{\rm \tiny Low}} \\
                e_{\text{\rm \tiny Mid}}(\zeta) & \hspace{5mm} \zeta_{\text{\rm \tiny Low}} < \zeta < \zeta_{\text{\rm \tiny High}} \\
                e_{\text{\rm \tiny High}}(\zeta) & \hspace{5mm} \zeta \geq \zeta_{\text{\rm \tiny High}} \\
                \end{array} \right.
\end{align}
The $e_{\text{\rm \tiny Low}}(\zeta)$ and $e_{\text{\rm \tiny High}}(\zeta)$ are found by inverting Eq.~\eqref{eq:zetae} in the $\zeta << 1$ and $\zeta >> 1$ limits through a Taylor expansion, while $e_{\text{\rm \tiny Mid}}(\zeta)$ is built through a numerical fit. The bounds $\zeta_{\text{\rm \tiny Low}}$ and $\zeta_{\text{\rm \tiny High}}$ correspond to the $\zeta$ at which the two inversions exceed $\mathcal{O}(10^{-6})$ relative error with an exact solution.  The details of the construction of this function and the values of the coefficients are given in Appendix \ref{app:F(e)}.

Although these analytic representations are sufficiently accurate for our purposes, they are not the fastest and most accurate solution to the problem. As we indicated above, Eq.~\eqref{eq:zetae} is system-independent, so any inversion that is accurate and fast will do. In particular, due to the presence of the unit step functions in Eq.~\eqref{eq:SPA}, each harmonic samples $e(f)$ at frequencies $f \in [jF_0, jF_{\text{\rm \tiny LSO}}]$, and since by the stationary phase condition $f=jF$, any harmonic samples $e(F)$ at orbital frequencies $F \in [F_0, F_{LSO}]$. Thus one only needs to solve for $e(F)$ once, and the result can be used to sample $e(f)$ for any harmonic. A fully-numerical possibility is to either solve $de/dF$ numerically (constructed by differentiating Eqs.~\eqref{eq:dFdt} and \eqref{eq:dedt}) or to sample $F(e)$ sufficiently discretely, swapping the columns and then interpolating the result with a cubic spline. Either of these two methods is extremely fast and it can be carried out to double precision. 

Figure \ref{fig:fits} shows the relative error between the numerical solution for $e(\zeta)$, the fit developed in Eq.~\eqref{eq:legendrefit}, and the piecewise representation of $e(\zeta)$ in Eq.~\eqref{eq:piecewise1}. Both analytic representations of $e(\zeta)$ maintain a relative error below $\mathcal{O}(10^{-6})$ which is what we will show in Sec.~\ref{sec:Err} is needed for applications in which one is concerned with keeping a high value of the match. However, the evaluation of the Legendre (piecewise) representations is roughly $100$ ($3$)  times slower than the evaluation of the numerical solution. For this reason, we employ the numerical solution when constructing the NeF model henceforth due to its computational speed advantages.   
 
\subsection{NeT Model}
\label{subsec:T4}

The validation of the NeF model requires its comparison to another model that we consider more accurate or exact. We will take this to be an eccentric extension of the quasi-circular TaylorT4 approximant~\cite{2009PhRvD..80h4043B}, which we will refer to as the \emph{Newtonian eccentric Time domain model} (NeT). This is obtained by numerically solving the set of differential equations in Eqs.~\eqref{eq:phidot}, \eqref{eq:dFdt} , and \eqref{eq:dedt} for $\phi(t)$, $F(t)$, and $e(t)$, respectively, and then inserting these expressions in the plus and cross polarizations given by Eqs.~\eqref{eq:plus_strain} and \eqref{eq:cross_strain}. The time-domain polarizations are then discretely Fourier transformed to obtain a representation in the frequency domain. 

The integration of the equations of motion is done as follows. We start all numerical integrations with initial conditions $\phi(t_0)=-\pi/2$, $F(t_0)=F_0$, and $e(t_0)=e_0$. This choice is such that binary is initially moving towards and close to pericenter passage. For highly eccentric systems ($e\approx 0.9$), much of the GW power is emitted during pericenter passage, and the eccentricity can decrease significantly between passages if the system is compact enough. Thus, our choice of initial conditions is such that the binary begins emitting GWs with significant power at an initial orbital eccentricity of $e_0$. For systems with small initial eccentricity or large initial separation (low initial mean orbital frequency) the choice of $\phi_0$ becomes less important. We stop our numerical evolution at the last stable orbit (LSO), which is given by Eq.~\eqref{eq:FLSO}. Two independent implementations of this integration were developed, one in \texttt{Mathematica} with precision and accuracy goals of 13, and one in \texttt{C} using the Implicit Bulirsch-Stoer method with a supplied Jacobian for adaptive step-size control. These implementations lead to $h(t)$, $n(t)$, and $e(t)$ that agree to $\sim 10^{-8}$ relative error.

The discrete Fourier transform is done as follows. First, the time-domain functions are sampled at a rate of 8192~Hz, leading to $N$ samples. Then, this discretely sampled time-series is zero padded on both edges with pads of equal length, such that the new time-series is of length $2^{p}$, with $p$ the smallest integer such that $2^{p} > N$. The new time series is then discrete Fourier transformed. 

\section{Error Analysis of SPA Ingredients}
\label{sec:Err}

In this section, we investigate the loss in match, maximized via the prescription given in Sec.~\ref{sec:maximization}, due to errors in the inversion of $F(e)$ and in the calculation of $\psi_{j}(e)$ to inform analytic models on tolerable errors. We then show that the standard low eccentricity inversion of $F(e)$ leads to a significant loss of match when $e_0 \sim 0.3$ for a BBH system.
\begin{table*}
\begin{tabular}{||c|c|c|c|c||}
\hline
$\log_{10}(\epsilon)$ & $e_0=0.1$ & $e_0=0.3$ & $e_0=0.6$ & $e_0=0.9$ \\
\hline
-16 & 0.9877 \, \, (0.9877) & 0.9847 \, \, (0.9847)& 0.9825 \, \, (0.9825)& 0.9828 \, \, (0.9828) \\
\hline
-8 & 0.9877 \, \, (0.9877) & 0.9847 \, \, (0.9847)& 0.9825 \, \, (0.9825)& 0.9828 \, \, (0.9828)\\
\hline 
-6 & 0.9877 \, \, (0.9877) & 0.9847 \, \, (0.9847)& 0.9825 \, \, (0.9825)& 0.9824 \, \, (0.9825)\\
\hline 
-5 & 0.9877 \, \, (0.9877)& 0.9847 \, \, (0.9847)& 0.9820 \, \, (0.9822)& 0.9743 \, \, (0.9761)\\
\hline 
-4 & 0.9872 \, \, (0.9876) & 0.9844 \, \, (0.9846)& 0.9765 \, \, (0.9790)& 0.9046 \, \, (0.9134)\\
\hline 
-3.5 & 0.9854 \, \, (0.9871) & 0.9790 \, \, (0.9826)& 0.9312 \, \, (0.9546)& 0.8459 \, \, (0.8852)\\
\hline 
-3 & 0.9670 \, \, (0.9833) & 0.9469 \, \, (0.9705)& 0.8641 \, \, (0.9189)& 0.7623 \, \, (0.8389)\\
\hline 
-2.75 & 0.9395 \, \, (0.9773)& 0.9108 \, \, (0.9605)& 0.8185 \, \, (0.8952)& 0.7189 \, \, (0.8086)\\
\hline 
-2.5 & 0.8776 \, \, (0.9639)& 0.8500 \, \, (0.9447)& 0.7614 \, \, (0.8682)& 0.6532 \, \, (0.7783)\\
\hline 
-2.25 & 0.7886 \, \, (0.9438)& 0.7633 \, \, (0.9223)& 0.6853 \, \, (0.8444)& 0.5887  \, \, (0.7502)\\
\hline 
-2 & 0.6703 \, \, (0.9168)& 0.6506 \, \, (0.8923)& 0.5890 \, \, (0.8105)& 0.5137  \, \, (0.7130)\\
\hline
\end{tabular}
\caption{\label{tab:einverr} Match for different values of $\epsilon$ when using $e_{1}$ (not in parenthesis) and $e_{2}$ (in parenthesis) as approximate inversions. The first column lists the value of $\epsilon$, the other columns list the value of the match for a given $e_0$. For each of the values shown here, the corresponding system is a BBH system with an initial dimensionless semi-latus rectum of $p_0=50$. For large eccentricities, the match begins to decrease as $\epsilon \sim 10^{-5}$.}
\end{table*}
\begin{table*}
\begin{tabular}{||c|c|c|c|c||}
\hline
$\log_{10}(\epsilon)$ & $e_0=0.1$ & $e_0=0.3$ & $e_0=0.6$ & $e_0=0.9$ \\
\hline
-16 & 0.9877 \, \, (0.9877) & 0.9847 \, \, (0.9847)& 0.9825 \, \, (0.9825)& 0.9828 \, \, (0.9828) \\
\hline
-8 & 0.9877 \, \, (0.9877) & 0.9847 \, \, (0.9847)& 0.9825 \, \, (0.9825)& 0.9828 \, \, (0.9828) \\
\hline 
-6 & 0.9877 \, \, (0.9877) & 0.9847 \, \, (0.9847)& 0.9825 \, \, (0.9825)& 0.9827 \, \, (0.9827) \\
\hline 
-5 & 0.9877 \, \, (0.9877)& 0.9847 \, \, (0.9847)& 0.9823 \, \, (0.9824)& 0.9809 \, \, (0.9812) \\
\hline 
-4 & 0.9874 \, \, (0.9877) & 0.9847 \, \, (0.9847)& 0.9820 \, \, (0.9823)& 0.9311 \, \, (0.9416) \\
\hline 
-3.5 & 0.9874 \, \, (0.9874) & 0.9830 \, \, (0.9841)& 0.9670 \, \, (0.9743)& 0.8904 \, \, (0.9057) \\
\hline 
-3 & 0.9862 \, \, (0.9858) & 0.9680 \, \, (0.9781)& 0.9067 \, \, (0.9399)& 0.8276 \, \, (0.8693) \\
\hline 
-2.75 & 0.9655 \, \, (0.9825)& 0.9431 \, \, (0.9695)& 0.8647 \, \, (0.9202)& 0.7714 \, \, (0.8459) \\
\hline 
-2.5 & 0.9299 \, \, (0.9753)& 0.9028 \, \, (0.9587)& 0.8156 \, \, (0.8957)& 0.7269 \, \, (0.8158) \\
\hline 
-2.25 & 0.8617 \, \, (0.9603)& 0.8369 \, \, (0.9419)& 0.7561 \, \, (0.8724)& 0.6540 \, \, (0.7819) \\
\hline 
-2 & 0.7661 \, \, (0.9390)& 0.7445 \, \, (0.9180)& 0.6755 \, \, (0.8423)& 0.5887 \, \, (0.7522) \\
\hline
\end{tabular}
\caption{\label{tab:psierr} Match for different values of $\epsilon$ when approximating the Fourier phase with $I_{1}$ (not in parenthesis) and $I_{2}$ in parenthesis. The first column lists the value of $\epsilon$, and the other columns list the value of the match for a given $e_0$. For each of the values shown here, the corresponding system is a BBH system with an initial dimensionless semi-latus rectum of $p_0=50$.}
\end{table*} 

In order to investigate the loss in match due to these inaccuracies, we require an exact solution for the analytic Fourier response. To create this exact model we interpolate the eccentricity dependent term of the phase, $I(e)$, given in Eq.~\eqref{eq:psij}, and we generate a numerical solution for $e(F)$ by numerically solving Eq.~\eqref{eq:e_F}. With these solutions in hand, we have an exact (to machine precision), numerical solution for the Fourier response in the SPA, and it is this exact solution that we will use to investigate inaccuracies in $\psi_{j}$ and $e(F)$. For consistency with our faithfulness study in Sec.~\ref{subsec:faith}, we truncate the the sum over harmonics in Eq.~\eqref{eq:SPA} at a sufficiently high harmonic index such that the match between it and the NeT model is $\sim 0.99$, as detailed in Sec.\ref{sub:jmax} and~\ref{subsec:faith}. 

\subsection{Loss in match due to inaccurate $e(F)$ and $\psi_j$}
\label{sub:eferr}

To investigate the loss in match due to the inaccurate inversion of $F(e)$, we begin by numerically solving Eq.~\eqref{eq:e_F} and refer to this solution as $e(F)$. To simulate inaccuracies, we construct two different expressions for $e(F)$ using the exact numerical solution:
\begin{align}
e_{1} &=e(F) \; (1-\epsilon) \nonumber \, , \\
e_{2} &=e(F) \left[1-\epsilon \left(\frac{e(F)}{e_0}\right)^2\right] \,,
\end{align}
with $\epsilon$ a constant. In $e_1$, the relative error with respect to the exact solution is a constant $\epsilon$, while in $e_2$ it is $\epsilon \; [e(F)/e_0]^{2}$, decreasing as $e(F)^2$. As before, by relative error we here mean $1-[e_{\text{\rm \tiny app}}/e(F)]$, where $e_{\text{\rm \tiny app}} = e_{1}$ or $e_{2}$ is the approximate solution. 

We calculate the value of the match for different values of $\epsilon$, given a BBH system with initial dimensionless semi-latus rectum of $p_0=50$ and at various different initial eccentricities ($e_0 = 0.1, 0.3, 0.6, ~\text{and} ~0.9$). We increase epsilon to reflect larger values of relative error between the approximate and exact solutions. The values of these matches are shown in Table \ref{tab:einverr}. In the $e_1$ case, the low eccentricity cases begin to show a significant decrease in match when $\epsilon \sim 10^{-3.5}$. This implies that a more accurate inversion of $F(e)$ is required than that quoted by many PC models if one is interested in applications that require such high values of the match, such as parameter estimation. In the moderately eccentric case, the match starts to decrease significantly when $\epsilon \sim 10^{-4}$, while in the large eccentricity case even when $\epsilon \sim 10^{-5}$. In the $e_2$ case, the trend is similar, but the decrease in match is less sharp for larger $\epsilon$. The match will be more sensitive for systems that are longer lived (lower mass), but these results set roughly tolerable inaccuracy in the inversion of $F(e)$.

We take a similar approach to analyze the loss in match due to inaccuracies in the Fourier phase, $\psi_j(e)$. We interpolate the eccentricity dependent part of the phase appearing in Eq.~\eqref{eq:psij} and refer to this as $I(e)$. To simulate inaccuracies, we construct two approximate expressions for $I(e)$ using the exact solution
\begin{align}
I_{1}&=I(e) (1-\epsilon) \nonumber \, , \\
I_{2}&=I(e) \left[1-\epsilon \left(\frac{e}{e_0}\right)^2\right] \,,
\end{align}
with $\epsilon$ a constant. In the above, we use a numerical solution for $e(F)$ to isolate inaccuracies in the latter from those in the functional solution for $I(e)$. As in the $e(F)$ case, the relative error in $I_{1}$ is a constant $\epsilon$, while that in $I_{2}$ decreases as $e^2$.

Table \ref{tab:psierr} shows the match when using the approximate $I_{1}$ and $I_{2}$ expressions in the Fourier phase.
Observe that the Fourier phase can tolerate more inaccuracy than the inversion of $e(F)$. This is not surprising given that $I(e)$ is multiplied by an overall factor of $e^{30/19}$. In the low eccentricity case, the match begins to decay significantly when $\epsilon \sim 10^{-2.75}$, while in the large eccentricity case, the match deteriorates even when $\epsilon \sim 10^{-4}$. The greater sensitivity to error in the higher eccentricity case can be explained by the appearance of the overall factor of harmonic index $j$ multiplying $I(e)$, and thus, also multiplying the error. As the higher harmonics become more important (i.e. for moderate to large eccentricities), the phase must be approximated more accurately.

\subsection{Error in the low eccentricity inversion of $F(e)$}
\label{sub:lowefinv}

\begin{figure}[htp]
	\begin{centering}
		\includegraphics[clip=true,angle=0,width=0.475\textwidth]{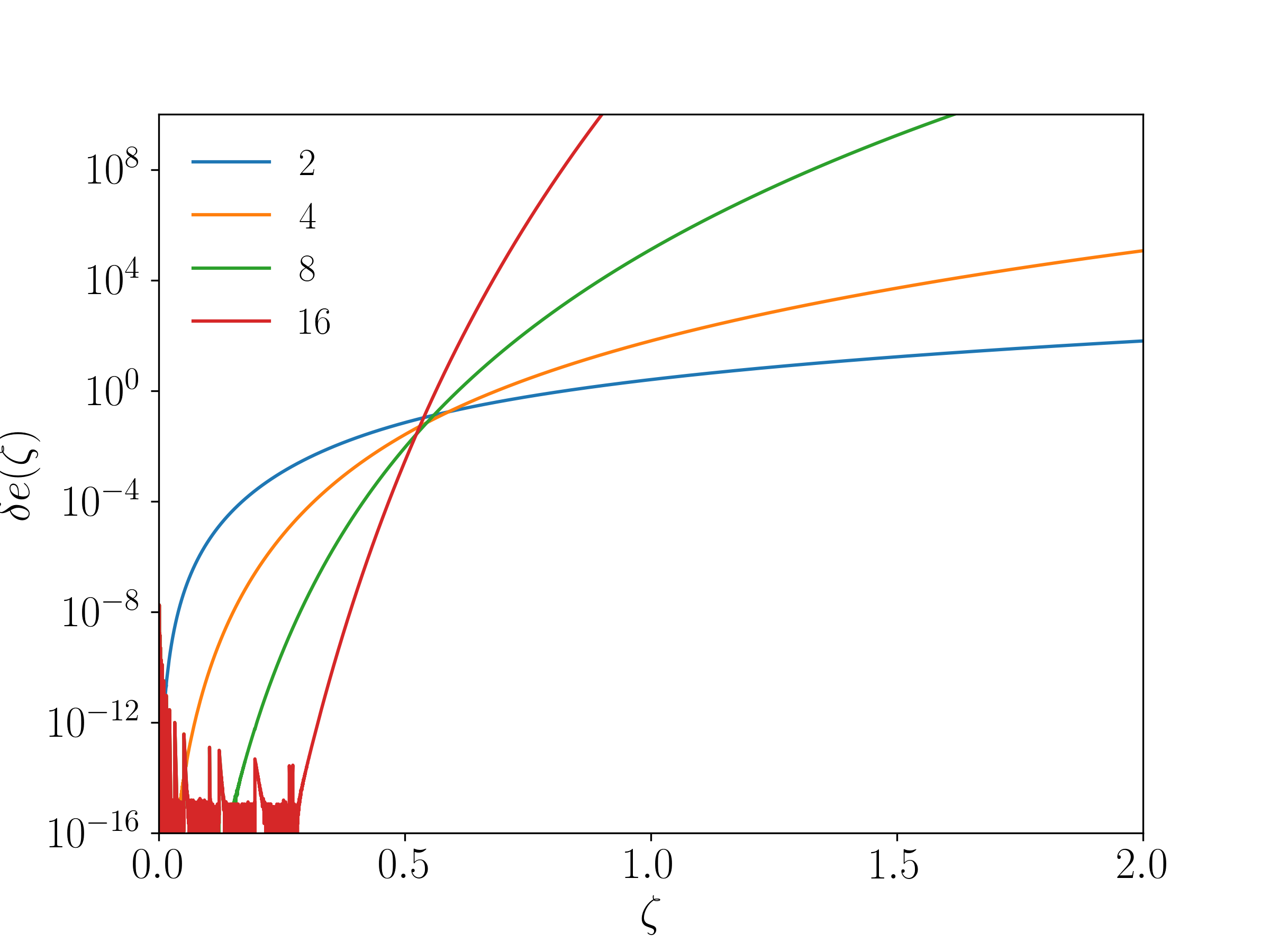} 
		\caption{\label{fig:efinverr}(Color Online) The relative error between a numerically solved $e(\zeta)$ and an $e(\zeta)$ obtained by inverting $\zeta(e)$ in the low eccentricity limit. Here $\delta e(\zeta) = 1-e_{\text{\rm \tiny app}}(\zeta)/e(\zeta)$, where $e_{\text{\rm \tiny app}}(\zeta)$ is the approximate inversion of $F(e)$ and $e(\zeta)$ is the exact solution. The different colored lines represent different values of $n_{\maxtxt}$ (number of terms kept in the Taylor expansion). For $\zeta \sim 0.5$ the error in $e(\zeta)$ becomes considerable which corresponds to $e_0 \sim 0.3$ regardless of the number of terms kept. This suggests that even keeping even more terms in the approximate inversion of $\zeta(e)$ will not increase its domain of validity.}
	\end{centering}
\end{figure}

Let us conclude by reviewing the standard low eccentricity inversion of $F(e)$ as given in Eq.~\eqref{eq:e_F}, and investigate its associated loss in match. Our task is then to solve Eq.~\eqref{eq:zetae} for $e(F)$ perturbatively in $e_{0} \ll 1$. We then wish to invert Eq.~\eqref{eq:zetae} in the limit of small $\zeta$, which corresponds to a low eccentricity and large frequency expansion, since $\zeta << 1$ requires $e_0 << 1$ and $(F_0/F) << 1$. For a system with initial eccentricity $e_0$, $\zeta \in [\sigma(e_0)(F_0/F_{\text{\rm \tiny LSO}})^{2/3},\sigma(e_0)]$  . We invert Eq.~\eqref{eq:zetae} by proposing a solution of the form
\begin{equation}
\label{eq:ezeta}
e(\zeta) \approx \zeta^{19/12}\sum_{n=0}^{n_{\maxtxt}}M_n\zeta^{19n/6}. 
\end{equation}
This proposed solution for $e(\zeta)$ is then substituted into $\sigma(e)$ appearing in Eq.~\eqref{eq:zetae} and expanded in small $\zeta$. The coefficients, $M_n$, are determined by demanding that $\sigma[e(\zeta)] = \zeta$.

Figure~\ref{fig:efinverr} shows the relative error between the solution in Eq.~\eqref{eq:ezeta} with different values of $n_{\maxtxt}$ and a numerical solution for $e(\zeta)$. Regardless of the number of terms kept in the approximate inversion, the error becomes considerable as $\zeta \sim 0.5$ which corresponds to an initial orbital eccentricity of $e_0 \sim 0.3$. This suggests that the PC models (where all quantities are expanded in low eccentricity) will become unfaithful near $e_0 \sim 0.3$ due to the approximate inversion of $F(e)$. Table \ref{tab:pcef} shows the match when using the low-eccentricity expansion of $e(\zeta)$ for different values of $n_{\maxtxt}$. Increasing $n_{\maxtxt}$ mitigates the loss in match until the inversion is pushed past $e_0=0.3$, at which point the value of the match drops to nearly 0 in one case as the inversion becomes unphysical. While keeping more terms may increase the accuracy at low $\zeta$, the solution becomes useless for $e_0 \gtrsim 0.35$. 

\begin{table*}
\begin{tabular}{||c|c|c|c|c|c|c|c||}
\hline
$i_{\maxtxt}$ & $e_0=0.05$ & $e_0=0.10$ & $e_0=0.15$ & $e_0=0.20$ & $e_0=0.25$ & $e_0=0.30$ & $e_0=0.35$ \\
\hline
2 & 0.9911 & 0.9870 & 0.9766 & 0.9531 & 0.9242 & 0.8921 & 0.8527\\
\hline
4 & 0.9911 & 0.9877 & 0.9908 & 0.9878 & 0.9799 & 0.9671 & 0.9465\\
\hline
8 & 0.9911 & 0.9877 & 0.9908 & 0.9903 & 0.9875 & 0.9806 & 0.9724\\
\hline
16 & 0.9911 & 0.9877 & 0.9908 & 0.9904 & 0.9884 & 0.9831 & 0.0057\\
\hline
exact & 0.9911 & 0.9877 & 0.9908 & 0.9904 & 0.9884 & 0.9841 & 0.9886\\
\hline
\end{tabular}
\caption{\label{tab:pcef}The value of the match for different values of $i_{\maxtxt}$ when inverting $F(e)$ in a low eccentricity approximation. The first column lists the value of $n_{\maxtxt}$ (the ``exact" in the last row indicates a numerical solution for $e(F)$), and the other columns list the value of the match for a given $e_0$. For each of the values shown here, the corresponding system is a BBH system with an initial semilatus rectum of $p_0=50$. The value of 0.0057 for the 16 term inversion at an $e_0$ of 0.35 is due to $e(\zeta)$ returning eccentricities less than 0. Presumably, as more terms are kept in the inversion $e(\zeta)$ is very accurate for small $\zeta$ and then becomes unusable for $\zeta \sim 0.5$ as indicated in Fig.~\ref{fig:efinverr} which corresponds to $e_0 \sim 0.3$. The loss in match becomes fairly large for initial eccentricities greater than $e_0 \sim 0.35$ regardless of the number of terms kept in the inversion.}
\end{table*} 

\section{Faithfulness Analysis of NeF Model}
\label{sec:faithharms}

In this section, we will determine the faithfulness of the NeF model by comparing it to the NeT model. All matches in this paper are computed with the response function $h(t)=F_{+}h_{+}(t)+F_{\times}h_{\times}(t)$, where the plus and cross polarizations are given by the different waveform models (either NeF or NeT), while $F_{+,\times}$ are the antenna patterns of the detector 
\begin{subequations}
\begin{align}
F_{+}(\theta,\Phi,\psi) &= \frac{1}{2}(1+\cos^{2}\theta)\cos 2\Phi \cos 2\psi
\nonumber \\ &
- \cos \theta \sin 2\Phi \sin 2\psi, \\
F_{\times}(\theta,\Phi,\psi) &= F_{+}(\theta,\Phi,\psi-\pi/4).
\end{align}
\end{subequations}
Here $\theta$ and $\Phi$ are the sky angles associated with the orientation of the detector and $\psi$ is the polarization angle that defines the rotation from the wave's polarization basis to that defined by the arms of the detector. All throughout, we choose $\theta=\Phi=\psi=\beta=\iota=3\pi/7$ as an arbitrary location and source orientation, where recall that $\beta$ and $\iota$ are the polar angles describing the polarization axes.

Before we can study the faithfulness of the NeF model of Sec.~\ref{sub:TayF2}, we must first decide the appropriate number of harmonics that ought to be kept in Eq.~\eqref{eq:SPA}. Section~\ref{sub:jmax} addresses this in detail. With this at hand, Section~\ref{subsec:faith} studies the faithfulness of the NeF model with the optimal number of harmonics kept against the NeT model.

\subsection{Maximum Number of Harmonics Needed}
\label{sub:jmax}

\begin{figure}[htp]
		\includegraphics[clip=true,angle=0,width=0.475\textwidth]{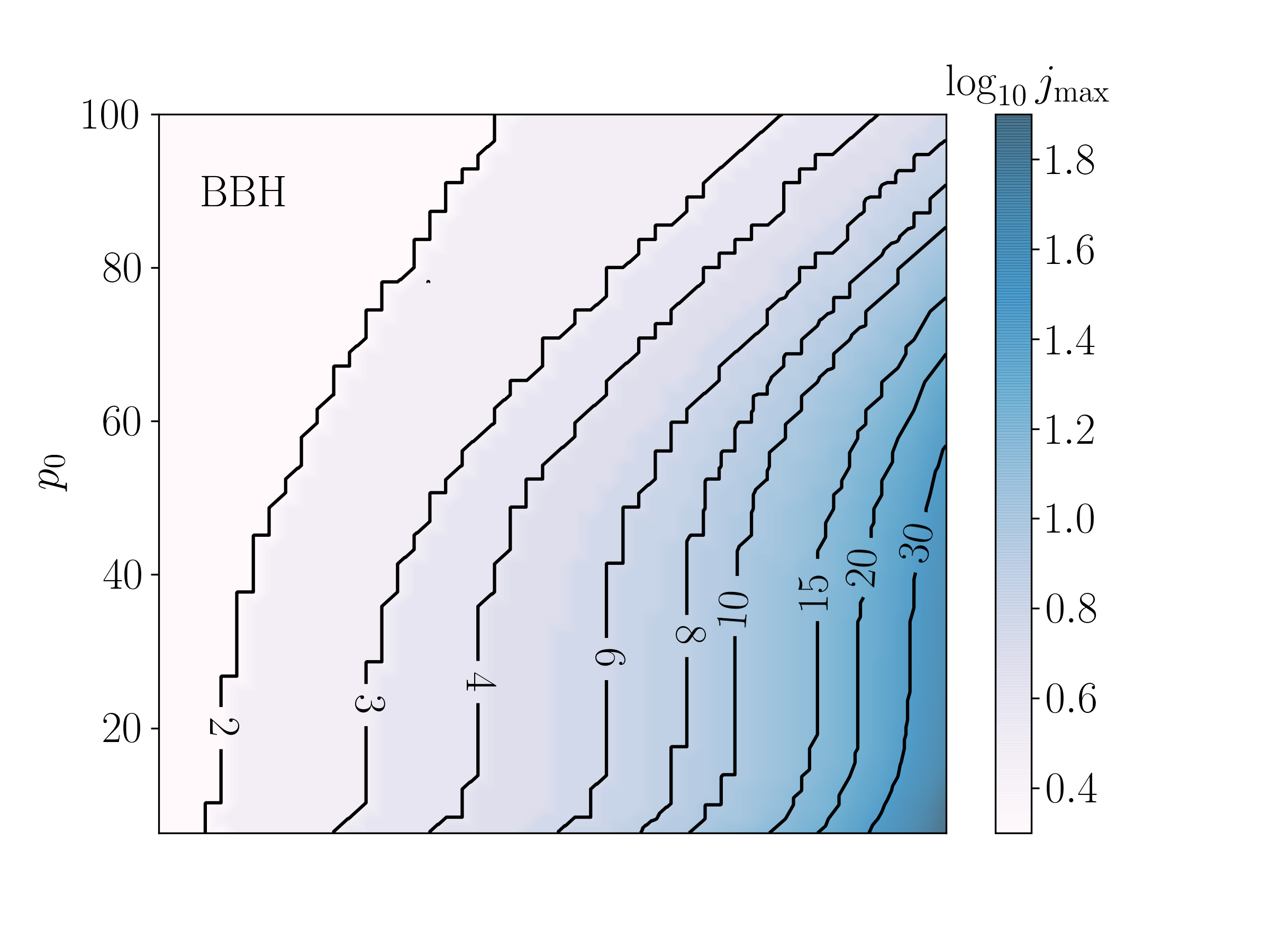} 
		\includegraphics[clip=true,angle=0,width=0.475\textwidth]{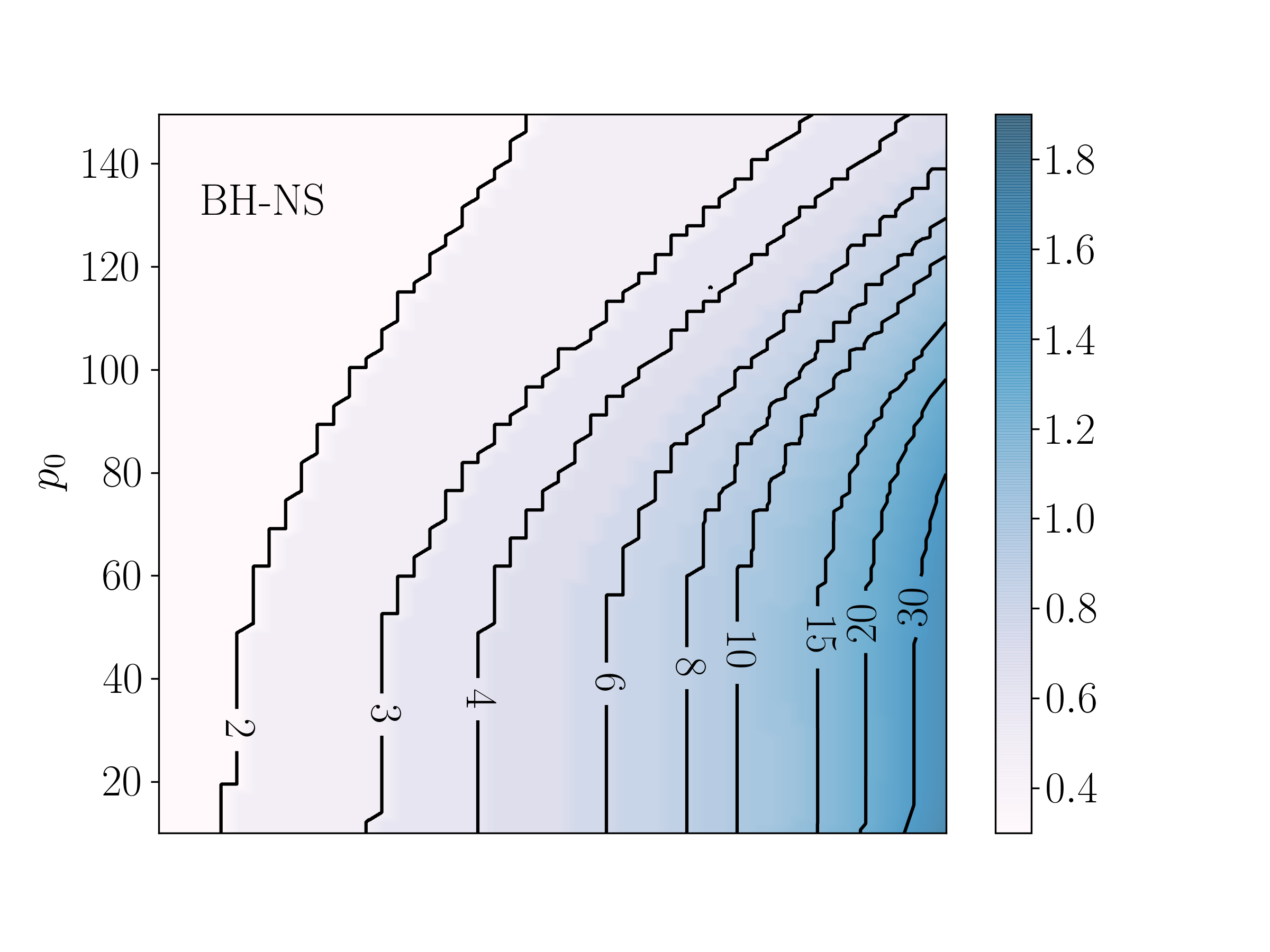} 
		\includegraphics[clip=true,angle=0,width=0.475\textwidth]{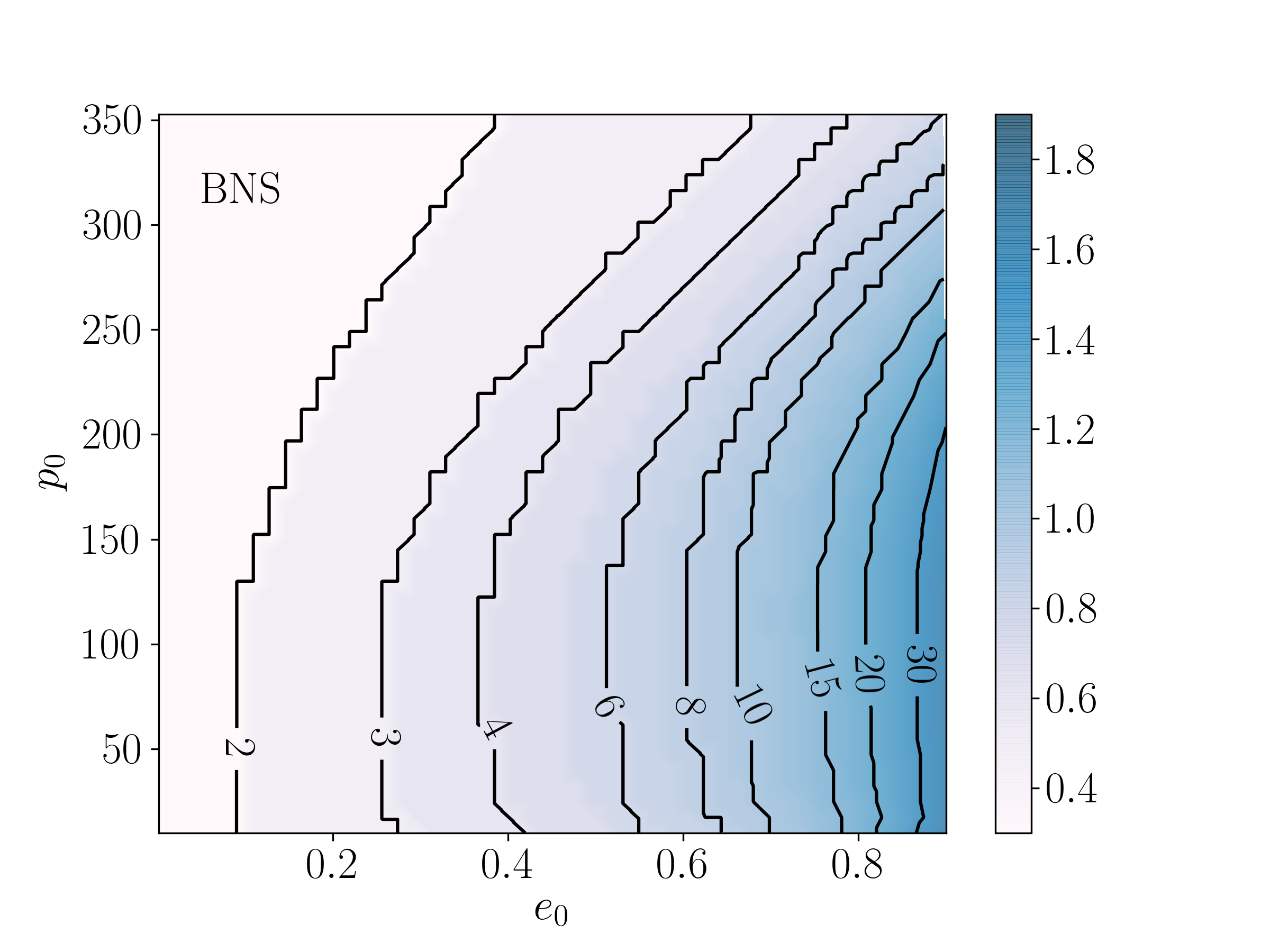} 
		\caption{\label{fig:numharm} (Color Online) The number of harmonics required to achieve a match greater than 0.99 as a function of $e_{0}$ and $p_{0}$ (the initial orbital eccentricity and dimensionless semilatus rectum) for a BBH (top), BH-NS (middle), and BNS (bottom) system. For systems with initial eccentricities less than $\sim 0.6$ (and even larger eccentricities at larger initial separations) we only require $10$ or less harmonics to achieve a match of 0.99.}
\end{figure}

We begin by studying the optimal number of harmonics that need to be kept in order for the match to be above 99\%. We do so by computing the match between the NeT model (which contains all harmonics) and a harmonically-decomposed time domain model (obtained by solving $\dot{F}$, $\dot{e}$, $\dot{l}$ and using Eq.~\eqref{eq:timedomdecomp}). We then demand that we keep as many harmonics in the latter such that its match against NeT exceeds 0.99. This value of the match ensures the statistical error exceeds the systematic error due to mis-modeling for sources of roughly SNR 20 or less, as discussed in Sec.~\ref{sec:maximization}. 

Figure~\ref{fig:numharm} shows the number of harmonics needed in the time domain harmonic decomposition for the match against NeT to exceed 0.99. Results are shown for three different fiducial systems, a $(1.4,1.4)M_{\odot}$ binary neutron star (BNS) system, a $(1.4,10)M_{\odot}$ neutron star - black hole (BH-NS) system, and a $(10,10)M_{\odot}$ binary black hole (BBH) system. For systems with initial eccentricities less than $\sim 0.6$ (and for even larger eccentricities at larger initial separations), we only require $10$ or less harmonics to achieve this threshold match. We thus expect that one need only keep just as many harmonics in the NeF model. This result is very promising as the biggest computational expense of the NeF model is the inclusion of higher harmonics.

Attempts have been made to analytically estimate the number of harmonics that one must keep in order to recover $99\%$ \textit{signal power}. Reference~\cite{2009MNRAS.395.2127O} argued that since the power in GWs is peaked near $\omega_{p}$ (orbital frequency at pericenter) and the fractional power in GWs is less than $10^{-3}$ past $5\omega_p$ \cite{1977ApJ...216..610T}, one can then approximate the number of harmonics needed to capture $99\%$ of signal power via
\begin{equation}
\label{eq:jmaxoleary}
j_{\maxtxt }=\frac{5\omega_p}{\omega_{orb}}=5\frac{(1+e_0)^{1/2}}{(1-e_0)^{3/2}}.
\end{equation}
Reference~\cite{2003ApJ...598..419W} takes a different approach and uses~\cite{PetersMathews} to calculate the power in a given harmonic. With this at hand, they then compute the harmonic with the most power, $j_{\text{\rm \tiny m}}$, for a collection of different eccentricities, which they finally fit to obtain
\begin{equation}
\label{eq:Wen}
j_{\text{\rm \tiny m}}=2\frac{(1+e)^{1.1954}}{(1-e^2)^{3/2}} \, .
\end{equation}
Equation \eqref{eq:Wen} correctly limits to the $j_{\text{\rm \tiny m}}=2$ quasi-circular harmonic as $e \rightarrow 0$, while Eq.~\eqref{eq:jmaxoleary} reduces to $j_{\maxtxt}=5$ in this limit. 

\begin{figure}[htp]
		\includegraphics[clip=true,angle=0,width=0.5\textwidth]{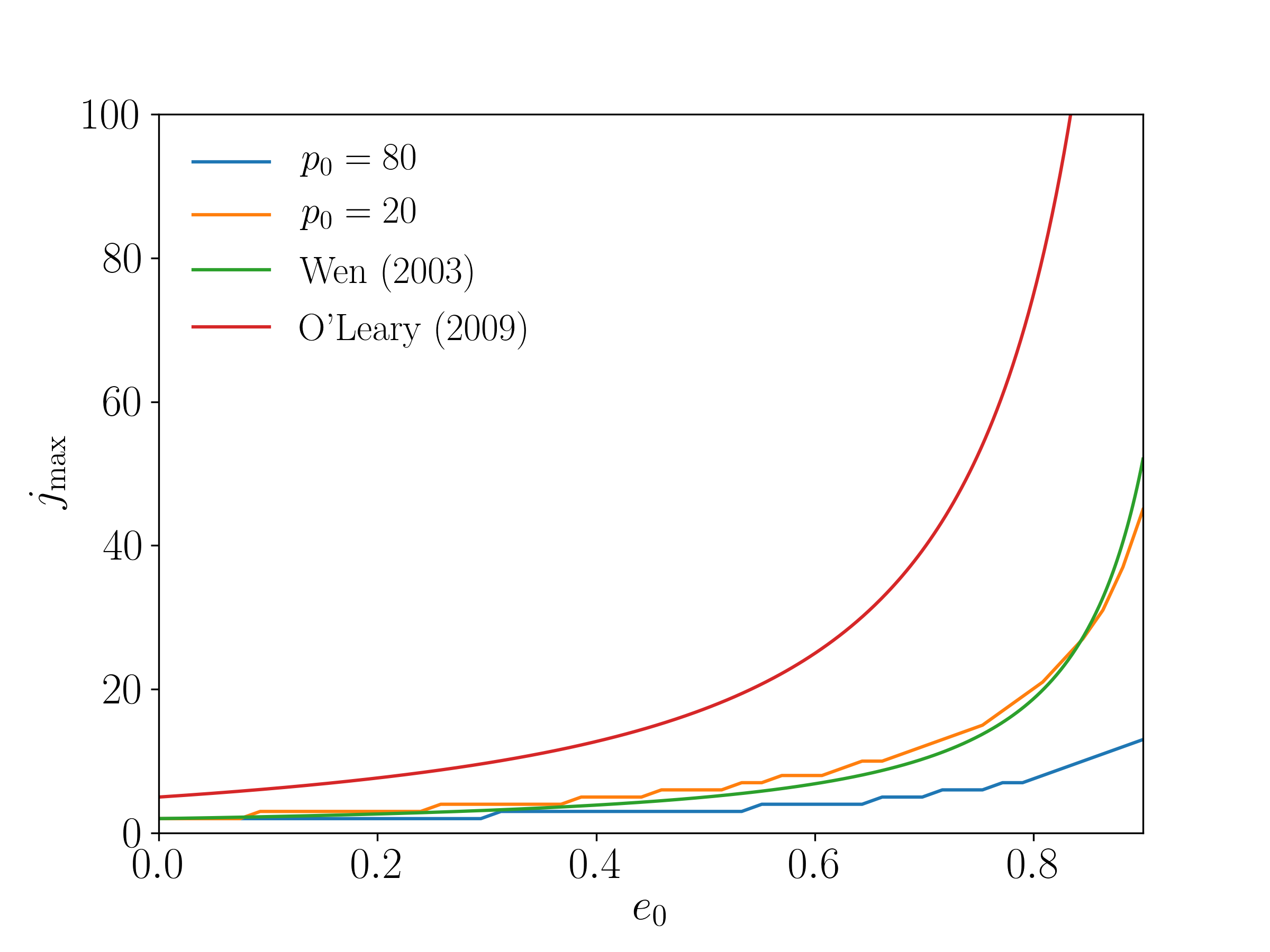} 
		\caption{\label{fig:jmaxoleary} (Color Online) Number of harmonics needed as predicted by Eq.~\eqref{eq:jmaxoleary} in red and Eq.~\eqref{eq:Wen} in green. In orange and blue we plot the number of harmonics needed for a fixed $p_0$ using the BBH results of Fig.~\ref{fig:numharm} for a close initial dimensionless semi-latus rectum ($20$) and a wider initial dimensionless semi-latus rectum ($80$). For close initial separation, Eq.~\eqref{eq:Wen} gives a good estimate for the number of harmonics needed. As the initial separation increases, one needs even less harmonics than predicted by Eq.~\eqref{eq:Wen}. Equation \eqref{eq:jmaxoleary} overestimates the number of harmonics needed in all cases.}
\end{figure}

Figure~\ref{fig:jmaxoleary} compares Eq.~\eqref{eq:jmaxoleary} to Eq.~\eqref{eq:Wen} and to a cross-section of our results for a fixed $p_0 = 20$ and $p_{0} = 80$, using a BBH system. Observe that Eq.~\eqref{eq:jmaxoleary} overestimates the number of harmonics needed regardless of initial separation. This is because this equation is derived in the limit of an unbound orbit, and thus, it behaves poorly for moderate eccentricities. Also, this equation is designed to recover $99\%$ of the signal \textit{power}, whereas the match, and the inner product, are weighted by the spectral noise density of the detector; therefore, $99\%$ signal power does not imply a $99\%$ match (the threshold for our choice of $j_{\maxtxt}$). 
	
Equation \eqref{eq:Wen} does remarkably well at predicting the number of harmonics needed when the initial separation is small. This result is particularly surprising because Eq.~\eqref{eq:Wen} predicts the harmonic at which the power spectrum peaks, but referring back to Fig.~\ref{fig:no_RR}, although the amplitude peaks at the $4^{\rm \tiny th}$ harmonic, the higher harmonics are still relatively strong. One would therefore reasonably expect that significantly more harmonics should be kept than just the peak harmonic. However, the eccentricity of the binary is rapidly decaying, and as a result, the signal power in higher harmonics also rapidly decays. Coincidentally, the harmonic at which the power spectrum peaks at the initial orbital eccentricity gives a fair estimate for the number of harmonics needed for faithful matches. However, as the initial separation increases, even Eq.~\eqref{eq:Wen} overestimates the number of harmonics needed. For such large initial separations, the power of the higher harmonics has sufficiently decreased by the time they enter the sensitivity band, and thus, they affect the match even less. It is also worth reiterating that this comparison was done requiring a 0.99 match (which is driven by the desire to reduce systematic biases for signals with roughly SNR 20 or less), but to achieve higher matches (considering sources with higher SNR) one would need to keep more harmonics.

\subsection{Faithfulness of the SPA}
\label{subsec:faith}

Finally, in this section we estimate the faithfulness and the domain of validity of the NeF model of Sec.~\ref{sub:TayF2}. To do so, we truncate the NeF model at the maximum harmonic index found in Sec.~\ref{sub:jmax}. We then compute the match between the NeF and the NeT models. Since the harmonic index at which we choose to truncate our frequency domain model corresponds to the number of harmonics needed to achieve a match of at least $0.99$ in the numerical time domain harmonic decomposition, the best match the NeF model can possibly achieve is also $0.99$. Of course, this maximum could be improved by keeping more harmonics, but we expect the results we obtain will not qualitatively change. 

The NeT and NeF models are discussed in detail in Sec.~\ref{subsec:T4} and Sec.~\ref{sub:TayF2} respectively, but we go over a few key details here related to the match calculations. Recall that the NeT model is evolved from an initial eccentricity and semi-latus rectum $(e_0,p_0)$ to the LSO, which corresponds to a mean orbital frequency $F_0$ to $F_{LSO}$. This time series is then zero padded, such that the length of the new time series is a power of 2, with the result then discrete Fourier transformed. The NeF model, on the other hand, is sampled at the frequencies associated with the discrete Fourier transform of the NeT model. In order to do, so we numerically invert $F(e)$ in order to map the Fourier frequency, $f$, to the mean orbital frequency, $F$, which is then further mapped to the orbital eccentricity, $e$, which appears explicitly in the harmonic amplitudes and phases.

Once both models are sampled we compute the match with all integrals over frequency evaluated as discrete sums from $f\in [0,f_{ny}]$, and then maximized as described in Sec.~\ref{sec:maximization}. We do not window the NeF model, but we do use a square window for the NeT model. Two independent versions of this code were implemented, one in \texttt{Mathematica}, and one in \texttt{C} and found to agree in the match to within $\sim 0.1 \%$ absolute error. We take this to be an estimate of the numerical error of the match presented here.
\begin{figure*}[htp]
		\includegraphics[clip=true,angle=0,width=0.475\textwidth]{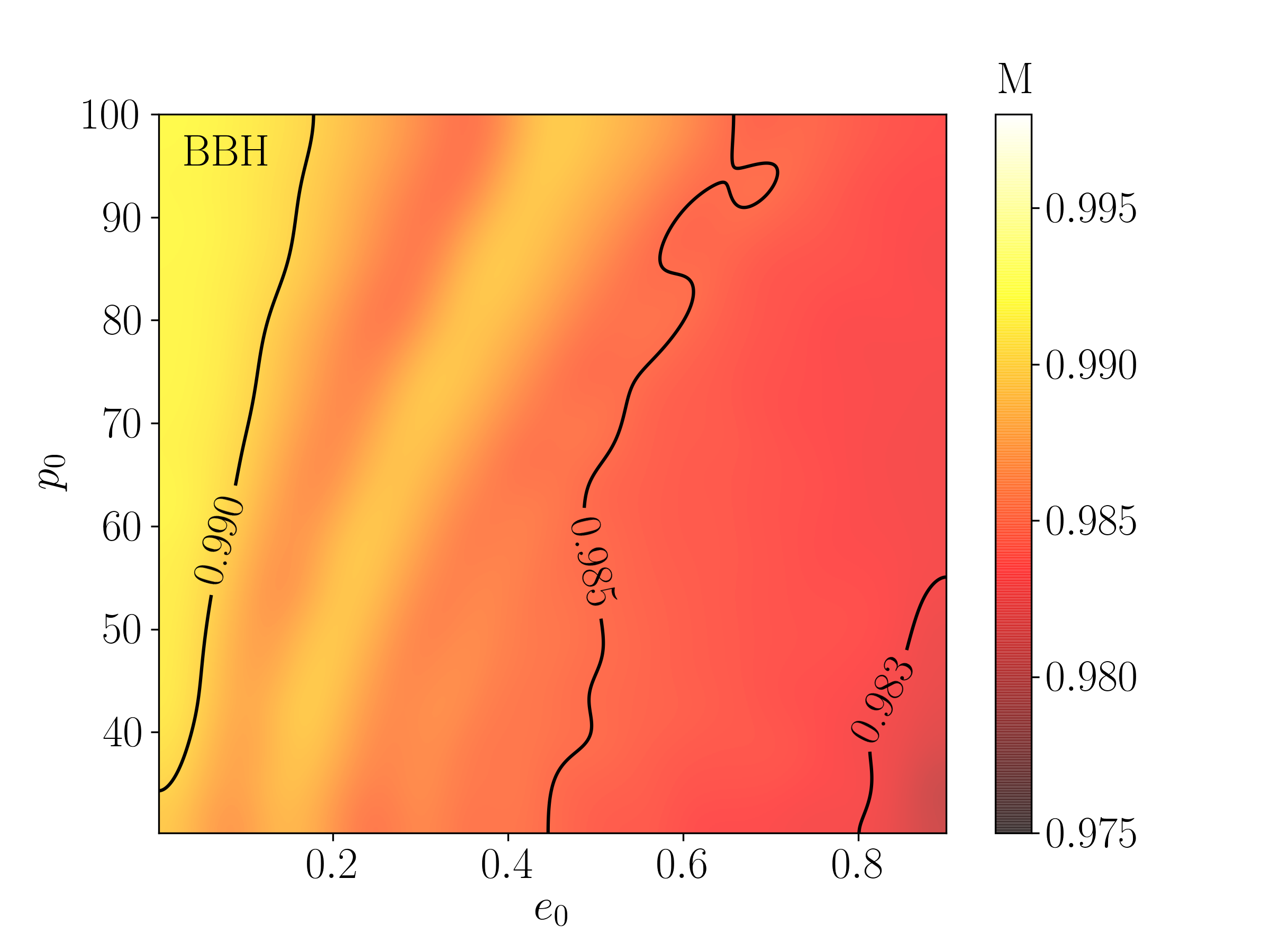} 
		\includegraphics[clip=true,angle=0,width=0.475\textwidth]{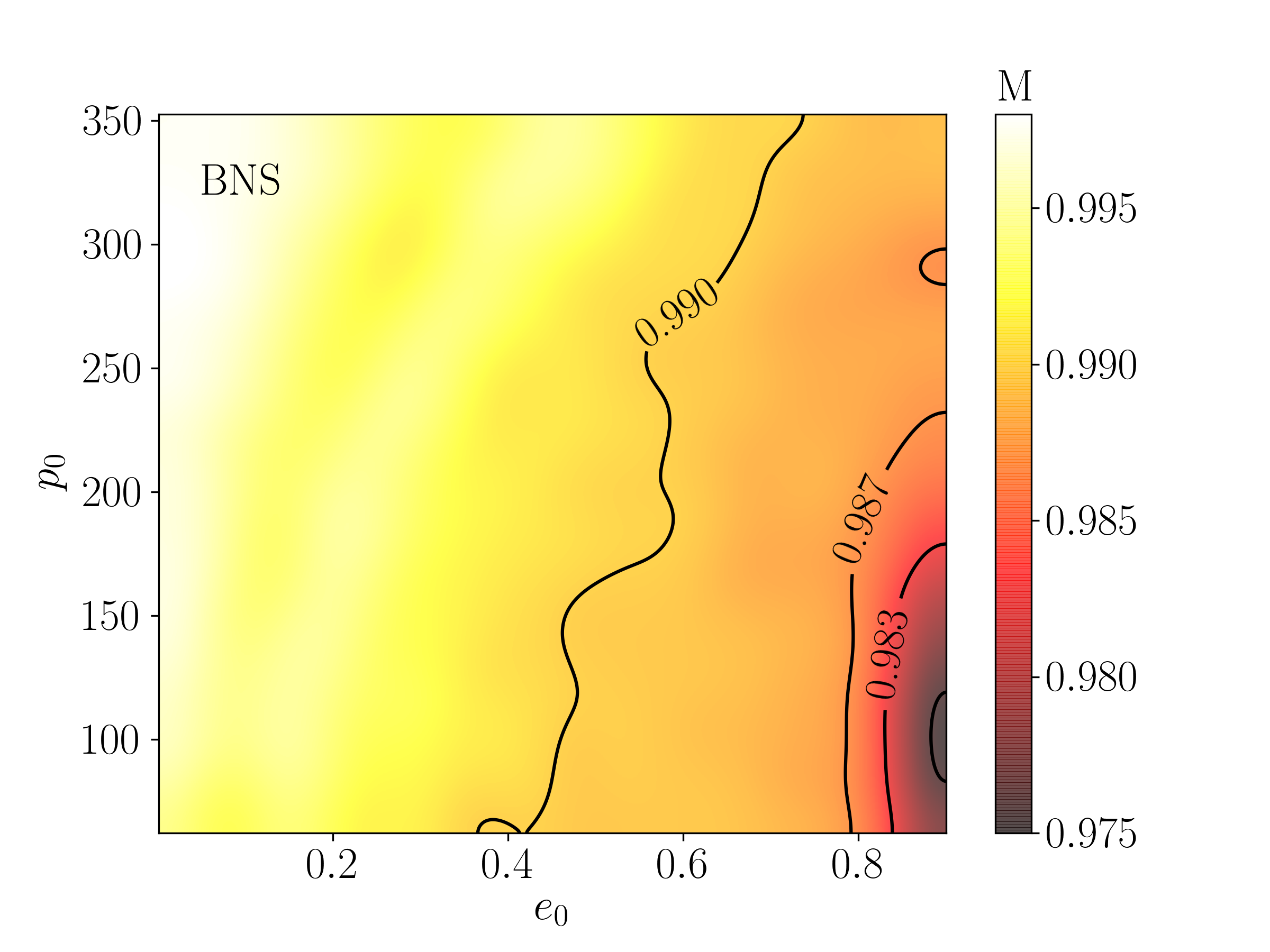} 
		\caption{\label{fig:bbhmatch} (Color Online) Match between the NeT model described in Sec.~\ref{subsec:T4} and the NeF model as described in Sec.~\ref{sub:TayF2} for a BBH (top) and BNS (bottom) system (a BH-NS case is presented in Fig.~\ref{fig:nsbhmatch}). The match is given as a function of the initial condition $(e_0,p_0)$ on the $x$ and $y$ axis. The maximum match is $~0.99$, and for a vast majority of the parameter space, NeF very faithfully represents NeT. For the shorter lived source (BBH), the match behaves more poorly than its longer lived counterparts due to finite time effects. The boundaries of the fringes correspond to a boundary where the model demands including more harmonics and thus the match displays a sharp increase as the next harmonic is included.} 
\end{figure*}	

Figure~\ref{fig:bbhmatch} shows the match for two of the three systems considered in the last section (BNS, BBH, BH-NS), for different initial eccentricities and dimensionless semi-latus recta. Figure \ref{fig:nsbhmatch} shows the corresponding figure for the BH-NS case. All $\sim 2,500$ matches are Gaussian filtered in order to obtain smooth contours in the figure. In reality, some of the matches presented in Fig.~\ref{fig:bbhmatch} may be above or below the value of the corresponding contour, but the contours represent the general features.

Observe that the NeF model matches the NeT model extremely well in nearly all cases. As the initial eccentricity increases and the initial semi-latus rectum decreases, however, the match begins to deteriorate from 0.99 (which is nearly perfect due to the truncation of the harmonic sum) to 0.985 and 0.98. This is the result of finite time effects in the DFT of NeT and windowing. Finite time effects result in a loss of match (see \cite{Droz:1999qx} and \cite{2000PhRvD..62h4036D}) and this is exacerbated by the fact that eccentric binaries evolve faster than their quasi-circular counterparts (for the same initial mean orbital frequency). This reasoning is supported by the fact that the match is higher for longer lived sources (BH-NS and BNS) than shorter lived ones (BBH) for the same initial eccentricity and mean orbital frequency.  If one were interested in a model with a higher value of the match (i.e.~one faithful for higher SNR sources), one could include more harmonics in the SPA sum to obtain a higher match than that required in Sec.~\ref{sub:jmax}. One could also attempt to incorporate finite time effects into the SPA itself, as done in \cite{2000PhRvD..62h4036D} or minimize finite time effects by applying the same window functions to both models.

In the BNS system, the match begins to drop steeply to as low as $0.975$ near $e_0 \sim 0.85$ even for relatively large initial separations. The loss of match here is not due to finite time effects, but rather it is due to the inaccuracy of the approximation to the Fourier phase in Eq.~\eqref{eq:chebapprox}. One could improve this match by increasing the number of terms kept in the Chebyshev resummation and in the Taylor expansion. Alternatively, one could choose a different number of terms in the representation of the Fourier phase for sources with different lifetimes, as longer lived sources are more sensitive to error in the representation of the phase. For shorter lived sources, the phase could likely be approximated even more cheaply than done here. We leave a more thorough investigation of such an implementation to future work. 

The sudden jumps and drops in the value of the match that appear most prominently near the low eccentricity region of Fig.~\ref{fig:bbhmatch} are due to the harmonic nature of the model. For a fixed number of harmonics kept, the value of the match begins to decrease as the initial eccentricity increases. When the model demands the inclusion of another harmonic (informed by the results of Sec.~\ref{sub:jmax}), the value of the match sharply increases. As such, the sudden jumps correspond to the contours of $j_{\maxtxt}$ shown in Fig.~\ref{fig:numharm}.

\section{Conclusions \& Future Work}
\label{sec:futwork}

We have taken the first steps toward the construction of a Fourier domain model that is valid for parameter estimation to arbitrary eccentricity. In particular, we explicitly calculated an efficient analytic frequency domain model to Newtonian order, the NeF model, which we show is faithful to very high initial eccentricity $(e_0 \sim 0.9)$ through match calculations against a numerical time-domain model to Newtonian order, the NeT model (see Fig.~\ref{fig:nsbhmatch} and \ref{fig:bbhmatch}). As byproducts of this analysis, we also derived new techniques to rapidly maximize the match over phase and time at coalescence for eccentric signals, and we analyzed the most important source of errors in the NeF model. Given its analytic features, we have analytic control over the NeF model, which thus allows us to straightforwardly improve it by keeping higher order terms in the analytic expansions. 

As a proof-of-concept, we have here presented the NeF model to leading PN order, so it is not yet useful for data analysis. Even once higher PN terms are included, the model as presented here is likely still not accurate enough for a LISA-like detector, which will require a higher match for high signal-to-noise ratio events. Both of these problems, however, can be straightforwardly solved by (i) extending this NeF model to higher PN order and (ii) retaining more terms in the Chebyshev representation of the Fourier phase and more harmonics in the SPA amplitude. All of this future work will make for very interesting extensions of the NeF model. 

The least straightforward extension is the inclusion of higher PN order terms. Much of the work required for this, however, already exists. The fluxes and time derivatives of the  orbital elements have been computed in \cite{Arun:2007rg, Arun:2007sg, Arun:2009mc, 2017CQGra..34d4003L} to 3PN order. The time domain harmonic decomposition becomes troublesome at higher PN order, as Kepler's equation are modified at 2PN order. A method to treat the time domain harmonic decomposition at 3PN, however, is introduced in \cite{2017PhRvD..96d4011B}. Therefore, the extension of the NeF model to higher PN order requires the careful assembly of pieces that are already present in the literature. 

With a higher PN waveform in hand, one could in the future conduct many useful studies. For example, one could investigate the ability of aLIGO, LISA, or 3G detectors to measure the parameters and any biases of the model given eccentric signals. Since eccentric binaries are thought to be the result of specific astrophysical scenarios, it could be possible to study whether the detection of eccentric GWs can constrain different formation scenarios by recovering $e_0$ and $F_0$. 

Another useful future study is to incorporate effects of modified gravity theories into the new model. If a modified gravity theory modifies the eccentricity evolution of a binary, such a theory could be constrained more stringently than with quasi-circular events only. This is because eccentricity excites additional harmonics in the signal, which contain useful information that could be leveraged to break degeneracies and constrain General Relativity. One such approach has already been taken by extending the parameterized post-Einsteinian (ppE) framework~\cite{PPE} to high-eccentricity bursts~\cite{2014PhRvD..90j4010L}.  

Perhaps the most difficult and rewarding goal is an inspiral-merger-ringdown (IMR) type hybrid that incorporates the effects of orbital eccentricity of moderate and large magnitude. Currently, the only IMR model that incorporates orbital eccentricity is that of~\cite{2018PhRvD..97b4031H}. This model is constructed in the time domain using PN theory, results of numerical relativity, and a machine learning algorithm to tune the model. It would be interesting to compare that model to this future hybrid model, once one is developed. 

\section*{Acknowledgments}
B.M. was supported by the Joan L. Dalton Memorial Fellowship in Astronomy from Montclair State University. T. R. appreciates the support of the NASA grant NNX16AB98G. N. Y. acknowledges support from NASA grants NNX16AB98G and 80NSSC17M0041. Computational efforts were performed on the Hyalite High Performance Computing system, which is supported by University Information Technology at Montana State University.

\appendix
\section{Application of the Stationary Phase Approximation}
\label{app:SPA}
In this Appendix we review the application of the stationary phase approximation (SPA) as applied to the time domain harmonic decomposition of the GW signal given in Eq.~\eqref{eq:timedomdecomp}. For a more detailed and general presentation of the approximation see \cite{Bender}. We begin by considering the time domain GW signal generated by the $j^{\rm th}$ harmonic: 
\begin{align}
h_j(t) \! = \! \left[A_j(t)e^{-ijl(t)} \! + \! \text{c.c.}\right] \! \Theta \! \left(F(t)-F_0\right) \! \Theta \! \left(F_{\text{\rm \tiny LSO}}-F(t)\right),
\end{align}
where $\text{c.c.}$ stands for the complex conjugate and the amplitude is given by
\begin{equation}
A_j(t)=-\frac{m \eta}{2R}(2\pi m F)^{2/3}\left[C_{+,\times}^{(j)}(t)+iS_{+,\times}^{(j)}(t)\right] \, .
\end{equation}
The unit step function, $\Theta(x)$, arises because the binary is assumed to be formed at a time $t_0$ with mean orbital frequency $F_0$ and the emission is terminated at the LSO when the binary has mean orbital frequency $F_{\text{\rm \tiny LSO}}$ (as is customary in the GW literature).

We wish to approximate the Fourier transform of the GW signal given by
\begin{align}
\label{eq:FFint}
\tilde{h}_j(f)&=\int_{-\infty}^{\infty}\left\{A_{j}(t)e^{i[2\pi f t-jl(t)]}+A_{j}^{\ast}(t)e^{i[2\pi f t+jl(t)]}\right\} \nonumber \\
& \times \Theta\left(F(t)-F_0\right)\Theta\left(F_{\text{\rm \tiny LSO}}-F(t)\right)dt \, .
\end{align}
In the above integrand, the amplitude varies much more slowly than the mean anomaly ($\dot{A_{j}}/A_{j}<<j\dot{l}$) appearing in the phase. Thus, for most values of $t$ this integrand is rapidly oscillating. However, there exists a time where the phase of the integrand is roughly constant and thus strongly contributes to the integral. This time is called the stationary time (denoted $t_j^{\ast}$) and it occurs when the stationary phase condition is satisfied, i.e.~when the first derivative of the argument of the complex exponential vanishes. Thus, the stationary phase condition is
\begin{equation}
2 \pi f \mp j \dot{l}(t_j^{\ast})=0 \, ,
\end{equation}
where the minus sign corresponds to the stationary phase condition of the first term in the integrand, and the plus corresponds to the second. Since the time derivative of the mean anomaly is the mean orbital frequency, a positive quantity, the stationary phase condition is never satisfied for the second term in the integrand and so it does not contribute strongly to the integral. The stationary phase condition that is satisfied provides a mapping between the Fourier frequency $f$ and the mean orbital frequency $F$ for the $j^{\rm th}$ harmonic: $f=jF$.

To approximate the integral in Eq.~\eqref{eq:FFint}, we Taylor expand the amplitude to leading order and the phase to first order (as the amplitude is more slowly varying than the phase) about the stationary time and evaluate the unit step functions at the stationary time:
\begin{align}
\tilde{h}_j(f) & \approx A_j(t_j^{\ast})e^{i(2\pi f t_j^{\ast}-jl(t_j^{\ast}))} \! \Theta \! \left(f-jF_0\right) \! \Theta \! \left(jF_{\text{\rm \tiny LSO}}-f\right) \nonumber \\
& \times \int_{-\infty}^{\infty}e^{-ij\frac{\ddot{l}}{2}(t_j^{\ast})(t-t_j^{\ast})^2}dt \, .
\end{align}
We carry out the above integral and obtain
\begin{align}
\tilde{h}_j(f) & \approx  A_j(t_j^{\ast}) \sqrt{\frac{2\pi}{|j\ddot{l}(t_j^{\ast})|}}e^{i(2\pi f t_j^{\ast}-jl(t_j^{\ast})-\pi/4)} \nonumber \\
& \times \Theta\left(f-jF_0\right)\Theta\left(jF_{\text{\rm \tiny LSO}}-f\right) \, .
\end{align}
Rewriting the above in terms of more familiar quantities we have
\begin{align}
\tilde{h}_j(f) & \approx -\frac{m \eta}{2 R} \frac{(2\pi m F(t^{\ast}_j))^{2/3}}{\sqrt{j \dot{F}(t_{j}^{\ast})}} \left[C^{(j)}_{+,\times}(t_{j}^{\ast}) + i S^{(j)}_{+,\times}(t_{j}^{\ast})\right] \nonumber \\
& \times e^{i\psi_j} \Theta\left(f-jF_0\right)\Theta\left(jF_{\text{\rm \tiny LSO}}-f\right) \, ,
\end{align}
with 
\begin{equation}
\label{eq:SPA-def-pieces}
\psi_j =2\pi f t_j^{\ast} - jl(t_j^{\ast})- \pi/4.
\end{equation}
To recover the full frequency response due to many harmonics, we simply sum the above over the harmonic index $j$.
\section{Chebyshev Resummation}
\label{app:phasecoef}

In this appendix we review the approximation of the Fourier phase and provide the coefficients of the resulting series in eccentricity to 16 digits of precision. Since the exact coefficients are unwieldy rationals, they are provided in a supplementary notebook. We begin by approximating $I(e)$ in Eq.~\eqref{eq:psij} through a Maclaurin series
\begin{equation}
\label{eq:tayexp}
I(e) \approx \sum_{n=0}^{n_{\maxtxt}}A_{2n} e^{2n}.
\end{equation}
We use an even basis function for the expansion about small eccentricity because $I(e)$ is even. The coefficients $A_{n}$ are simply the coefficients of the Taylor expansion, namely
\be
A_{n} = \frac{1}{n!} \left.\frac{d^{n} I}{de^{n}}\right|_{e=0}\,.
\ee
The first seventeen non-zero $A_{n}$ are given by 
\begin{align}
A_{0} &= \num{-0.2375000000000000} \, \,, \nonumber \\ 
A_{2} &= \num{-0.3006158763229894} \, \,, \nonumber \\ 
A_{4} &= \num{-0.008991117053859665} \, \,, \nonumber \\ 
A_{6} &= \num{-0.03785183403118392} \, \,, \nonumber \\ 
A_{8} &= \num{-0.007615853560080763} \, \,, \nonumber \\ 
A_{10} &= \num{-0.009769540332607544} \, \,, \nonumber \\ 
A_{12} &= \num{-0.004556366050156979} \, \,, \nonumber \\ 
A_{14} &= \num{-0.003968026824286266} \, \,, \nonumber \\ 
A_{16} &= \num{-0.002630835486794598} \, \,, \nonumber \\ 
A_{18} &= \num{-0.002115104845741527} \, \,, \nonumber \\ 
A_{20} &= \num{-0.001610593129877186} \, \,, \nonumber \\ 
A_{22} &= \num{-0.001304358739038224} \, \,, \nonumber \\ 
A_{24} &= \num{-0.001055578416146812} \, \,, \nonumber \\ 
A_{26} &= \num{-0.0008755814223244812} \, \,, \nonumber \\ 
A_{28} &= \num{-0.0007329639089438226} \, \,, \nonumber \\ 
A_{30} &= \num{-0.0006219751825605856} \, \,, \nonumber \\ 
A_{32} &= \num{-0.0005327237934609112} \, \,.  
\end{align} 

We then wish to resum this Taylor expansion using Chebyshev polynomials (see \cite{2007PhRvD..76j4002P} for a discussion of the convergence properties of Chebyshev series). The Chebyshev polynomials, $T_n(x)$, are defined for $x \in [-1,1]$, and thus, to make them a function of the eccentricity and preserve their orthogonality we rescale them as $T_n(s)$ where $s=2e^2-1$, which guarantees $s \in [-1,1]$ for $e \in [0,1]$. The first few Chebyshev polynomials are 
\begin{subequations}
\begin{align}
T_0(s) &=1 \,\,, \\
T_1(s) &=2e^2-1 \,\,, \\
T_2(s) &=8e^4-8e^2+1 \,\,, \\
T_3(s) &=32e^6-48e^4+18e^2-1 \,\,,
\end{align}
\end{subequations}
which can be inverted to find different powers of the eccentricity as functions of Chebyshev polynomials
\begin{subequations}
\label{eq:mono}
\begin{align}
e^0 &=T_0(s) \,\,, \\
e^2 &=\frac{1}{2}(T_1(s)+T_0(s)) \,\,, \\
e^4 &=\frac{1}{8}(T_2(s)+4T_1(s)+3T_0(s)) \,\,, \\
e^6 &=\frac{1}{32}(T_3(s)+6T_2(s)+15T_1(s)+10T_0(s)) \,\,.
\end{align}
\end{subequations}

We now seek a resummation of Eq.~\eqref{eq:tayexp} into the form 
\begin{equation}
\label{eq:cheb}
I(e) \approx \sum_{k=0}^{k_{\maxtxt}}B_k T_{k}(s)\,,
\end{equation}
for some set of coefficients $B_{k}$. We do so by inserting the monomials of $e$ in terms of the Chebyshev polynomials, Eq.~\eqref{eq:mono}, into Eq.~\eqref{eq:tayexp} and collecting terms order by order in the Chebyshev polynomials. The first terms for $B_{k}$ are 
\begin{align}
B_{0} &= \num{-0.4113459107529266} \, \,, \nonumber \\
B_{1} &= \num{-0.1867242969578177} \, \,, \nonumber \\
B_{2} &= \num{-0.01697087007811417} \, \,, \nonumber \\
B_{3} &= \num{-0.005185894691512303} \, \,, \nonumber \\
B_{4} &= \num{-0.001448030529666706} \, \,, \nonumber \\
B_{5} &= \num{-0.0004670829989068411} \, \,, \nonumber \\
B_{6} &= \num{-0.0001467244779335400} \, \,, \nonumber \\
B_{7} &= \num{-0.00004385063990056747} \, \,, \nonumber \\
B_{8} &= \num{-0.00001199101619561229} \, \,, \nonumber \\
B_{9} &= \num{-2.925622159845224} \times 10^{-6} \, \,, \nonumber \\
B_{10} &= \num{-6.210389911462248} \times 10^{-7}  \, \,, \nonumber \\
B_{11} &= \num{-1.117172525811035} \times 10^{-7}  \, \,, \nonumber \\
B_{12} &= \num{-1.649269175725618} \times 10^{-8} \, \,, \nonumber \\
B_{13} &= \num{-1.913379819810920} \times 10^{-9} \, \,, \nonumber \\
B_{14} &= \num{-1.632587267118429} \times 10^{-10} \, \,, \nonumber \\
B_{15} &= \num{-9.096722174897557} \times 10^{-12} \, \,, \nonumber \\
B_{16} &= \num{-2.480688474424701} \times 10^{-13} \, \,. 
\end{align}
Clearly if the Maclaurin series is truncated at $n_{\maxtxt}$, and the Chebyshev series is truncated at the same order, then the two approximations are identical. However, we find that we can keep less terms in the Chebyshev series than the Maclaurin series to accurately approximate the phase. Specifically, we find that we obtain a sufficiently accurate representation of $I(e)$ if we use a Maclaurin series truncated at $n_{\maxtxt}=16$ and a Chebyshev resummation truncated at $k_{\maxtxt}=12$.
  
After truncating the Chebyshev series in Eq.~\eqref{eq:cheb} at $k_{\maxtxt}=12$, it is more computationally efficient to collect like terms in eccentricity when implementing this function, lest one evaluates the same power of eccentricity many times. After collecting the like terms in eccentricity we are left with
 \begin{align}
 I(e)& \approx \sum_{n=0}^{n=12}C_n e^{2n}.
 \end{align}
We now provide the coefficients $C_{n>5}$ in decimal form to 16 digits: 
\begin{align}
C_5 &=\num{0.04199334008465763} \, \,, \nonumber \\ \nonumber
C_6 &=-\num{0.2231307406491055} \, \,, \nonumber \\
C_7 &=\num{0.6086652104241595} \, \,, \nonumber \\
C_8 &=-\num{1.161566561696015} \, \,, \nonumber \\
C_9 &=\num{1.469991268471017} \, \,, \nonumber \\
C_{10} &=-\num{1.216042897081895} \, \,, \nonumber \\
C_{11} &=\num{0.5958162964137532} \, \,, \nonumber \\
C_{12} &=-\num{0.1383507260164533} \, \,. 
\end{align}
The coefficients $C_{n\in(1,4)}$ were already provided in Eq.~\eqref{eq:C-coeffs}. All of the $A_n$, $B_k$, and $C_n$ are provided in their exact rational form in a supplemental notebook.

\section{Inversion of F(e)}
\label{app:F(e)}
Let us review our attempts at inverting 
\begin{equation}
\label{eq:sigzet}
\sigma(e)=\zeta,
\end{equation}
where 
\begin{equation}
\label{eq:sigmae}
\sigma(e)=\frac{e^{12/19}}{1-e^2}\left(1+\frac{121}{304}e^2\right)^{870/2299}.
\end{equation}
Recall that $\zeta = \sigma(e_0)(F_0/F)$ and sources with initial conditions $(e_0, F_0)$ will sample values of $\zeta \in [\sigma(e_0)(F_0/F_{\text{\rm \tiny LSO}}), \sigma(e_0)]$. In Sec.~\ref{sub:lowefinv} we invert Eq.~\eqref{eq:sigzet} in the low eccentricity and late frequency regime (i.e. $\zeta<<1$) and we find that this inversion leads to a significant loss in match for sources which sample $\zeta \sim 0.5$ which corresponds to sources with $e_0 \sim 0.3$. Since the focus of this work is to provide a model which is useful for parameter estimation (i.e. with matches $\sim 0.99$) and valid for arbitrary eccentricity, we must investigate other representations of $e(\zeta)$.

Two models for $e(\zeta)$ meet the accuracy goals laid out in Sec.~\ref{sec:Err}: (i) a model which is composed of a controlling factor obtained by introducing an approximate $\sigma(e)$ into Eq.~\eqref{eq:sigzet} and algebraically solving and then fitting the remaining error and (ii) a piecewise representation of $e(\zeta)$, which is composed of two Taylor expansions about $\zeta << 1$ and $\zeta >> 1$, and an efficient numerical fit of the function in the range in $\zeta$ for which the error in the Taylor expansions is too large for faithful modeling. The piecewise representation is the best of the two methods in speed, accuracy, and domain of validity. All of the coefficients listed in this Appendix are provided in a supplementary notebook.

\begin{figure*}[htp]
\includegraphics[clip=true,angle=0,width=0.475\textwidth]{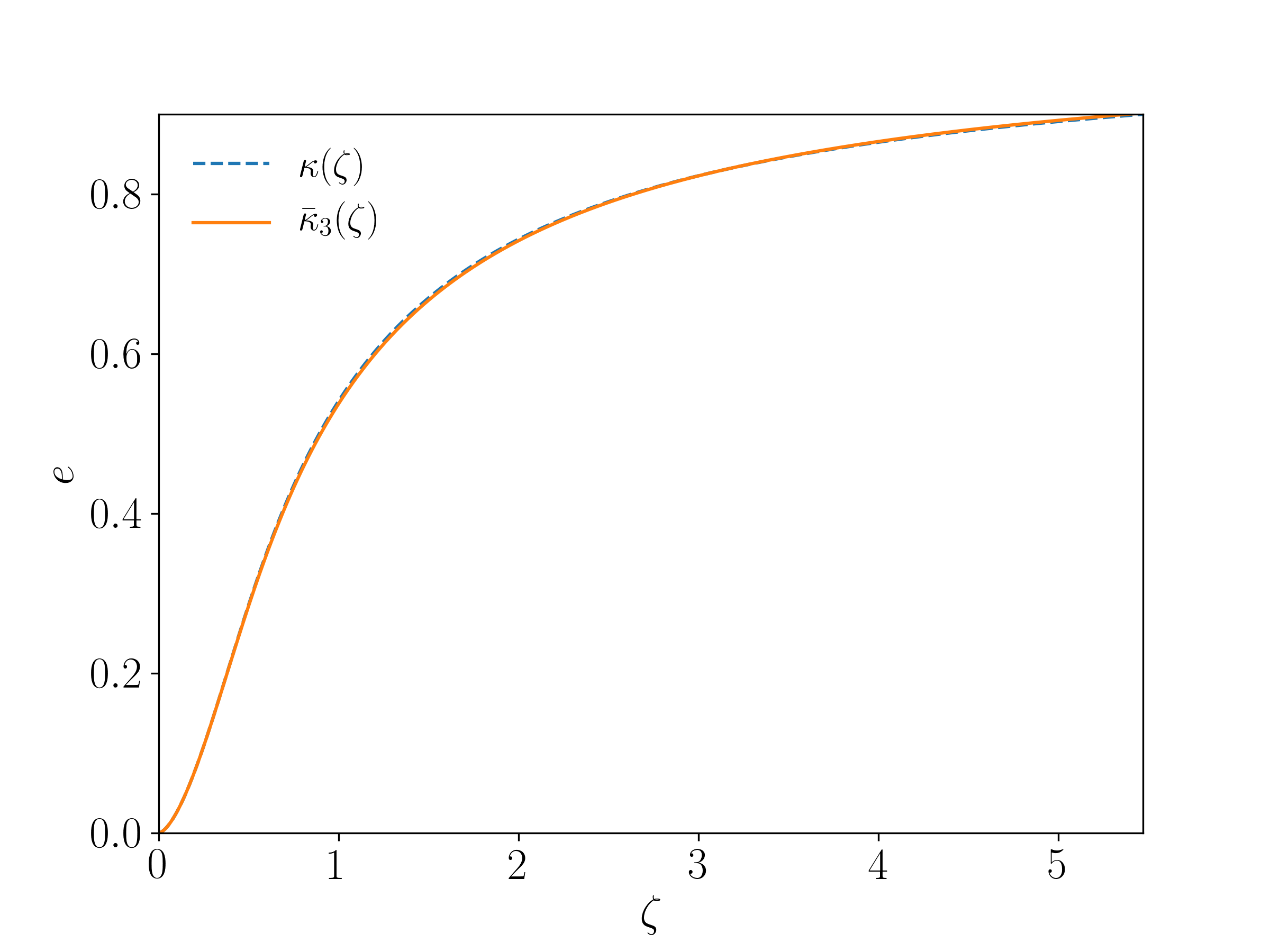} 
\includegraphics[clip=true,angle=0,width=0.475\textwidth]{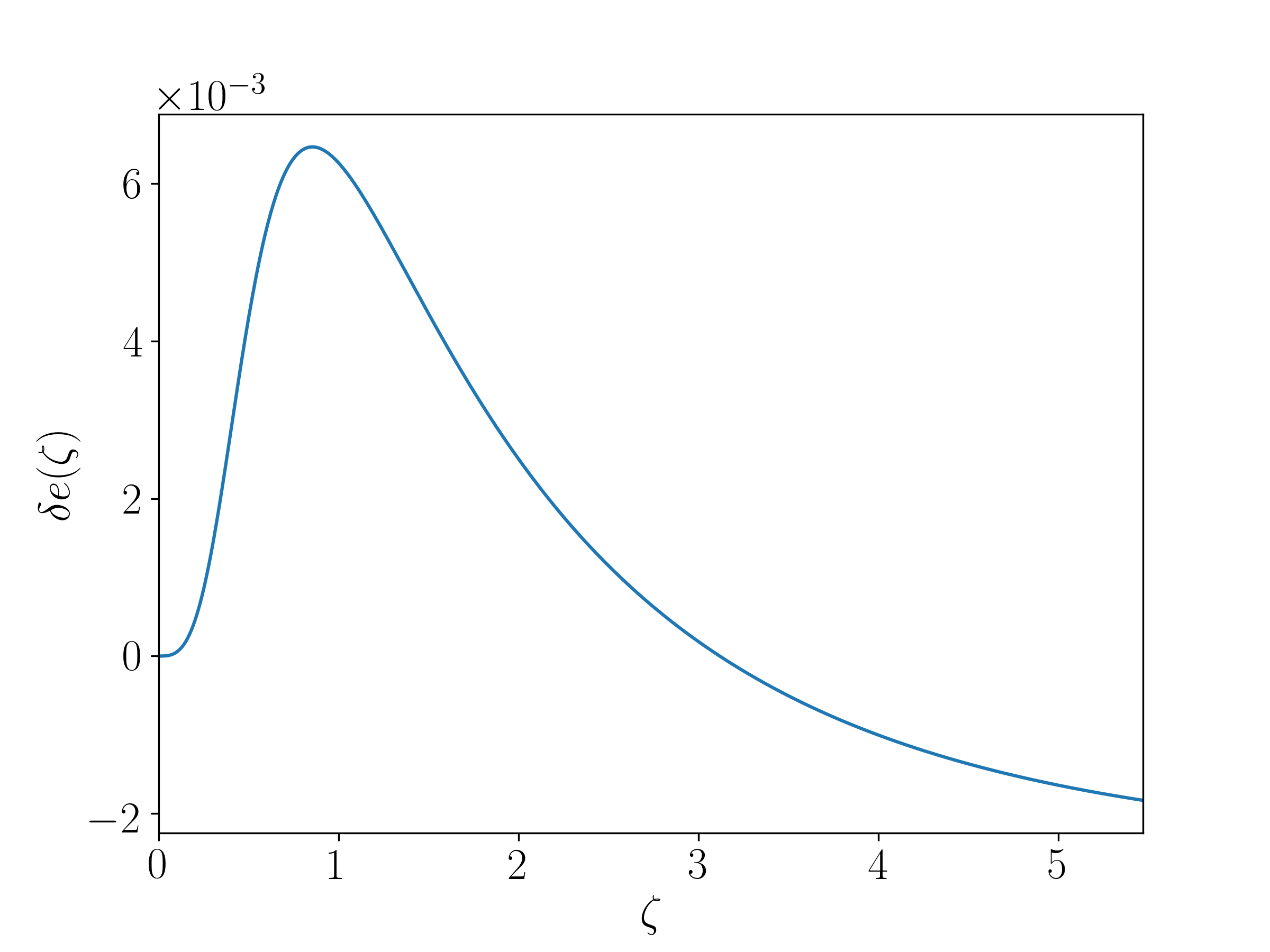} 
		\caption{\label{fig:approxinversion} (Color Online) The value of a numerical solution for $\kappa(\zeta)$, $\bar{\kappa}_{3}(\zeta)$ as given in Eq.~\eqref{eq:kappa3} (left), and the relative error between the exact solution and $\bar{\kappa}_{3}(\zeta)$ (right). We see that the two inversions are nearly indistinguishable by eye, but the relative error is $\mathcal{O}(10^{-3})$.}
\end{figure*}

As a first attempt to solve Eq.~\eqref{eq:sigzet}, we introduce
\begin{equation}
\label{eq:approxFe}
\bar{\sigma}(e)=\zeta,
\end{equation}
\begin{figure*}[htp]
\includegraphics[clip=true,angle=0,width=0.475\textwidth]{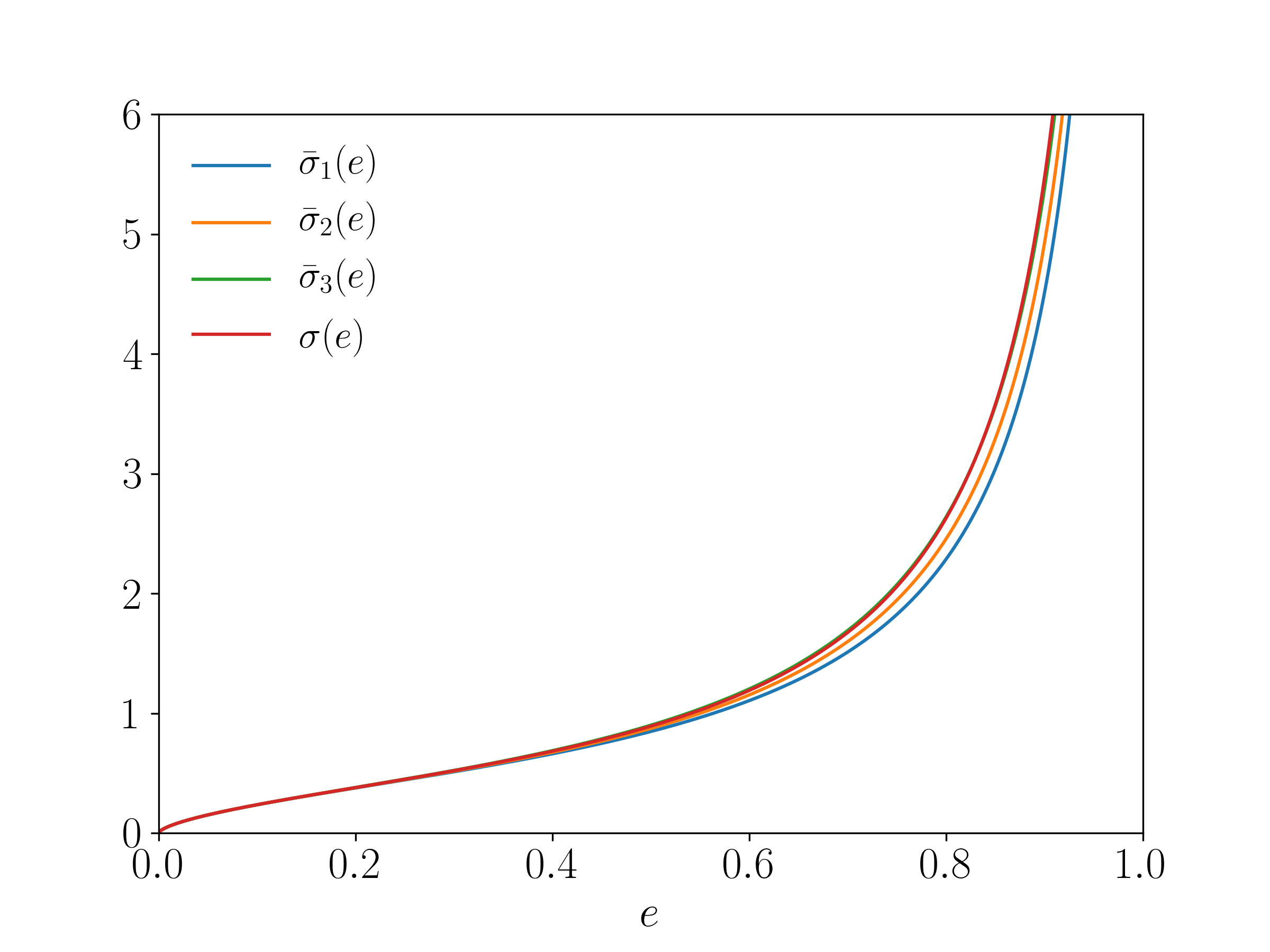} 
\includegraphics[clip=true,angle=0,width=0.475\textwidth]{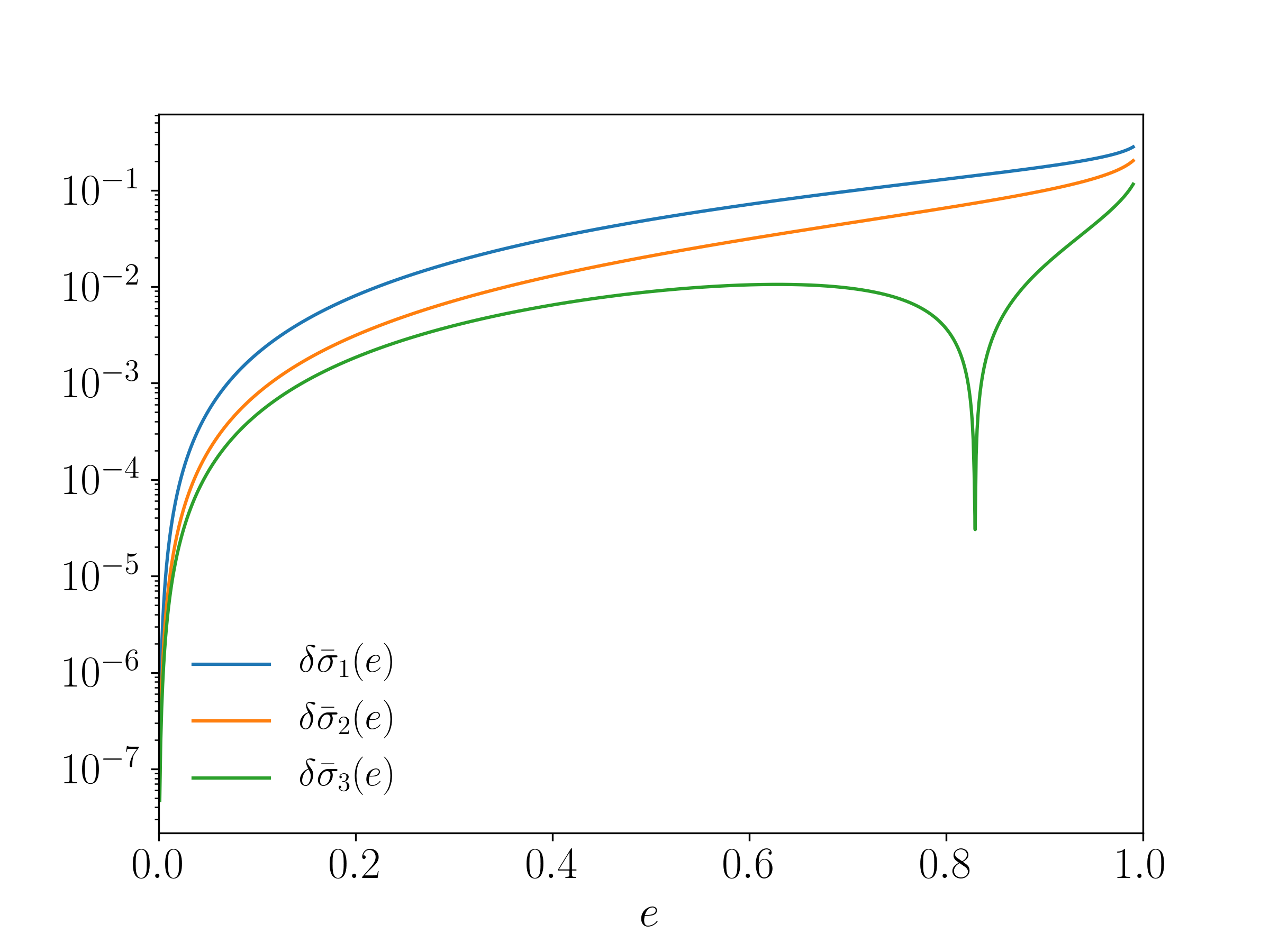} 
		\caption{\label{fig:approxsigma} (Color Online) The value of $\sigma(e)$, $\bar{\sigma}_{1}(e)$, $\bar{\sigma}_{2}(e)$, $\bar{\sigma}_{3}(e)$ (left), and their relative errors (right). Observe that $\bar{\sigma}_{3}(e)$ most closely represents $\sigma(e)$.}
\end{figure*}
where $\bar{\sigma}(e)$ is a simpler rational polynomial similar to $\sigma(e)$ which admits an exact solution. Let us define the inverse function $\bar{\kappa}(\zeta)$ such that $\bar{\kappa}[\bar{\sigma}(e)]=e$. We then seek to leverage this inverse function, which we can solve for exactly, to find an approximate solution for $e(\zeta)$ by multiplying by a Taylor series: 
\begin{equation}
\label{eq:contfac}
e(\zeta) \approx \bar{\kappa}(\zeta)\left(\sum_{k=0}^{k_{\maxtxt}}D_{k}(\zeta-\zeta_{0})^k\right) \, .
\end{equation}
The factor which is raised to the $870/2299$ power in Eq.~\eqref{eq:sigmae} varies from 1 to 1.135, so we first neglect this term. With this term set to 1, we observe that the resulting equation can be solved if the factor of $(1-e^2)$ appearing in the denominator is raised to the power $18/19$. We then raise the entire resulting equation to the power $19/6$ and we are left with a polynomial equation which can be readily algebraically solved. This corresponds to a $\bar{\sigma}(e)$
\begin{equation}
\bar{\sigma}_{1}(e)= \frac{e^{12/19}}{(1-e^2)^{18/19}} \, . \\
\end{equation}
We further note that we can incorporate the factor of $(1+\frac{121}{304}e^2)^{870/2299}$ if we approximate it as $(1+\frac{121}{304}e^2)^{3n/19}$. We generate two more $\bar{\sigma}(e)$ with different choices of $n$
\begin{align}
\bar{\sigma}_{2}(e)&= \frac{e^{12/19}}{(1-e^2)^{18/19}}\left(1+\frac{121}{304}e^2\right)^{6/19} \, , \\
\bar{\sigma}_{3}(e)&= \frac{e^{12/19}}{(1-e^2)^{18/19}}\left(1+\frac{121}{304}e^2\right)^{12/19} \, . \\
\end{align}
In Fig.~\ref{fig:approxsigma} we plot the values of $\sigma(e)$, the approximate $\bar{\sigma}(e)$, and the relative error associated with the $\bar{\sigma}(e)$. Since $\bar{\sigma}_{3}(e)$ most closely represents $\sigma(e)$, we provide its inverse function in Eq.~\eqref{eq:kappa3} of Sec.~\ref{sub:TayF2}. In Fig.~\ref{fig:approxinversion} we plot a numerical solution $\kappa(\zeta)$, $\bar{\kappa}_{3}(\zeta)$, and the associated relative error for $\zeta \in [0, 5.47]$ which corresponds to systems with $e_0$ as high as $0.9$. As shown in Sec.~\ref{sec:Err}, we require relative error of $\mathcal{O}(10^{-6})$ in order to faithfully model sources with arbitrary eccentricity. The relative error between $\bar{\kappa}_{3}(\zeta)$ and the numerical solution is $\mathcal{O}(10^{-3})$, so we conclude that we must further improve the inversion if we wish to have something accurate enough for our purposes. 

In order to decrease the relative error further, we multiply $\bar{\kappa}_{3}(\zeta)$ by a Taylor series as written in Eq.~\eqref{eq:contfac}. The coefficients $D_{k}$ are determined by substituting this $e(\zeta)$ into $\sigma[e(\zeta)]$ and demanding $\sigma[e(\zeta)]=\zeta$ upon Taylor expansion. After trying many Taylor expansions about different $\zeta_{0}$, we find that the resulting solution behaves better than $\bar{\kappa}_{3}(\zeta)$ in a small region near $\zeta_0$, but leads to considerably more error than $\bar{\kappa}_{3}(\zeta)$ for most of the domain. We conclude that this approach does not lead to an accurate enough inversion of $F(e)$ to be useful for applications that require a high value of the match.

After exhausting purely analytic techniques, we then explored semi-analytic methods to approximate $e(\zeta)$ which could be better than an interpolation in speed or domain. Inspection of the relative error, $\delta e(\zeta)$, plotted in Fig.~\ref{fig:approxinversion} and considering that
\begin{equation}
e^{\exact}(\zeta) = \bar{\kappa}_{3}(\zeta)\left(1+\delta e(\zeta) \right)
\end{equation}
suggests that we should seek a solution of the form
\begin{equation}
\label{eq:gaussfit}
e(\zeta) \approx \bar{\kappa}_{3}(\zeta)\left(1+\frac{ae^{-b\zeta^{-c}}}{\zeta^d}\right) \, .
\end{equation}
Using \texttt{Mathematica}'s \texttt{Nonlinearmodelfit} function, we are able to fit the difference in Fig.\ref{fig:approxinversion} with 
\begin{align}
a &= \num{2.9566794504712916} \times 10^{16} \, , \nonumber \\
b &= \num{42.97674936137399} \, , \nonumber \\
c &= \num{0.26033495384531447} \, ,\nonumber \\
d &= \num{11.681575454188149} \, .
\end{align} 
The relative error resulting from the use of Eq.~\eqref{eq:gaussfit} is $\mathcal{O}(10^{-4})$. Again, this level of relative error is too large to faithfully model higher eccentricity sources. We can further the remaining error using 30 Legendre polynomials to obtain
\begin{equation}
\label{eq:finfit}
e(\zeta) \approx \bar{\kappa}_{3}(\zeta)\left(1+\frac{ae^{-b\zeta^{-c}}}{\zeta^d} + \sum_{n}^{30}H_{n}L_n(\zeta)\right).
\end{equation}
After collecting like terms in $\zeta$ such that
\begin{equation}
\sum_{n}^{30}H_{n}L_n(\zeta) = \sum_{n}^{30}\bar{H}_{n}\zeta^n, 
\end{equation}
the $\bar{H_{n}}$ are 
\allowdisplaybreaks
\begin{align}
\bar{H}_{0} & = \num{ -1.7214990610968794826  }\times 10^{-6} \, , \nonumber \\
\bar{H}_{1} & = \num{ 0.00018524477170712685142  } \, , \nonumber \\
\bar{H}_{2} & = \num{ 0.0023111638941593870385  } \, , \nonumber \\
\bar{H}_{3} & = \num{  -0.27963187862683638956  } \, , \nonumber \\
\bar{H}_{4} & = \num{ 6.8872152213948038679  } \, , \nonumber \\
\bar{H}_{5} & = \num{ -64.860765273570804917  } \, , \nonumber \\
\bar{H}_{6} & = \num{ 344.56616572329823575  } \, , \nonumber \\
\bar{H}_{7} & = \num{ -1211.4465306416844664  } \, , \nonumber \\
\bar{H}_{8} & = \num{ 3064.1962215204634475  } \, , \nonumber \\
\bar{H}_{9} & = \num{  -5864.1457583432321719 } \, , \nonumber \\
\bar{H}_{10} & = \num{ 8779.6100533336715562  } \, , \nonumber \\
\bar{H}_{11} & = \num{  -10527.533182057502189 } \, , \nonumber \\
\bar{H}_{12} & = \num{ 10284.888719037608893  } \, , \nonumber \\
\bar{H}_{13} & = \num{ -8291.4807907813173717  } \, , \nonumber \\
\bar{H}_{14} & = \num{ 5568.8218377213417903 }  \, , \nonumber \\
\bar{H}_{15} & = \num{  -3137.9293091918166972 } \, , \nonumber \\
\bar{H}_{16} & = \num{  1490.8174580335099360 } \, , \nonumber \\
\bar{H}_{17} & = \num{  -599.07143152106864930 } \, , \nonumber \\
\bar{H}_{18} & = \num{  203.91776770384057994 } \, , \nonumber \\
\bar{H}_{19} & = \num{  -58.788873758558224518 } \, , \nonumber \\
\bar{H}_{20} & = \num{ 14.328519642928394012 } \, , \nonumber \\
\bar{H}_{21} & = \num{ -2.9414187725194417881  } \, , \nonumber \\
\bar{H}_{22} & = \num{ 0.50556911887490046563  } \, , \nonumber \\
\bar{H}_{23} & = \num{  -0.072125242824691859339 } \, , \nonumber \\
\bar{H}_{24} & = \num{ 0.0084352578236945894203  } \, , \nonumber \\
\bar{H}_{25} & = \num{ -0.00079467622476636165015  } \, , \nonumber \\
\bar{H}_{26} & = \num{ 0.000058795867134862938221  } \, , \nonumber \\
\bar{H}_{27} & = \num{  -3.2882024625409486629  } \times 10^{-6} \, , \nonumber \\
\bar{H}_{28} & = \num{ 1.3062561630530294326  } \times 10^{-7} \, , \nonumber \\
\bar{H}_{29} & = \num{  -3.2836547675862438974 } \times 10^{-9}  \, , \nonumber \\
\bar{H}_{30} & = \num{ 3.9254696291117736871  } \times 10^{-11} \, . 
\end{align}
With this new fit in hand, we are able to obtain the required $\mathcal{O}(10^{-6})$ relative error in inverting $F(e)$. Using a simple timing study implemented in \texttt{Mathematica}, we find that Eq.~\eqref{eq:finfit} is $\sim 100$ times slower to evaluate than an interpolated numerical solution for $e(\zeta)$.

We can obtain a better approximation to $e(\zeta)$ than Eq.~\eqref{eq:finfit} in speed, accuracy, and domain of validity if we use a piecewise representation of $e(\zeta)$ composed of a Taylor expansion for $\zeta << 1$, a numerical fit, and a Taylor expansion for $\zeta >> 1$. This is constructed by imposing a maximum relative error (which we choose to be $10^{-6}$ informed by the results of Sec.~\ref{sec:Err}), and determining the domain of the components of the piecewise function according to where the Taylor expansions exceed this error. The domain in $\zeta$ that is not well approximated by either of the two Taylor expansions is then fit numerically, which finally results in 
  \begin{align}
  \label{eq:piecewise}
e(\zeta) \approx \left\{ \begin{array}{cc} 
                e_{\text{\rm \tiny Low}}(\zeta) & \hspace{5mm} \zeta \leq \zeta_{\text{\rm \tiny Low}} \\
                e_{\text{\rm \tiny Mid}}(\zeta) & \hspace{5mm} \zeta_{\text{\rm \tiny Low}} < \zeta < \zeta_{\text{\rm \tiny High}} \\
                e_{\text{\rm \tiny High}}(\zeta) & \hspace{5mm} \zeta \geq \zeta_{\text{\rm \tiny High}} \\
                \end{array} \right.
\end{align}
The advantage of this method is that it is valid for all $\zeta$ and thus all $e_0$, whereas both the numerical fit in Eq.~\eqref{eq:finfit} and an interpolated numerical solution are only valid on the domain on which they are constructed which could be computationally expensive as $e_0$ becomes large. For example, an $e_0$ of 0.95 (0.99) corresponds to a domain in $\zeta$ that is as high as $\sim$11 (57). 

For the Taylor expansion valid for $\zeta << 1$ we follow the prescription laid out in Sec.~\ref{sub:lowefinv} and we obtain 
\begin{equation}
\label{eq:ezeta}
e_{\text{\rm \tiny Low}}(\zeta) \approx \zeta^{19/12}\sum_{i=0}^{i_{\maxtxt}}M_n\zeta^{19n/6}. 
\end{equation}
In the spirit of a quick to evaluate representation of $e(\zeta)$ we keep only 9 terms in this Taylor expansion. They are 
\begin{align}
M_{0} &= 1 \, , \nonumber \\
M_{1} &= -1.821820175438596 \, , \nonumber \\
M_{2} &= 7.553367231984841 \, , \nonumber \\
M_{3} &= -40.43088984555313 \, , \nonumber \\
M_{4} &= 245.5334064974539 \, , \nonumber \\
M_{5} &= -1607.942508437681 \, , \nonumber \\
M_{6} &= 11072.13243739240 \, , \nonumber \\
M_{7} &= -79021.97843134498 \, , \nonumber \\
M_{8} &= 579327.2428554708 \, .
\end{align}
The value of $\zeta$ at which this Taylor expansion reaches a relative error of $\mathcal{O}(10^{-6})$ sets $\zeta_{\text{\rm \tiny Low}} = 0.363$.

In order to obtain $e_{\text{\rm \tiny High}}(\zeta)$ we need to introduce a small parameter $\delta = 1-e^2$. For $\zeta >> 1$, $e \sim 1$, and $\delta << 1$. We re-express $\sigma(e)$ as $\sigma(\delta)$ and propose a solution for $\delta$ (and thus $e$) as 
\begin{equation}
\delta(\xi) = \xi \sum_{n}^{n_{\maxtxt}}N_n \xi^{n} \,
\end{equation}
where $\xi= 1/\zeta$. Since $\sigma(\delta) \rightarrow \infty$ as $\delta \rightarrow 0$, we solve for the coefficients $N_n$ by demanding that $1/\sigma[\delta(\xi)]=\xi$ upon a Taylor expansion about small $\xi$. Finally we substitute this $\delta(\xi)$ into $e_{\text{\rm \tiny High}}(\xi) = \left[1+\delta(\xi)\right]^{1/2}$, expand once more in small $\xi$ and substitute $\xi= 1/\zeta$ to obtain
\begin{equation}
e_{\text{\rm \tiny High}}(\zeta) = 1+\frac{1}{\zeta}\sum_{n=0}^{8}X_n \frac{1}{\zeta^{n}} \, ,
\end{equation}
where the coefficients $X_n$ are
\begin{align}
X_{0} &= \num{ -0.5675925779937008 } \, , \nonumber \\
X_{1} &= \num{ 0.1118089337706976 } \, , \nonumber \\
X_{2} &= \num{ -0.0066334474512642514 } \, , \nonumber \\
X_{3} &= \num{ 0.004340040509348508 } \, , \nonumber \\
X_{4} &= \num{ -0.0002731834200518127 } \, , \nonumber \\
X_{5} &= \num{ -0.0005482671585155188 } \, , \nonumber \\
X_{6} &= \num{ -0.0001375254892966566} \, , \nonumber \\
X_{7} &= \num{ 5.9506405977837105 } \times 10^{-6}\, , \nonumber \\
X_{8} &= \num{ 0.000029451819538605622 } \, . 
\end{align}
The $e_{\text{\rm \tiny High}}(\zeta)$ reaches a relative error of $\mathcal{O}(10^{-6})$ around $\zeta = 1.34$, which sets $\zeta_{\text{\rm \tiny High}} = 1.34$.

Since both the Taylor expansions about $\zeta << 1 $ and $\zeta >> 1$ have relative errors that exceed our threshold of $\mathcal{O}(10^{-6})$ for $\zeta \in [0.363, 1.34]$, we choose to numerically fit $e(\zeta)$ in this regime using \texttt{Mathematica}'s \texttt{Nonlinearmodelfit}. We fit a model given by
\begin{equation}
e_{\text{\rm \tiny Mid}} = \zeta^{\beta}\sum_{n=0}^{n_{\maxtxt}}V_{n}\zeta^{n}+\gamma.
\end{equation}
We can fit this regime to $\mathcal{O}(10^{-6})$ accuracy in relative error by using an $n_{\maxtxt}=8$. Our fitted values are $\beta=1.983025291515884$, $\gamma=258.9923485636087$, and 
\begin{align}
V_{0} &= \num{ 0.017784933490872047 } \, , \nonumber \\
V_{1} &= \num{  -0.6471012098142174 } \, , \nonumber \\
V_{2} &= \num{ -266.05126463952934 } \, , \nonumber \\
V_{3} &= \num{ 16.342347642545178 } \, , \nonumber \\
V_{4} &= \num{ -14.73676142107974 } \, , \nonumber \\
V_{5} &= \num{ 10.203956907570282 } \, , \nonumber \\
V_{6} &= \num{ -4.701828427817006 } \, , \nonumber \\
V_{7} &= \num{ 1.2752146500776196 } \, , \nonumber \\
V_{8} &= \num{ -0.1536871497323068 } \, . 
\end{align}
The final piecewise function written in Eq.~\eqref{eq:piecewise} is 3 times slower to evaluate than an interpolation of the numerical $e(\zeta)$, but has the advantage that it covers the entire domain of $\zeta \in [0, \infty]$. In Fig.~\ref{fig:fits} of Sec.~\ref{sub:TayF2} we plot the relative error of both the Legendre fit and the piecewise representations of $e(\zeta)$. The piecewise representation of $e(\zeta)$ is more accurate and faster than the fit using Legendre polynomials displayed in Eq.~\eqref{eq:finfit}. This is the best analytic approximation to $e(\zeta)$ which we were able to obtain. 
\bibliography{master}
\end{document}